%% file: root.tex
\def\compileforpublish{1}

\documentclass[journal,twoside]{IEEEtran}  %

\IEEEoverridecommandlockouts                              %

\input{packages}

\input{macros}

\newcommand{\titletext}{Actuator Fault-Tolerant Vehicle Motion Control: A Survey}

\begin{document}

\title{\LARGE \bf
	\titletext%
}

\author{Torben Stolte$^{1}$

\thanks{Manuscript received AAAAAA BB, CCCC; revised DDDDDD EE, FFFF. This research is accomplished within the project ``UNICAR\emph{agil}'' (FKZ 16EMO0285)~\cite{woopen_2018}. 
	We acknowledge the financial support for the project by the Federal Ministry of Education and Research of Germany (BMBF).} 

\thanks{\hspace{-1em}$^{1}$The author is with the Institute of Control Engineering at Technische Universität Braunschweig, Braunschweig, Germany. {\tt\small stolte@ifr.ing.tu-bs.de}\vspace{0.3em}}%
}

\markboth{Transactions on Intelligent Vehicles,~Vol.~AA, No.~B, MONTH~YEAR}%
{Stolte \MakeLowercase{\textit{et al.}}: \titletext}

\maketitle

\begin{abstract}%
	\input{abstract}
\end{abstract}%

\IEEEpeerreviewmaketitle

\copyrightnotice

\section{Introduction}
\input{introduction}

\section{Focus of the survey}
\input{focus}

\section{Degradations}
\input{degradations}

\section{Actuator topologies}
\input{topologies}

\section{Control approaches}
\input{approaches}

\section{Experiments}
\input{experiments}

\section{Conclusion and outlook}
\input{conclusion}

\section*{Acknowledgment}
The author would like to thank Dr.~Marcus Grobe, Dr.~Sonja Luther, Christina Harms, Marcus Nolte, Liren Wu, and Prof.~Markus Maurer for their support.

\renewcommand*{\bibfont}{\footnotesize} 
\printbibliography 
\begin{IEEEbiography}{Torben Stolte}
	\input{stolte}
\end{IEEEbiography}

\end{document}

%% file: packages.tex
\usepackage[utf8]{inputenc}
\usepackage{tubscolors}

\usepackage{tikz}
\usetikzlibrary{shapes,
				automata,
				arrows,
				arrows.meta,
				matrix,
				backgrounds,
				fit,
				patterns,
				decorations.markings, 
				svg.path, 
				shapes.multipart, 
				shapes.geometric, 
				external, 
				shadows, 
				patterns,
				positioning, 
				calc, 
				decorations, 
				scopes,
				fadings,
				bending,
				scopes}
\usetikzlibrary{external}

\usepackage{pgfplots}
\pgfplotsset{compat=newest} 
\usepackage{pgfplotstable}

\usepackage{enumitem}

\usepackage{siunitx}
\usepackage[%
			bibstyle=ieee,
			citestyle=numeric-comp,
			backend=biber,
			minnames=1,
			maxcitenames=2,
			maxbibnames=6,
			isbn=false,
			url=false,
			natbib=true,
			clearlang=true,
			autolang=hyphen,
			date=year
			]{biblatex} 

\DeclareSourcemap{
	\maps[datatype=biber]{
		\map{
			\step[fieldsource=note, final]
			\step[fieldset=addendum, origfieldval, final]
			\step[fieldset=note, null]
		}
	}
}

\input{IEEEpatch.tex}

\usepackage{xpatch}
\xpatchbibdriver{online}
{\printtext[parens]{\usebibmacro{date}}}
{\iffieldundef{year}
	{}
	{\printtext[parens]{\usebibmacro{date}}}}
{}
{\typeout{There was an error patching biblatex-ieee (specifically, ieee.bbx's @online driver)}}

\usepackage[caption=false, font=footnotesize]{subfig}
\usepackage{xifthen}
\usepackage{lipsum}

\usepackage{amsmath}
\let\vec\mathbf
\DeclareMathOperator{\diag}{diag}  
\usepackage{amssymb} 	%
\usepackage{bm} 		%
\usepackage{stackengine}
\usepackage{booktabs} 	%
\usepackage{multirow}   %
\usepackage{array}
\usepackage{lmodern}
\usepackage{subfig}
\usepackage{footmisc}
\usepackage{microtype}
\usepackage{fmtcount}
\usepackage{xifthen}
\usepackage{flushend} 
\usepackage[all]{nowidow}

\usepackage[hidelinks]{hyperref}

\AtBeginDocument{} 
\AtBeginDocument{}

%% file: IEEEpatch.tex
\usepackage{xpatch}
\usepackage{xstring}

\DeclareSourcemap{
	\maps[datatype=bibtex]{
		\map{
			\step[fieldsource=langid, match=\regexp{\A(n)?((swiss)?german|austrian)\Z}, final]
			\step[fieldset=language, fieldvalue={(in German)}]
		}
	}
}

\xpatchbibmacro{publisher+location+date}
	{\usebibmacro{date}}
		{	\ifentrytype{inproceedings}{
				\IfSubStr{\strfield{booktitle}}{\strfield{year}}{}{\usebibmacro{date}} %
			}
		}
	{}
	{\typeout{There was an error patching biblatex-ieee (specifically, ieee.bbx's @inproceedings driver)}}

\xpatchbibdriver{online}
{\printtext[parens]{\usebibmacro{date}}}
	{\iffieldundef{year}
		{}
		{\printtext[parens]{\usebibmacro{date}}}}
	{}
{\typeout{There was an error patching biblatex-ieee (specifically, ieee.bbx's @online driver)}}

\DeclareSourcemap{
	\maps[datatype=biber]{
		\map{
			\step[fieldsource=note, final]
			\step[fieldset=addendum, origfieldval, final]
			\step[fieldset=note, null]
		}
	}
}

\DeclareBibliographyDriver{standard}{%
	\usebibmacro{bibindex}%
	\usebibmacro{begentry}%
	\usebibmacro{maintitle+title}%
	\setunit{\addcomma\addspace}%
	\usebibmacro{publisher+type+number}%
	\setunit{\labelnamepunct}\newblock
	\IfSubStr{\strfield{number}}{\strfield{year}}{}{\usebibmacro{date}} %
	\newunit\newblock
	\usebibmacro{finentry}%
}

\newbibmacro*{publisher+type+number}{%
	\printtext{%
		\printlist{publisher}
		\iflistundef{publisher}
		{\space}{}%
		\iffieldundef{type}
		{Standard}
		{\printfield{type}}
		\printfield{number}
	}%
}

\DeclareSourcemap{
	\maps[datatype=bibtex]{
		\map[overwrite, foreach={booktitle,journaltitle,eventtitle,series,publisher,institution,type}]{
			\step[fieldsource=\regexp{$MAPLOOP}, match={XXX}, replace={Acad. Serbe Sci. Arts Glas Cl. Sci. Tech.}]
			\step[fieldsource=\regexp{$MAPLOOP}, match={XXX}, replace={Acta Acust.}]
			\step[fieldsource=\regexp{$MAPLOOP}, match={XXX}, replace={Acta Astron. (Poland)}]
			\step[fieldsource=\regexp{$MAPLOOP}, match={XXX}, replace={Acta Astron. Sin. (China)}]
			\step[fieldsource=\regexp{$MAPLOOP}, match={XXX}, replace={Acta Astronaut. (U.K.)}]
			\step[fieldsource=\regexp{$MAPLOOP}, match={XXX}, replace={Acta Astrophys. Sin. (China)}]
			\step[fieldsource=\regexp{$MAPLOOP}, match={XXX}, replace={Acta Autom. Sin.}]
			\step[fieldsource=\regexp{$MAPLOOP}, match={XXX}, replace={Acta Cienc. Indica Math.}]
			\step[fieldsource=\regexp{$MAPLOOP}, match={XXX}, replace={Acta Cienc. Indica Phys.}]
			\step[fieldsource=\regexp{$MAPLOOP}, match={XXX}, replace={Acta Crystallogr. A, Found. Crystallogr.}]
			\step[fieldsource=\regexp{$MAPLOOP}, match={XXX}, replace={Acta Crystallogr. B, Struct. Sci.}]
			\step[fieldsource=\regexp{$MAPLOOP}, match={XXX}, replace={Acta Crystallogr. C, Cryst. Struct. Commun.}]
			\step[fieldsource=\regexp{$MAPLOOP}, match={XXX}, replace={Acta Electron. (France)}]
			\step[fieldsource=\regexp{$MAPLOOP}, match={XXX}, replace={Acta Cyberno}]
			\step[fieldsource=\regexp{$MAPLOOP}, match={XXX}, replace={Acta Electron. Sin. (China)}]
			\step[fieldsource=\regexp{$MAPLOOP}, match={XXX}, replace={Acta Geod. Geophys. Montan. Hung.}]
			\step[fieldsource=\regexp{$MAPLOOP}, match={XXX}, replace={Acta Geophys. Pol.}]
			\step[fieldsource=\regexp{$MAPLOOP}, match={XXX}, replace={Acta Geophys. Sin. (China)}]
			\step[fieldsource=\regexp{$MAPLOOP}, match={XXX}, replace={Acta Geophys. Sin. (USA)}]
			\step[fieldsource=\regexp{$MAPLOOP}, match={XXX}, replace={Acta Metall.}]
			\step[fieldsource=\regexp{$MAPLOOP}, match={XXX}, replace={Acta Mex. Cienc. Tecnol.}]
			\step[fieldsource=\regexp{$MAPLOOP}, match={XXX}, replace={Acta Phys. Hung.}]
			\step[fieldsource=\regexp{$MAPLOOP}, match={XXX}, replace={Acta Phys. Pol. A}]
			\step[fieldsource=\regexp{$MAPLOOP}, match={XXX}, replace={Acta Phys. Pol. B}]
			\step[fieldsource=\regexp{$MAPLOOP}, match={XXX}, replace={Acta Phys. Sin.}]
			\step[fieldsource=\regexp{$MAPLOOP}, match={XXX}, replace={Acta Phys. Slovaca}]
			\step[fieldsource=\regexp{$MAPLOOP}, match={XXX}, replace={Acta Politec. Mex.}]
			\step[fieldsource=\regexp{$MAPLOOP}, match={XXX}, replace={Acta Polytech. Scand. Appl. Phys. Ser.}]
			\step[fieldsource=\regexp{$MAPLOOP}, match={XXX}, replace={Acta Polytech. Scand. Chem. Technol.}]
			\step[fieldsource=\regexp{$MAPLOOP}, match={XXX}, replace={Metall. Ser.}]
			\step[fieldsource=\regexp{$MAPLOOP}, match={XXX}, replace={Acta Polytech. Scand. Electr. Eng. Ser.}]
			\step[fieldsource=\regexp{$MAPLOOP}, match={XXX}, replace={Acta Polytech. Scand. Math. Comput. Sci. Ser.}]
			\step[fieldsource=\regexp{$MAPLOOP}, match={XXX}, replace={Acta Polytech. Scand. Mech. Eng. Ser.}]
			\step[fieldsource=\regexp{$MAPLOOP}, match={XXX}, replace={Acta Seismol. Sin.}]
			\step[fieldsource=\regexp{$MAPLOOP}, match={XXX}, replace={Acta Tech. Acad. Sci. Hung.}]
			\step[fieldsource=\regexp{$MAPLOOP}, match={XXX}, replace={Acta Tech. CSAV}]
			\step[fieldsource=\regexp{$MAPLOOP}, match={XXX}, replace={Acustica}]
			\step[fieldsource=\regexp{$MAPLOOP}, match={XXX}, replace={(AEU) Arch. Elektr. Ubertragung}]
			\step[fieldsource=\regexp{$MAPLOOP}, match={XXX}, replace={Akust. Zh.}]
			\step[fieldsource=\regexp{$MAPLOOP}, match={XXX}, replace={Algorithmica}]
			\step[fieldsource=\regexp{$MAPLOOP}, match={XXX}, replace={Alta Freq.}]
			\step[fieldsource=\regexp{$MAPLOOP}, match={XXX}, replace={An. Acad. Bras. Cienc.}]
			\step[fieldsource=\regexp{$MAPLOOP}, match={XXX}, replace={An. Fis.}]
			\step[fieldsource=\regexp{$MAPLOOP}, match={XXX}, replace={An. Mec. Electr.}]
			\step[fieldsource=\regexp{$MAPLOOP}, match={XXX}, replace={Angew. Inform.}]
			\step[fieldsource=\regexp{$MAPLOOP}, match={XXX}, replace={Ann. Inst. Henri Poincare Phys. Theor.}]
			\step[fieldsource=\regexp{$MAPLOOP}, match={XXX}, replace={Ann. Soc. Sci. Brux. I, Sci. Math. Astron. Phys.}]
			\step[fieldsource=\regexp{$MAPLOOP}, match={XXX}, replace={Arch. Elektr. Uebertrag. (AEU)}] %
			\step[fieldsource=\regexp{$MAPLOOP}, match={XXX}, replace={Arch. Elektron. Uebertrag. Tech.}] %
			\step[fieldsource=\regexp{$MAPLOOP}, match={XXX}, replace={Arch. Elektrotech. (Poland)}]
			\step[fieldsource=\regexp{$MAPLOOP}, match={XXX}, replace={Arch. Elektrotech. (Germany)}]
			\step[fieldsource=\regexp{$MAPLOOP}, match={XXX}, replace={Ark. Fys. Semin. Trondheim}]
			\step[fieldsource=\regexp{$MAPLOOP}, match={XXX}, replace={Astrofizika}]
			\step[fieldsource=\regexp{$MAPLOOP}, match={XXX}, replace={Astron. Nachr. (Germany)}]
			\step[fieldsource=\regexp{$MAPLOOP}, match={XXX}, replace={Astron. Tidsskr.}]
			\step[fieldsource=\regexp{$MAPLOOP}, match={XXX}, replace={Astron. Vestn.}]
			\step[fieldsource=\regexp{$MAPLOOP}, match={XXX}, replace={Astron. Zh.}]
			\step[fieldsource=\regexp{$MAPLOOP}, match={XXX}, replace={Atomwirtsch.-Atomtech.}]
			\step[fieldsource=\regexp{$MAPLOOP}, match={XXX}, replace={Atti Accad. Sci. Ist. Bologna CI. Sci. Fis.}]
			\step[fieldsource=\regexp{$MAPLOOP}, match={XXX}, replace={Rend. XIII}]
			\step[fieldsource=\regexp{$MAPLOOP}, match={XXX}, replace={Atti Accad. Sci. Torino I, CI. Sci. Fis. Math. Nat.}]
			\step[fieldsource=\regexp{$MAPLOOP}, match={XXX}, replace={Autom. Strum.}]
			\step[fieldsource=\regexp{$MAPLOOP}, match={XXX}, replace={Autom. Tech. Prax.}]
			\step[fieldsource=\regexp{$MAPLOOP}, match={XXX}, replace={Automatica}]
			\step[fieldsource=\regexp{$MAPLOOP}, match={XXX}, replace={Automatie}]
			\step[fieldsource=\regexp{$MAPLOOP}, match={XXX}, replace={Automatika}]
			\step[fieldsource=\regexp{$MAPLOOP}, match={XXX}, replace={Automatisierungstechnik}]
			\step[fieldsource=\regexp{$MAPLOOP}, match={XXX}, replace={Automatizace}]
			\step[fieldsource=\regexp{$MAPLOOP}, match={XXX}, replace={Automedica}]
			\step[fieldsource=\regexp{$MAPLOOP}, match={XXX}, replace={Avtom. Telemekh.}]
			\step[fieldsource=\regexp{$MAPLOOP}, match={XXX}, replace={Avtom. Vychisl. Tekh.}]
			\step[fieldsource=\regexp{$MAPLOOP}, match={XXX}, replace={Avtomatika}]
			\step[fieldsource=\regexp{$MAPLOOP}, match={XXX}, replace={Avtometriya}]
			\step[fieldsource=\regexp{$MAPLOOP}, match={XXX}, replace={Ber. Bunsenges. Phys. Chem.}]
			\step[fieldsource=\regexp{$MAPLOOP}, match={XXX}, replace={Biofizika}]
			\step[fieldsource=\regexp{$MAPLOOP}, match={XXX}, replace={Biometrika}]
			\step[fieldsource=\regexp{$MAPLOOP}, match={XXX}, replace={Boll. Geofis. Teor. Appl.}]
			\step[fieldsource=\regexp{$MAPLOOP}, match={XXX}, replace={Bull. Acad. Serbe Sci. Arts cl. Sci. Tech.}]
			\step[fieldsource=\regexp{$MAPLOOP}, match={XXX}, replace={Bull. Annu. Soc. Suisse Chronom. Lab. Suisse.}]
			\step[fieldsource=\regexp{$MAPLOOP}, match={XXX}, replace={Rech. Horlog.}]
			\step[fieldsource=\regexp{$MAPLOOP}, match={XXX}, replace={Bull. Cl. Sci. Acad. R. Belg.}]
			\step[fieldsource=\regexp{$MAPLOOP}, match={XXX}, replace={Bull. Dir. Etud. Rech. A}]
			\step[fieldsource=\regexp{$MAPLOOP}, match={XXX}, replace={Bull. Dir. Etud. Rech. B}]
			\step[fieldsource=\regexp{$MAPLOOP}, match={XXX}, replace={Bull. Dir. Etud. Rech. C}]
			\step[fieldsource=\regexp{$MAPLOOP}, match={XXX}, replace={Bull. Liaison Rech. Inform. Autom.}]
			\step[fieldsource=\regexp{$MAPLOOP}, match={XXX}, replace={Bur. Etud. Autom.}]
			\step[fieldsource=\regexp{$MAPLOOP}, match={XXX}, replace={CFI-Ceram. Forum Int.-Ber. Dtsch. Keram. Ges.}]
			\step[fieldsource=\regexp{$MAPLOOP}, match={XXX}, replace={Chem. Scr.}]
			\step[fieldsource=\regexp{$MAPLOOP}, match={XXX}, replace={Ciel Terre}]
			\step[fieldsource=\regexp{$MAPLOOP}, match={XXX}, replace={Cybernetica}]
			\step[fieldsource=\regexp{$MAPLOOP}, match={XXX}, replace={Deut. Hydrogr. Z.}]
			\step[fieldsource=\regexp{$MAPLOOP}, match={XXX}, replace={Dokl. Akad. Nauk SSSR}]
			\step[fieldsource=\regexp{$MAPLOOP}, match={XXX}, replace={Electroacoustique}]
			\step[fieldsource=\regexp{$MAPLOOP}, match={XXX}, replace={Electrochim. Acta}]
			\step[fieldsource=\regexp{$MAPLOOP}, match={XXX}, replace={Electrochim. Metal.}]
			\step[fieldsource=\regexp{$MAPLOOP}, match={XXX}, replace={Elektor Electron.}]
			\step[fieldsource=\regexp{$MAPLOOP}, match={XXX}, replace={Elektr. Bahnen}]
			\step[fieldsource=\regexp{$MAPLOOP}, match={XXX}, replace={Elektr. Energ.-Tech.}]
			\step[fieldsource=\regexp{$MAPLOOP}, match={XXX}, replace={Elektr. Masch.}]
			\step[fieldsource=\regexp{$MAPLOOP}, match={XXX}, replace={Elektr. Stn.}]
			\step[fieldsource=\regexp{$MAPLOOP}, match={XXX}, replace={Elektrichestvo}]
			\step[fieldsource=\regexp{$MAPLOOP}, match={XXX}, replace={Elektrie}]
			\step[fieldsource=\regexp{$MAPLOOP}, match={XXX}, replace={Elektrizitaetswirtschaft}]
			\step[fieldsource=\regexp{$MAPLOOP}, match={XXX}, replace={Elektro}]
			\step[fieldsource=\regexp{$MAPLOOP}, match={XXX}, replace={Elektro-Anz.}]
			\step[fieldsource=\regexp{$MAPLOOP}, match={XXX}, replace={Elektro-Jahr}]
			\step[fieldsource=\regexp{$MAPLOOP}, match={XXX}, replace={Elektrokhimiya}]
			\step[fieldsource=\regexp{$MAPLOOP}, match={XXX}, replace={Elektron}]
			\step[fieldsource=\regexp{$MAPLOOP}, match={XXX}, replace={Elektron Int.}]
			\step[fieldsource=\regexp{$MAPLOOP}, match={XXX}, replace={Elektron. Entwitkl.}]
			\step[fieldsource=\regexp{$MAPLOOP}, match={XXX}, replace={Elektron. Ind}]
			\step[fieldsource=\regexp{$MAPLOOP}, match={XXX}, replace={Elektron. J.}]
			\step[fieldsource=\regexp{$MAPLOOP}, match={XXX}, replace={Elektron. Prax.}]
			\step[fieldsource=\regexp{$MAPLOOP}, match={XXX}, replace={Elektron. Tekh.}]
			\step[fieldsource=\regexp{$MAPLOOP}, match={XXX}, replace={Elektronica}]
			\step[fieldsource=\regexp{$MAPLOOP}, match={XXX}, replace={Elektronik}]
			\step[fieldsource=\regexp{$MAPLOOP}, match={XXX}, replace={Elektronika}]
			\step[fieldsource=\regexp{$MAPLOOP}, match={XXX}, replace={Elektroniker}]
			\step[fieldsource=\regexp{$MAPLOOP}, match={XXX}, replace={Elektronikschau}]
			\step[fieldsource=\regexp{$MAPLOOP}, match={XXX}, replace={Elektrosvyaz}]
			\step[fieldsource=\regexp{$MAPLOOP}, match={XXX}, replace={Elektrotech. Cas.}]
			\step[fieldsource=\regexp{$MAPLOOP}, match={Elektrotechnik und Informationstechnik}, replace={Elektrotech. Inf. Tech.}]
			\step[fieldsource=\regexp{$MAPLOOP}, match={XXX}, replace={Elektrotech. Obz.}]
			\step[fieldsource=\regexp{$MAPLOOP}, match={XXX}, replace={Elektrotechniek}]
			\step[fieldsource=\regexp{$MAPLOOP}, match={XXX}, replace={Elektrotechnik (Czechoslovakia)}]
			\step[fieldsource=\regexp{$MAPLOOP}, match={XXX}, replace={Elektrotechnik (Switzerland)}]
			\step[fieldsource=\regexp{$MAPLOOP}, match={XXX}, replace={Elektrotechnik (Germany)}]
			\step[fieldsource=\regexp{$MAPLOOP}, match={XXX}, replace={Elektrotechnika}]
			\step[fieldsource=\regexp{$MAPLOOP}, match={XXX}, replace={Elektrotehnika, Zagreb}]
			\step[fieldsource=\regexp{$MAPLOOP}, match={XXX}, replace={Elektrotekhnika}]
			\step[fieldsource=\regexp{$MAPLOOP}, match={XXX}, replace={Elektroteknikeren}]
			\step[fieldsource=\regexp{$MAPLOOP}, match={XXX}, replace={Elektrowaerme Int. B.}]
			\step[fieldsource=\regexp{$MAPLOOP}, match={XXX}, replace={Elettrificazione}]
			\step[fieldsource=\regexp{$MAPLOOP}, match={XXX}, replace={Elettron. Oggi}]
			\step[fieldsource=\regexp{$MAPLOOP}, match={XXX}, replace={Elettron. Telecomun.}]
			\step[fieldsource=\regexp{$MAPLOOP}, match={XXX}, replace={Elettrotecnica}]
			\step[fieldsource=\regexp{$MAPLOOP}, match={XXX}, replace={Elek. Med Aktuell Elektron.}]
			\step[fieldsource=\regexp{$MAPLOOP}, match={XXX}, replace={Elteknik}]
			\step[fieldsource=\regexp{$MAPLOOP}, match={XXX}, replace={Energ. Atomtech.}]
			\step[fieldsource=\regexp{$MAPLOOP}, match={XXX}, replace={Energ. Elettr.}]
			\step[fieldsource=\regexp{$MAPLOOP}, match={XXX}, replace={Energetica}]
			\step[fieldsource=\regexp{$MAPLOOP}, match={XXX}, replace={Energetik}]
			\step[fieldsource=\regexp{$MAPLOOP}, match={XXX}, replace={Energetika}]
			\step[fieldsource=\regexp{$MAPLOOP}, match={XXX}, replace={Energetyka}]
			\step[fieldsource=\regexp{$MAPLOOP}, match={XXX}, replace={Energia Nuclear}]
			\step[fieldsource=\regexp{$MAPLOOP}, match={XXX}, replace={Energie Technik (Germany)}]
			\step[fieldsource=\regexp{$MAPLOOP}, match={XXX}, replace={Energie Technik (Switzerland)}]
			\step[fieldsource=\regexp{$MAPLOOP}, match={XXX}, replace={Entropie}]
			\step[fieldsource=\regexp{$MAPLOOP}, match={XXX}, replace={ETZ}]
			\step[fieldsource=\regexp{$MAPLOOP}, match={XXX}, replace={ETZ Arch.}]
			\step[fieldsource=\regexp{$MAPLOOP}, match={XXX}, replace={Feingeraetetechnik}]
			\step[fieldsource=\regexp{$MAPLOOP}, match={XXX}, replace={Feinw. Tech. Messtech.}]
			\step[fieldsource=\regexp{$MAPLOOP}, match={XXX}, replace={Fert. Tech. Betr.}]
			\step[fieldsource=\regexp{$MAPLOOP}, match={XXX}, replace={Fis. Tecnol.}]
			\step[fieldsource=\regexp{$MAPLOOP}, match={XXX}, replace={Fiz. Khim. Obrab. Mater.}]
			\step[fieldsource=\regexp{$MAPLOOP}, match={XXX}, replace={Fiz. Met. Metalloved}]
			\step[fieldsource=\regexp{$MAPLOOP}, match={XXX}, replace={Fiz. Nizk. Temp.}]
			\step[fieldsource=\regexp{$MAPLOOP}, match={XXX}, replace={Fiz. Plazmy}]
			\step[fieldsource=\regexp{$MAPLOOP}, match={XXX}, replace={Fiz. Tekh. Poluprovodn.}]
			\step[fieldsource=\regexp{$MAPLOOP}, match={XXX}, replace={Fiz. Tverd. Tela}]
			\step[fieldsource=\regexp{$MAPLOOP}, match={XXX}, replace={Fiz.-Khim. Mekh. Mater.}]
			\step[fieldsource=\regexp{$MAPLOOP}, match={XXX}, replace={Fizika}]
			\step[fieldsource=\regexp{$MAPLOOP}, match={XXX}, replace={Forsch.-Ber. Landes Nordrh.-Westfal.}]
			\step[fieldsource=\regexp{$MAPLOOP}, match={XXX}, replace={Frequenz}]
			\step[fieldsource=\regexp{$MAPLOOP}, match={XXX}, replace={Fys. Tidsskr.}]
			\step[fieldsource=\regexp{$MAPLOOP}, match={XXX}, replace={G. Fis.}]
			\step[fieldsource=\regexp{$MAPLOOP}, match={XXX}, replace={Geliotekhnika}]
			\step[fieldsource=\regexp{$MAPLOOP}, match={XXX}, replace={Geochim. Cosmochim. Acta}]
			\step[fieldsource=\regexp{$MAPLOOP}, match={XXX}, replace={Haerterei-Tech. Mitt.}]
			\step[fieldsource=\regexp{$MAPLOOP}, match={XXX}, replace={Helv. Chim. Acta}]
			\step[fieldsource=\regexp{$MAPLOOP}, match={XXX}, replace={Helv. Med. Acta}]
			\step[fieldsource=\regexp{$MAPLOOP}, match={XXX}, replace={Helv. Phys. Acta}]
			\step[fieldsource=\regexp{$MAPLOOP}, match={XXX}, replace={Hochfreq. Electroakust}]
			\step[fieldsource=\regexp{$MAPLOOP}, match={XXX}, replace={Hoppe-Seylers Z. Physiol. Chem.}]
			\step[fieldsource=\regexp{$MAPLOOP}, match={XXX}, replace={Inf. Elektron.}]
			\step[fieldsource=\regexp{$MAPLOOP}, match={XXX}, replace={Inf. Elettron.}]
			\step[fieldsource=\regexp{$MAPLOOP}, match={XXX}, replace={Inform. Forsch. Entwickl.}]
			\step[fieldsource=\regexp{$MAPLOOP}, match={XXX}, replace={Inform. Spektrum}]
			\step[fieldsource=\regexp{$MAPLOOP}, match={XXX}, replace={Inform.-Fachber.}]
			\step[fieldsource=\regexp{$MAPLOOP}, match={XXX}, replace={Informatie}]
			\step[fieldsource=\regexp{$MAPLOOP}, match={XXX}, replace={Informatik}]
			\step[fieldsource=\regexp{$MAPLOOP}, match={XXX}, replace={Informatologia Yugosl.}]
			\step[fieldsource=\regexp{$MAPLOOP}, match={XXX}, replace={Informatyka}]
			\step[fieldsource=\regexp{$MAPLOOP}, match={XXX}, replace={Infowelt}]
			\step[fieldsource=\regexp{$MAPLOOP}, match={XXX}, replace={Ing. Electr. Mec.}]
			\step[fieldsource=\regexp{$MAPLOOP}, match={XXX}, replace={Ing. Mec. Electr.}]
			\step[fieldsource=\regexp{$MAPLOOP}, match={XXX}, replace={Ing.-Arch.}]
			\step[fieldsource=\regexp{$MAPLOOP}, match={XXX}, replace={Inzh.-Fiz. Zh.}]
			\step[fieldsource=\regexp{$MAPLOOP}, match={XXX}, replace={Izmer. Tekh.}]
			\step[fieldsource=\regexp{$MAPLOOP}, match={XXX}, replace={Izv. Akad. Nauk Arm. SSR Ser. Tekh. Nauk}]
			\step[fieldsource=\regexp{$MAPLOOP}, match={XXX}, replace={Izv. Akad. Nauk SSSR Energ. Transp.}]
			\step[fieldsource=\regexp{$MAPLOOP}, match={XXX}, replace={Izv. Akad. Nauk SSSR Fiz. Atmos. Okeana}]
			\step[fieldsource=\regexp{$MAPLOOP}, match={XXX}, replace={Izv. Akad. Nauk SSSR Fiz. Zemli}]
			\step[fieldsource=\regexp{$MAPLOOP}, match={XXX}, replace={Izv. Akad. Nauk SSSR Ser. Fiz.}]
			\step[fieldsource=\regexp{$MAPLOOP}, match={XXX}, replace={Izv. Vyssh. Uchebn. Zaved. Elektromekh.}]
			\step[fieldsource=\regexp{$MAPLOOP}, match={XXX}, replace={Izv. Vyssh. Uchebn. Zaved. Radioelektron.}]
			\step[fieldsource=\regexp{$MAPLOOP}, match={XXX}, replace={Izv. Vyssh. Uchebn. Zaved. Radiofiz.}]
			\step[fieldsource=\regexp{$MAPLOOP}, match={XXX}, replace={J. Chim Phys. Phys.-Chim Biol.}]
			\step[fieldsource=\regexp{$MAPLOOP}, match={XXX}, replace={Kernenergie}]
			\step[fieldsource=\regexp{$MAPLOOP}, match={XXX}, replace={Kerntechnik}]
			\step[fieldsource=\regexp{$MAPLOOP}, match={XXX}, replace={Khim. Fiz.}]
			\step[fieldsource=\regexp{$MAPLOOP}, match={XXX}, replace={Kibern. Vychisl. Tekh.}]
			\step[fieldsource=\regexp{$MAPLOOP}, match={XXX}, replace={Kibernetika}]
			\step[fieldsource=\regexp{$MAPLOOP}, match={XXX}, replace={Kristallografiya}]
			\step[fieldsource=\regexp{$MAPLOOP}, match={XXX}, replace={Kvantovaya Elektron. Mosk.}]
			\step[fieldsource=\regexp{$MAPLOOP}, match={XXX}, replace={Kybernetes}]
			\step[fieldsource=\regexp{$MAPLOOP}, match={XXX}, replace={Kybernetika}]
			\step[fieldsource=\regexp{$MAPLOOP}, match={XXX}, replace={Med. Tek.}]
			\step[fieldsource=\regexp{$MAPLOOP}, match={XXX}, replace={Mekh. Avtom. Proizvod.}]
			\step[fieldsource=\regexp{$MAPLOOP}, match={XXX}, replace={Meres Autom.}]
			\step[fieldsource=\regexp{$MAPLOOP}, match={XXX}, replace={Mesures}]
			\step[fieldsource=\regexp{$MAPLOOP}, match={XXX}, replace={Metallofizika}]
			\step[fieldsource=\regexp{$MAPLOOP}, match={XXX}, replace={Metalloved. Term. Obrab. Met.}]
			\step[fieldsource=\regexp{$MAPLOOP}, match={XXX}, replace={Meteorol. Gidrol.}]
			\step[fieldsource=\regexp{$MAPLOOP}, match={XXX}, replace={Meteorol. Rundsch.}]
			\step[fieldsource=\regexp{$MAPLOOP}, match={XXX}, replace={Metrol. Apl.}]
			\step[fieldsource=\regexp{$MAPLOOP}, match={XXX}, replace={Metrologia}]
			\step[fieldsource=\regexp{$MAPLOOP}, match={XXX}, replace={Medel. Simul.}]
			\step[fieldsource=\regexp{$MAPLOOP}, match={XXX}, replace={Nachr. Dok.}]
			\step[fieldsource=\regexp{$MAPLOOP}, match={XXX}, replace={Nachr.tech. Elektron.}]
			\step[fieldsource=\regexp{$MAPLOOP}, match={XXX}, replace={Naturwissentchaften}]
			\step[fieldsource=\regexp{$MAPLOOP}, match={XXX}, replace={Neue Tech.}]
			\step[fieldsource=\regexp{$MAPLOOP}, match={XXX}, replace={Neue Tech. Buero}]
			\step[fieldsource=\regexp{$MAPLOOP}, match={XXX}, replace={Nukleonika}]
			\step[fieldsource=\regexp{$MAPLOOP}, match={XXX}, replace={Numer. Math.}]
			\step[fieldsource=\regexp{$MAPLOOP}, match={XXX}, replace={Nuovo Cimento A}]
			\step[fieldsource=\regexp{$MAPLOOP}, match={XXX}, replace={Nuovo Cimento B}]
			\step[fieldsource=\regexp{$MAPLOOP}, match={XXX}, replace={Nouvo Cimento C}]
			\step[fieldsource=\regexp{$MAPLOOP}, match={XXX}, replace={Nuovo Cimento D}]
			\step[fieldsource=\regexp{$MAPLOOP}, match={XXX}, replace={Okeanologiya}]
			\step[fieldsource=\regexp{$MAPLOOP}, match={XXX}, replace={Opt. Spektrosk.}]
			\step[fieldsource=\regexp{$MAPLOOP}, match={XXX}, replace={Opt.-Mekh. Prom.}]
			\step[fieldsource=\regexp{$MAPLOOP}, match={XXX}, replace={Optik}]
			\step[fieldsource=\regexp{$MAPLOOP}, match={XXX}, replace={Photogrammetria}]
			\step[fieldsource=\regexp{$MAPLOOP}, match={XXX}, replace={Photonics Spectra}]
			\step[fieldsource=\regexp{$MAPLOOP}, match={XXX}, replace={Pis’ma Astron. Zh. }]
			\step[fieldsource=\regexp{$MAPLOOP}, match={XXX}, replace={Pis’ma Zh. Eksp. Teor. Fiz. }]
			\step[fieldsource=\regexp{$MAPLOOP}, match={XXX}, replace={Pis’ma Zh. Tekh. Fiz. }]
			\step[fieldsource=\regexp{$MAPLOOP}, match={XXX}, replace={Poverkhn., Fiz. Khim. Mekh.}]
			\step[fieldsource=\regexp{$MAPLOOP}, match={XXX}, replace={Pr. Inst. Elektrotech.}]
			\step[fieldsource=\regexp{$MAPLOOP}, match={XXX}, replace={Prib. Sist. Upr.}]
			\step[fieldsource=\regexp{$MAPLOOP}, match={XXX}, replace={Prib. Tekh. Eksp.}]
			\step[fieldsource=\regexp{$MAPLOOP}, match={XXX}, replace={Prikl. Mat. Mekh.}]
			\step[fieldsource=\regexp{$MAPLOOP}, match={XXX}, replace={Prikl. Mekh.}]
			\step[fieldsource=\regexp{$MAPLOOP}, match={XXX}, replace={Probl. Kibern.}]
			\step[fieldsource=\regexp{$MAPLOOP}, match={XXX}, replace={Probl. Peredachi Inf.}]
			\step[fieldsource=\regexp{$MAPLOOP}, match={XXX}, replace={Proc. K. Ned. Akad. Wet. B, Palaeontol.}]
			\step[fieldsource=\regexp{$MAPLOOP}, match={XXX}, replace={Anthropol. }]
			\step[fieldsource=\regexp{$MAPLOOP}, match={XXX}, replace={Programmirovanie}]
			\step[fieldsource=\regexp{$MAPLOOP}, match={XXX}, replace={Prz. Elektrotech.}]
			\step[fieldsource=\regexp{$MAPLOOP}, match={XXX}, replace={Prz. Telekomun.}]
			\step[fieldsource=\regexp{$MAPLOOP}, match={XXX}, replace={PT/Elektrotech. Elektron.}]
			\step[fieldsource=\regexp{$MAPLOOP}, match={XXX}, replace={Radio Fernsehen Elektron.}]
			\step[fieldsource=\regexp{$MAPLOOP}, match={XXX}, replace={Radiotekh. Elektron.}]
			\step[fieldsource=\regexp{$MAPLOOP}, match={XXX}, replace={Radiotekhnika Mosk.}]
			\step[fieldsource=\regexp{$MAPLOOP}, match={XXX}, replace={Rev. Acad. Cienc. Zaragoza}]
			\step[fieldsource=\regexp{$MAPLOOP}, match={XXX}, replace={Rev. Electrotec. (Argentina)}]
			\step[fieldsource=\regexp{$MAPLOOP}, match={XXX}, replace={Rev. Electrotec. (Spain)}]
			\step[fieldsource=\regexp{$MAPLOOP}, match={XXX}, replace={Rev. Energ.}]
			\step[fieldsource=\regexp{$MAPLOOP}, match={XXX}, replace={Rev. Esp. Electron. Rev. Geofis.}]
			\step[fieldsource=\regexp{$MAPLOOP}, match={XXX}, replace={Ric. Autom.}]
			\step[fieldsource=\regexp{$MAPLOOP}, match={XXX}, replace={Ric. Spettrosc.}]
			\step[fieldsource=\regexp{$MAPLOOP}, match={XXX}, replace={Robotersysteme}]
			\step[fieldsource=\regexp{$MAPLOOP}, match={XXX}, replace={Rozpr. Electrotech.}]
			\step[fieldsource=\regexp{$MAPLOOP}, match={XXX}, replace={Sadhana}]
			\step[fieldsource=\regexp{$MAPLOOP}, match={XXX}, replace={Schweiz. Tech. Z.}]
			\step[fieldsource=\regexp{$MAPLOOP}, match={XXX}, replace={Scientia}]
			\step[fieldsource=\regexp{$MAPLOOP}, match={XXX}, replace={Siemens Forsch. Entwickl. Ber.}]
			\step[fieldsource=\regexp{$MAPLOOP}, match={XXX}, replace={Sist. Autom.}]
			\step[fieldsource=\regexp{$MAPLOOP}, match={XXX}, replace={Sitzungsber. Oester. Akad. Wiss. Math.-}]
			\step[fieldsource=\regexp{$MAPLOOP}, match={XXX}, replace={Naturwiss. Kl. Abt. II (Austria)}]
			\step[fieldsource=\regexp{$MAPLOOP}, match={XXX}, replace={Spectrochim. Acta A, Mol. Spectrosc.}]
			\step[fieldsource=\regexp{$MAPLOOP}, match={XXX}, replace={Spectrochim. Acta B, At. Spectrosc.}]
			\step[fieldsource=\regexp{$MAPLOOP}, match={XXX}, replace={Sprache Datenverarb.}]
			\step[fieldsource=\regexp{$MAPLOOP}, match={XXX}, replace={Stanki Instrum.}]
			\step[fieldsource=\regexp{$MAPLOOP}, match={XXX}, replace={Steklo Keram.}]
			\step[fieldsource=\regexp{$MAPLOOP}, match={XXX}, replace={Svetotekhnika  TE Int.}]
			\step[fieldsource=\regexp{$MAPLOOP}, match={XXX}, replace={Tech. Bull. Vevey}]
			\step[fieldsource=\regexp{$MAPLOOP}, match={XXX}, replace={Tech. Mitt. Krupp (Engl. Ed.)}]
			\step[fieldsource=\regexp{$MAPLOOP}, match={XXX}, replace={Tech. Mitt. PTT}]
			\step[fieldsource=\regexp{$MAPLOOP}, match={XXX}, replace={Tech. Mitt. RFZ}]
			\step[fieldsource=\regexp{$MAPLOOP}, match={XXX}, replace={Technica}]
			\step[fieldsource=\regexp{$MAPLOOP}, match={XXX}, replace={Tecnica}]
			\step[fieldsource=\regexp{$MAPLOOP}, match={XXX}, replace={Teh. Fiz.}]
			\step[fieldsource=\regexp{$MAPLOOP}, match={XXX}, replace={Tehnika}]
			\step[fieldsource=\regexp{$MAPLOOP}, match={XXX}, replace={Tekh. Elektrodin.}]
			\step[fieldsource=\regexp{$MAPLOOP}, match={XXX}, replace={Tekh. Kibern.}]
			\step[fieldsource=\regexp{$MAPLOOP}, match={XXX}, replace={Tekh. Kino Telev.}]
			\step[fieldsource=\regexp{$MAPLOOP}, match={XXX}, replace={Tekh. Misul}]
			\step[fieldsource=\regexp{$MAPLOOP}, match={XXX}, replace={Telekomunikacije}]
			\step[fieldsource=\regexp{$MAPLOOP}, match={XXX}, replace={Telektronikk}]
			\step[fieldsource=\regexp{$MAPLOOP}, match={XXX}, replace={Teleteknik}]
			\step[fieldsource=\regexp{$MAPLOOP}, match={XXX}, replace={Teor. Mat. Fiz.}]
			\step[fieldsource=\regexp{$MAPLOOP}, match={XXX}, replace={Teploeoergetika}]
			\step[fieldsource=\regexp{$MAPLOOP}, match={XXX}, replace={Teplofiz. Vvs. Temp.}]
			\step[fieldsource=\regexp{$MAPLOOP}, match={XXX}, replace={Tidskr. Dok.}]
			\step[fieldsource=\regexp{$MAPLOOP}, match={XXX}, replace={TN Nachr.}]
			\step[fieldsource=\regexp{$MAPLOOP}, match={XXX}, replace={Toute Electron.}]
			\step[fieldsource=\regexp{$MAPLOOP}, match={XXX}, replace={Tr. Inst. Teor. Astron.}]
			\step[fieldsource=\regexp{$MAPLOOP}, match={XXX}, replace={Ukr. Fiz. Zh.}]
			\step[fieldsource=\regexp{$MAPLOOP}, match={XXX}, replace={Usp. Fiz. Nauk}]
			\step[fieldsource=\regexp{$MAPLOOP}, match={XXX}, replace={Vak.-Tech.}]
			\step[fieldsource=\regexp{$MAPLOOP}, match={XXX}, replace={VDE Fachiber.}]
			\step[fieldsource=\regexp{$MAPLOOP}, match={XXX}, replace={VDI Z.}]
			\step[fieldsource=\regexp{$MAPLOOP}, match={XXX}, replace={Vestn. Mashinostr.}]
			\step[fieldsource=\regexp{$MAPLOOP}, match={XXX}, replace={Vestn. Mosk. Univ. 15, Vychisl. Mat. Kibern.}]
			\step[fieldsource=\regexp{$MAPLOOP}, match={XXX}, replace={Vestn. Mosk. Univ. 3, Fiz. Astron.}]
			\step[fieldsource=\regexp{$MAPLOOP}, match={XXX}, replace={Vesti Akad. Navuk BSSR Ser. Fiz. Energ. Navuk}]
			\step[fieldsource=\regexp{$MAPLOOP}, match={XXX}, replace={VGB Kraftwerkstech. (Ger. Ed.)}]
			\step[fieldsource=\regexp{$MAPLOOP}, match={XXX}, replace={Vide Couches Minces}]
			\step[fieldsource=\regexp{$MAPLOOP}, match={XXX}, replace={Vistas Astron.}]
			\step[fieldsource=\regexp{$MAPLOOP}, match={XXX}, replace={Vopr. At. Nauki Tekh. Ser., Fiz. Radiats.}]
			\step[fieldsource=\regexp{$MAPLOOP}, match={XXX}, replace={Povrezhdenii Radiats. Materialoved.}]
			\step[fieldsource=\regexp{$MAPLOOP}, match={XXX}, replace={Vopr. At. Nauki Tekh. Ser., Obshch. Yad. Fiz.}]
			\step[fieldsource=\regexp{$MAPLOOP}, match={XXX}, replace={Vuoto Sci. Tecnol.}]
			\step[fieldsource=\regexp{$MAPLOOP}, match={XXX}, replace={Wiss. Z. Friedrich-Schiller-Univ. Jena}]
			\step[fieldsource=\regexp{$MAPLOOP}, match={XXX}, replace={Nat.wiss. Reihe}]
			\step[fieldsource=\regexp{$MAPLOOP}, match={XXX}, replace={Wiss. Z. Karl-Marx-Univ. Leipz. Math.-}]
			\step[fieldsource=\regexp{$MAPLOOP}, match={XXX}, replace={Nat.wiss. Reihe}]
			\step[fieldsource=\regexp{$MAPLOOP}, match={XXX}, replace={Wiss. Z. Tech. Hochsch. Ilmenau}]
			\step[fieldsource=\regexp{$MAPLOOP}, match={XXX}, replace={Wiss. Z. Tech. Univ. Dresd.}]
			\step[fieldsource=\regexp{$MAPLOOP}, match={XXX}, replace={Wiss. Z. Tech. Univ. Karl-Marx-Stadt}]
			\step[fieldsource=\regexp{$MAPLOOP}, match={XXX}, replace={Yad. Fiz.}]
			\step[fieldsource=\regexp{$MAPLOOP}, match={XXX}, replace={Z. Angew. Math. Mech.}]
			\step[fieldsource=\regexp{$MAPLOOP}, match={XXX}, replace={Z. Angew. Math. Phys.}]
			\step[fieldsource=\regexp{$MAPLOOP}, match={XXX}, replace={Z. Met.kd.}]
			\step[fieldsource=\regexp{$MAPLOOP}, match={XXX}, replace={Z. Nat. Forsch. A, Phys. Phys. Chem. Kosmophys.}]
			\step[fieldsource=\regexp{$MAPLOOP}, match={XXX}, replace={Z. Oper. Res. A, Theor.}]
			\step[fieldsource=\regexp{$MAPLOOP}, match={XXX}, replace={Z. Oper. Res. B, Prax.}]
			\step[fieldsource=\regexp{$MAPLOOP}, match={XXX}, replace={Z. Phys.}]
			\step[fieldsource=\regexp{$MAPLOOP}, match={XXX}, replace={Z. Phys. A, At. Nuclei}]
			\step[fieldsource=\regexp{$MAPLOOP}, match={XXX}, replace={Z. Phys. B, Condens. Matter}]
			\step[fieldsource=\regexp{$MAPLOOP}, match={XXX}, replace={Z. Phys. C, Part Fields}]
			\step[fieldsource=\regexp{$MAPLOOP}, match={XXX}, replace={Z. Phys. Chem. Neue Folge}]
			\step[fieldsource=\regexp{$MAPLOOP}, match={XXX}, replace={Z. Phys. Chem., Leipz.}]
			\step[fieldsource=\regexp{$MAPLOOP}, match={XXX}, replace={Z. Phys. D, At. Mol. Clusters}]
			\step[fieldsource=\regexp{$MAPLOOP}, match={XXX}, replace={Zavod. Lab.}]
			\step[fieldsource=\regexp{$MAPLOOP}, match={XXX}, replace={Zh. Eksp. Teor. Fiz.}]
			\step[fieldsource=\regexp{$MAPLOOP}, match={XXX}, replace={Zh. Fiz. Khim.}]
			\step[fieldsource=\regexp{$MAPLOOP}, match={XXX}, replace={Zh. Prikl. Mekh. Tekh. Fiz.}]
			\step[fieldsource=\regexp{$MAPLOOP}, match={XXX}, replace={Zh. Prikl. Spektrosk.}]
			\step[fieldsource=\regexp{$MAPLOOP}, match={XXX}, replace={Zh. Tekh. Fiz.}]
			\step[fieldsource=\regexp{$MAPLOOP}, match={XXX}, replace={Zh. Vychisl. Mat. Mat. Fiz.}]
			\step[fieldsource=\regexp{$MAPLOOP}, match={XXX}, replace={Zisin, J. Seismol. Soc. Jpn.}] 
			\step[fieldsource=\regexp{$MAPLOOP}, match={Production}, replace={Prod.}]
			\step[fieldsource=\regexp{$MAPLOOP}, match={Reliability}, replace={Rel.}]
			\step[fieldsource=\regexp{$MAPLOOP}, match={Report}, replace={Rep.}]
			\step[fieldsource=\regexp{$MAPLOOP}, match={Semiconductor}, replace={Semicond.}]
			\step[fieldsource=\regexp{$MAPLOOP}, match={Research}, replace={Res.}]
			\step[fieldsource=\regexp{$MAPLOOP}, match={Sensing}, replace={Sens.}]
			\step[fieldsource=\regexp{$MAPLOOP}, match={Resonance}, replace={Reson.}]
			\step[fieldsource=\regexp{$MAPLOOP}, match={Series}, replace={Ser.}]
			\step[fieldsource=\regexp{$MAPLOOP}, match={Resources}, replace={Resour.}]
			\step[fieldsource=\regexp{$MAPLOOP}, match={Simulation}, replace={Simul.}]
			\step[fieldsource=\regexp{$MAPLOOP}, match={Reviews}, replace={Rev.}]
			\step[fieldsource=\regexp{$MAPLOOP}, match={Review}, replace={Rev.}]
			\step[fieldsource=\regexp{$MAPLOOP}, match={Singapore}, replace={Singap.}]
			\step[fieldsource=\regexp{$MAPLOOP}, match={Robotics}, replace={Robot.}]
			\step[fieldsource=\regexp{$MAPLOOP}, match={Sistema}, replace={Sist.}]
			\step[fieldsource=\regexp{$MAPLOOP}, match={Royal}, replace={Roy.}]
			\step[fieldsource=\regexp{$MAPLOOP}, match={Society}, replace={Soc.}]
			\step[fieldsource=\regexp{$MAPLOOP}, match={Safety}, replace={Saf.}]
			\step[fieldsource=\regexp{$MAPLOOP}, match={Sociological}, replace={Sociol.}]
			\step[fieldsource=\regexp{$MAPLOOP}, match={Satellite}, replace={Satell.}]
			\step[fieldsource=\regexp{$MAPLOOP}, match={Software}, replace={Softw.}]
			\step[fieldsource=\regexp{$MAPLOOP}, match={Scandinavian}, replace={Scand.}]
			\step[fieldsource=\regexp{$MAPLOOP}, match={Solar}, replace={Sol.}]
			\step[fieldsource=\regexp{$MAPLOOP}, match={Sciences}, replace={Sci.}]
			\step[fieldsource=\regexp{$MAPLOOP}, match={Science}, replace={Sci.}]
			\step[fieldsource=\regexp{$MAPLOOP}, match={Soviet}, replace={Sov.}]
			\step[fieldsource=\regexp{$MAPLOOP}, match={Section}, replace={Sect.}]
			\step[fieldsource=\regexp{$MAPLOOP}, match={Spectroscopy}, replace={Spectrosc.}]
			\step[fieldsource=\regexp{$MAPLOOP}, match={Security}, replace={Secur.}]
			\step[fieldsource=\regexp{$MAPLOOP}, match={Spectrum}, replace={Spectr.}]
			\step[fieldsource=\regexp{$MAPLOOP}, match={Seismology}, replace={Seismol.}]
			\step[fieldsource=\regexp{$MAPLOOP}, match={Speculations}, replace={Specul.}]
			\step[fieldsource=\regexp{$MAPLOOP}, match={Selected}, replace={Sel.}]
			\step[fieldsource=\regexp{$MAPLOOP}, match={Statistics}, replace={Statist.}]
			\step[fieldsource=\regexp{$MAPLOOP}, match={Structures}, replace={Struct.}]
			\step[fieldsource=\regexp{$MAPLOOP}, match={Structure}, replace={Struct.}]
			\step[fieldsource=\regexp{$MAPLOOP}, match={Terrestrial}, replace={Terr.}]
			\step[fieldsource=\regexp{$MAPLOOP}, match={Studies}, replace={Stud.}]
			\step[fieldsource=\regexp{$MAPLOOP}, match={Theoretical}, replace={Theor.}]
			\step[fieldsource=\regexp{$MAPLOOP}, match={Superconductivity}, replace={Supercond.}]
			\step[fieldsource=\regexp{$MAPLOOP}, match={Transactions}, replace={Trans.}]
			\step[fieldsource=\regexp{$MAPLOOP}, match={Supplement}, replace={Suppl.}]
			\step[fieldsource=\regexp{$MAPLOOP}, match={Translation}, replace={Transl.}]
			\step[fieldsource=\regexp{$MAPLOOP}, match={Surface}, replace={Surf.}]
			\step[fieldsource=\regexp{$MAPLOOP}, match={Transmission}, replace={Transmiss.}]
			\step[fieldsource=\regexp{$MAPLOOP}, match={Survey}, replace={Surv.}]
			\step[fieldsource=\regexp{$MAPLOOP}, match={Transportation}, replace={Transp.}]
			\step[fieldsource=\regexp{$MAPLOOP}, match={Sustainable}, replace={Sustain.}]
			\step[fieldsource=\regexp{$MAPLOOP}, match={Tutorials}, replace={Tut.}]
			\step[fieldsource=\regexp{$MAPLOOP}, match={Symposium}, replace={Symp.}]
			\step[fieldsource=\regexp{$MAPLOOP}, match={Ultrasonic}, replace={Ultrason.}]
			\step[fieldsource=\regexp{$MAPLOOP}, match={Systems}, replace={Syst.}]
			\step[fieldsource=\regexp{$MAPLOOP}, match={System}, replace={Syst.}]
			\step[fieldsource=\regexp{$MAPLOOP}, match={University}, replace={Univ.}]
			\step[fieldsource=\regexp{$MAPLOOP}, match={\detokenize{Universität}}, replace={Univ.}]
			\step[fieldsource=\regexp{$MAPLOOP}, match={Technical}, replace={Tech.}]
			\step[fieldsource=\regexp{$MAPLOOP}, match={Technische}, replace={Tech.}]
			\step[fieldsource=\regexp{$MAPLOOP}, match={Vacuum}, replace={Vac.}]
			\step[fieldsource=\regexp{$MAPLOOP}, match={Techniques}, replace={Techn.}]
			\step[fieldsource=\regexp{$MAPLOOP}, match={Vehicles}, replace={Veh.}]
			\step[fieldsource=\regexp{$MAPLOOP}, match={Vehicle}, replace={Veh.}]
			\step[fieldsource=\regexp{$MAPLOOP}, match={Vehicular}, replace={Veh.}]
			\step[fieldsource=\regexp{$MAPLOOP}, match={Technology}, replace={Technol.}]
			\step[fieldsource=\regexp{$MAPLOOP}, match={Vibration}, replace={Vib.}]
			\step[fieldsource=\regexp{$MAPLOOP}, match={Telecommunications}, replace={Telecommun.}]
			\step[fieldsource=\regexp{$MAPLOOP}, match={Visual}, replace={Vis.}]
			\step[fieldsource=\regexp{$MAPLOOP}, match={Television}, replace={Telev.}]
			\step[fieldsource=\regexp{$MAPLOOP}, match={Welding}, replace={Weld.}]
			\step[fieldsource=\regexp{$MAPLOOP}, match={Temperature}, replace={Temp.}]
			\step[fieldsource=\regexp{$MAPLOOP}, match={Working}, replace={Work.}]
			\step[fieldsource=\regexp{$MAPLOOP}, match={Learning}, replace={Learn.}]
			\step[fieldsource=\regexp{$MAPLOOP}, match={Measurement}, replace={Meas.}]
			\step[fieldsource=\regexp{$MAPLOOP}, match={Letters}, replace={Lett.}]
			\step[fieldsource=\regexp{$MAPLOOP}, match={Letter}, replace={Lett.}]
			\step[fieldsource=\regexp{$MAPLOOP}, match={Mechanical}, replace={Mech.}]
			\step[fieldsource=\regexp{$MAPLOOP}, match={Mechanics}, replace={Mech.}]
			\step[fieldsource=\regexp{$MAPLOOP}, match={Mechanic}, replace={Mech.}]
			\step[fieldsource=\regexp{$MAPLOOP}, match={Lightwave}, replace={Lightw.}]
			\step[fieldsource=\regexp{$MAPLOOP}, match={Medical}, replace={Med.}]
			\step[fieldsource=\regexp{$MAPLOOP}, match={Logic, Logical}, replace={Log.}]
			\step[fieldsource=\regexp{$MAPLOOP}, match={Metals}, replace={Met.}]
			\step[fieldsource=\regexp{$MAPLOOP}, match={Luminescence}, replace={Lumin.}]
			\step[fieldsource=\regexp{$MAPLOOP}, match={Metallurgy}, replace={Metall.}]
			\step[fieldsource=\regexp{$MAPLOOP}, match={Machines}, replace={Mach.}]
			\step[fieldsource=\regexp{$MAPLOOP}, match={Machine}, replace={Mach.}]
			\step[fieldsource=\regexp{$MAPLOOP}, match={Meteorology}, replace={Meteorol.}]
			\step[fieldsource=\regexp{$MAPLOOP}, match={Magazine}, replace={Mag.}]
			\step[fieldsource=\regexp{$MAPLOOP}, match={Metropolitan}, replace={Metrop.}]
			\step[fieldsource=\regexp{$MAPLOOP}, match={Magnetics}, replace={Magn.}]
			\step[fieldsource=\regexp{$MAPLOOP}, match={Mexican, Mexico}, replace={Mex.}]
			\step[fieldsource=\regexp{$MAPLOOP}, match={Management}, replace={Manage.}]
			\step[fieldsource=\regexp{$MAPLOOP}, match={Microelectromechanical}, replace={Microelectromech.}]
			\step[fieldsource=\regexp{$MAPLOOP}, match={Managing}, replace={Manag.}]
			\step[fieldsource=\regexp{$MAPLOOP}, match={Microgravity}, replace={Microgr.}]
			\step[fieldsource=\regexp{$MAPLOOP}, match={Manufacturing}, replace={Manuf.}]
			\step[fieldsource=\regexp{$MAPLOOP}, match={Microscopy}, replace={Microsc.}]
			\step[fieldsource=\regexp{$MAPLOOP}, match={Marine}, replace={Mar.}]
			\step[fieldsource=\regexp{$MAPLOOP}, match={Microwaves}, replace={Microw.}]
			\step[fieldsource=\regexp{$MAPLOOP}, match={Microwave}, replace={Microw.}]
			\step[fieldsource=\regexp{$MAPLOOP}, match={Materials}, replace={Mater.}]
			\step[fieldsource=\regexp{$MAPLOOP}, match={Material}, replace={Mater.}]
			\step[fieldsource=\regexp{$MAPLOOP}, match={Military}, replace={Mil.}]
			\step[fieldsource=\regexp{$MAPLOOP}, match={Modeling}, replace={Model.}]
			\step[fieldsource=\regexp{$MAPLOOP}, match={Modelling}, replace={Model.}]
			\step[fieldsource=\regexp{$MAPLOOP}, match={Oceanic}, replace={Ocean.}]
			\step[fieldsource=\regexp{$MAPLOOP}, match={Molecular}, replace={Mol.}]
			\step[fieldsource=\regexp{$MAPLOOP}, match={Oceanography}, replace={Oceanogr.}]
			\step[fieldsource=\regexp{$MAPLOOP}, match={Monitoring}, replace={Monit.}]
			\step[fieldsource=\regexp{$MAPLOOP}, match={Occupation}, replace={Occupat.}]
			\step[fieldsource=\regexp{$MAPLOOP}, match={Multiphysics}, replace={Multiphys.}]
			\step[fieldsource=\regexp{$MAPLOOP}, match={Operational}, replace={Oper.}]
			\step[fieldsource=\regexp{$MAPLOOP}, match={Nanobioscience}, replace={Nanobiosci.}]
			\step[fieldsource=\regexp{$MAPLOOP}, match={Optical}, replace={Opt.}]
			\step[fieldsource=\regexp{$MAPLOOP}, match={Nanotechnology}, replace={Nanotechnol.}]
			\step[fieldsource=\regexp{$MAPLOOP}, match={Optics}, replace={Opt.}]
			\step[fieldsource=\regexp{$MAPLOOP}, match={National}, replace={Nat.}]
			\step[fieldsource=\regexp{$MAPLOOP}, match={Optimization}, replace={Optim.}]
			\step[fieldsource=\regexp{$MAPLOOP}, match={Naval}, replace={Nav.}]
			\step[fieldsource=\regexp{$MAPLOOP}, match={Organization}, replace={Org.}]
			\step[fieldsource=\regexp{$MAPLOOP}, match={Networking}, replace={Netw.}]
			\step[fieldsource=\regexp{$MAPLOOP}, match={Networked}, replace={Netw.}]
			\step[fieldsource=\regexp{$MAPLOOP}, match={Network}, replace={Netw.}]
			\step[fieldsource=\regexp{$MAPLOOP}, match={Packaging}, replace={Packag.}]
			\step[fieldsource=\regexp{$MAPLOOP}, match={Newsletter}, replace={Newslett.}]
			\step[fieldsource=\regexp{$MAPLOOP}, match={Particle}, replace={Part.}]
			\step[fieldsource=\regexp{$MAPLOOP}, match={Nondestructive}, replace={Nondestruct.}]
			\step[fieldsource=\regexp{$MAPLOOP}, match={Patent}, replace={Pat.}]
			\step[fieldsource=\regexp{$MAPLOOP}, match={Nuclear}, replace={Nucl.}]
			\step[fieldsource=\regexp{$MAPLOOP}, match={Performance}, replace={Perform.}]
			\step[fieldsource=\regexp{$MAPLOOP}, match={Numerical}, replace={Numer.}]
			\step[fieldsource=\regexp{$MAPLOOP}, match={Personal}, replace={Pers.}]
			\step[fieldsource=\regexp{$MAPLOOP}, match={Observations}, replace={Observ.}]
			\step[fieldsource=\regexp{$MAPLOOP}, match={Philosophical}, replace={Philos.}]
			\step[fieldsource=\regexp{$MAPLOOP}, match={Photonics}, replace={Photon.}]
			\step[fieldsource=\regexp{$MAPLOOP}, match={Productivity}, replace={Productiv.}]
			\step[fieldsource=\regexp{$MAPLOOP}, match={Photovoltaics}, replace={Photovolt.}]
			\step[fieldsource=\regexp{$MAPLOOP}, match={Programming}, replace={Program.}]
			\step[fieldsource=\regexp{$MAPLOOP}, match={Physics}, replace={Phys.}]
			\step[fieldsource=\regexp{$MAPLOOP}, match={Progress}, replace={Prog.}]
			\step[fieldsource=\regexp{$MAPLOOP}, match={Physiology}, replace={Physiol.}]
			\step[fieldsource=\regexp{$MAPLOOP}, match={Propagation}, replace={Propag.}]
			\step[fieldsource=\regexp{$MAPLOOP}, match={Planetary}, replace={Planet.}]
			\step[fieldsource=\regexp{$MAPLOOP}, match={Psychology}, replace={Psychol.}]
			\step[fieldsource=\regexp{$MAPLOOP}, match={Pneumatics}, replace={Pneum.}]
			\step[fieldsource=\regexp{$MAPLOOP}, match={Quality}, replace={Qual.}]
			\step[fieldsource=\regexp{$MAPLOOP}, match={Pollution}, replace={Pollut.}]
			\step[fieldsource=\regexp{$MAPLOOP}, match={Quarterly}, replace={Quart.}]
			\step[fieldsource=\regexp{$MAPLOOP}, match={Polymer}, replace={Polym.}]
			\step[fieldsource=\regexp{$MAPLOOP}, match={Radiation}, replace={Radiat.}]
			\step[fieldsource=\regexp{$MAPLOOP}, match={Polytechnic}, replace={Polytech.}]
			\step[fieldsource=\regexp{$MAPLOOP}, match={Radiology}, replace={Radiol.}]
			\step[fieldsource=\regexp{$MAPLOOP}, match={Practice}, replace={Pract.}]
			\step[fieldsource=\regexp{$MAPLOOP}, match={Reactor}, replace={React.}]
			\step[fieldsource=\regexp{$MAPLOOP}, match={Precision}, replace={Precis.}]
			\step[fieldsource=\regexp{$MAPLOOP}, match={Receivers}, replace={Receiv.}]
			\step[fieldsource=\regexp{$MAPLOOP}, match={Principles}, replace={Princ.}]
			\step[fieldsource=\regexp{$MAPLOOP}, match={Recognition}, replace={Recognit.}]
			\step[fieldsource=\regexp{$MAPLOOP}, match={Proceedings}, replace={Proc.}]
			\step[fieldsource=\regexp{$MAPLOOP}, match={Record}, replace={Rec.}]
			\step[fieldsource=\regexp{$MAPLOOP}, match={Processing}, replace={Process.}]
			\step[fieldsource=\regexp{$MAPLOOP}, match={Rehabilitation}, replace={Rehabil.}]
			\step[fieldsource=\regexp{$MAPLOOP}, match={Conversion}, replace={Convers.}]
			\step[fieldsource=\regexp{$MAPLOOP}, match={Digital}, replace={Digit.}]
			\step[fieldsource=\regexp{$MAPLOOP}, match={Convention}, replace={Conv.}]
			\step[fieldsource=\regexp{$MAPLOOP}, match={Disclosure}, replace={Discl.}]
			\step[fieldsource=\regexp{$MAPLOOP}, match={Correspondence}, replace={Corresp.}]
			\step[fieldsource=\regexp{$MAPLOOP}, match={Discussions}, replace={Discuss.}]
			\step[fieldsource=\regexp{$MAPLOOP}, match={Critical}, replace={Crit.}]
			\step[fieldsource=\regexp{$MAPLOOP}, match={Dissertations}, replace={Diss.}]
			\step[fieldsource=\regexp{$MAPLOOP}, match={Crystal}, replace={Cryst.}]
			\step[fieldsource=\regexp{$MAPLOOP}, match={Distributed}, replace={Distrib.}]
			\step[fieldsource=\regexp{$MAPLOOP}, match={Crystallography}, replace={Crystallogr.}]
			\step[fieldsource=\regexp{$MAPLOOP}, match={Dynamics}, replace={Dyn.}]
			\step[fieldsource=\regexp{$MAPLOOP}, match={Cybernetics}, replace={Cybern.}]
			\step[fieldsource=\regexp{$MAPLOOP}, match={Earthquake}, replace={Earthq.}]
			\step[fieldsource=\regexp{$MAPLOOP}, match={Decision}, replace={Decis.}]
			\step[fieldsource=\regexp{$MAPLOOP}, match={Economics}, replace={Econ.}]
			\step[fieldsource=\regexp{$MAPLOOP}, match={Economic}, replace={Econ.}]
			\step[fieldsource=\regexp{$MAPLOOP}, match={Economical}, replace={Econ.}]
			\step[fieldsource=\regexp{$MAPLOOP}, match={Edition}, replace={Ed.}]
			\step[fieldsource=\regexp{$MAPLOOP}, match={Evolutionary}, replace={Evol.}]
			\step[fieldsource=\regexp{$MAPLOOP}, match={Education}, replace={Educ.}]
			\step[fieldsource=\regexp{$MAPLOOP}, match={Exhibition}, replace={Exhib.}]
			\step[fieldsource=\regexp{$MAPLOOP}, match={Electrical}, replace={Elect.}]
			\step[fieldsource=\regexp{$MAPLOOP}, match={Experimental}, replace={Exp.}]
			\step[fieldsource=\regexp{$MAPLOOP}, match={Electrification}, replace={Electrific.}]
			\step[fieldsource=\regexp{$MAPLOOP}, match={Exploratory}, replace={Explor.}]
			\step[fieldsource=\regexp{$MAPLOOP}, match={Electromagnetic}, replace={Electromagn.}]
			\step[fieldsource=\regexp{$MAPLOOP}, match={Exposition}, replace={Expo.}]
			\step[fieldsource=\regexp{$MAPLOOP}, match={Electroacoustic}, replace={Electroacoust.}]
			\step[fieldsource=\regexp{$MAPLOOP}, match={Express}, replace={Express}]
			\step[fieldsource=\regexp{$MAPLOOP}, match={Electronics}, replace={Electron.}]
			\step[fieldsource=\regexp{$MAPLOOP}, match={Electronic}, replace={Electron.}]
			\step[fieldsource=\regexp{$MAPLOOP}, match={Fabrication}, replace={Fabr.}]
			\step[fieldsource=\regexp{$MAPLOOP}, match={Emerging}, replace={Emerg.}]
			\step[fieldsource=\regexp{$MAPLOOP}, match={Faculty}, replace={Fac.}]
			\step[fieldsource=\regexp{$MAPLOOP}, match={Engineering}, replace={Eng.}]
			\step[fieldsource=\regexp{$MAPLOOP}, match={Engineers}, replace={Eng.}]
			\step[fieldsource=\regexp{$MAPLOOP}, match={Engineer}, replace={Eng.}]
			\step[fieldsource=\regexp{$MAPLOOP}, match={Ferroelectrics}, replace={Ferroelect.}]
			\step[fieldsource=\regexp{$MAPLOOP}, match={Environment}, replace={Environ.}]
			\step[fieldsource=\regexp{$MAPLOOP}, match={Francais, French}, replace={Fr.}]
			\step[fieldsource=\regexp{$MAPLOOP}, match={Equations}, replace={Equ.}]
			\step[fieldsource=\regexp{$MAPLOOP}, match={Frequency}, replace={Freq.}]
			\step[fieldsource=\regexp{$MAPLOOP}, match={Equipment}, replace={Equip.}]
			\step[fieldsource=\regexp{$MAPLOOP}, match={Foundation}, replace={Found.}]
			\step[fieldsource=\regexp{$MAPLOOP}, match={Ergonomics}, replace={Ergonom.}]
			\step[fieldsource=\regexp{$MAPLOOP}, match={Fundamental}, replace={Fundam.}]
			\step[fieldsource=\regexp{$MAPLOOP}, match={European}, replace={Eur.}]
			\step[fieldsource=\regexp{$MAPLOOP}, match={Generation}, replace={Gener.}]
			\step[fieldsource=\regexp{$MAPLOOP}, match={Evaluation}, replace={Eval.}]
			\step[fieldsource=\regexp{$MAPLOOP}, match={Geology}, replace={Geol.}]
			\step[fieldsource=\regexp{$MAPLOOP}, match={Geophysics}, replace={Geophys.}]
			\step[fieldsource=\regexp{$MAPLOOP}, match={Innovations}, replace={Innov.}]
			\step[fieldsource=\regexp{$MAPLOOP}, match={Innovation}, replace={Innov.}]
			\step[fieldsource=\regexp{$MAPLOOP}, match={Geoscience}, replace={Geosci.}]
			\step[fieldsource=\regexp{$MAPLOOP}, match={Institute}, replace={Inst.}]
			\step[fieldsource=\regexp{$MAPLOOP}, match={Institution}, replace={Inst.}]
			\step[fieldsource=\regexp{$MAPLOOP}, match={Graphics}, replace={Graph.}]
			\step[fieldsource=\regexp{$MAPLOOP}, match={Instrument}, replace={Instrum.}]
			\step[fieldsource=\regexp{$MAPLOOP}, match={Guidance}, replace={Guid.}]
			\step[fieldsource=\regexp{$MAPLOOP}, match={Instrumentation}, replace={Instrum.}]
			\step[fieldsource=\regexp{$MAPLOOP}, match={Harmonics}, replace={Harmon.}]
			\step[fieldsource=\regexp{$MAPLOOP}, match={Harmonic}, replace={Harmon.}]
			\step[fieldsource=\regexp{$MAPLOOP}, match={Insulation}, replace={Insul.}]
			\step[fieldsource=\regexp{$MAPLOOP}, match={History}, replace={Hist.}]
			\step[fieldsource=\regexp{$MAPLOOP}, match={Integrated}, replace={Integr.}]
			\step[fieldsource=\regexp{$MAPLOOP}, match={Horizon}, replace={Horiz.}]
			\step[fieldsource=\regexp{$MAPLOOP}, match={Intelligence}, replace={Intell.}]
			\step[fieldsource=\regexp{$MAPLOOP}, match={Hungary}, replace={Hung.}]
			\step[fieldsource=\regexp{$MAPLOOP}, match={Hungarian}, replace={Hung.}]
			\step[fieldsource=\regexp{$MAPLOOP}, match={Intelligent}, replace={Intell.}]
			\step[fieldsource=\regexp{$MAPLOOP}, match={Hydraulics}, replace={Hydraul.}]
			\step[fieldsource=\regexp{$MAPLOOP}, match={Interactions}, replace={Interact.}]
			\step[fieldsource=\regexp{$MAPLOOP}, match={Hydrology}, replace={Hydrol.}]
			\step[fieldsource=\regexp{$MAPLOOP}, match={Internationales}, replace={Int.}]
			\step[fieldsource=\regexp{$MAPLOOP}, match={International}, replace={Int.}]
			\step[fieldsource=\regexp{$MAPLOOP}, match={Illuminating}, replace={Illum.}]
			\step[fieldsource=\regexp{$MAPLOOP}, match={Isotopes}, replace={Isot.}]
			\step[fieldsource=\regexp{$MAPLOOP}, match={Imaging}, replace={Imag.}]
			\step[fieldsource=\regexp{$MAPLOOP}, match={Israel}, replace={Isr.}]
			\step[fieldsource=\regexp{$MAPLOOP}, match={Industrial}, replace={Ind.}]
			\step[fieldsource=\regexp{$MAPLOOP}, match={Japan}, replace={Jpn.}]
			\step[fieldsource=\regexp{$MAPLOOP}, match={Information}, replace={Inf.}]
			\step[fieldsource=\regexp{$MAPLOOP}, match={Journal}, replace={J.}]
			\step[fieldsource=\regexp{$MAPLOOP}, match={Informatics}, replace={Inform.}]
			\step[fieldsource=\regexp{$MAPLOOP}, match={Knowledge}, replace={Knowl.}]
			\step[fieldsource=\regexp{$MAPLOOP}, match={Laboratory}, replace={Lab.}]
			\step[fieldsource=\regexp{$MAPLOOP}, match={Laboratories}, replace={Lab.}]
			\step[fieldsource=\regexp{$MAPLOOP}, match={Mathematical}, replace={Math.}]
			\step[fieldsource=\regexp{$MAPLOOP}, match={Language}, replace={Lang.}]
			\step[fieldsource=\regexp{$MAPLOOP}, match={Mathematics}, replace={Math.}]
			\step[fieldsource=\regexp{$MAPLOOP}, match={Abstracts}, replace={Abstr.}]
			\step[fieldsource=\regexp{$MAPLOOP}, match={Analysis}, replace={Anal.}]
			\step[fieldsource=\regexp{$MAPLOOP}, match={Academy}, replace={Acad.}]
			\step[fieldsource=\regexp{$MAPLOOP}, match={Annals}, replace={Ann.}]
			\step[fieldsource=\regexp{$MAPLOOP}, match={Accelerator}, replace={Accel.}]
			\step[fieldsource=\regexp{$MAPLOOP}, match={Annual}, replace={Annu.}]
			\step[fieldsource=\regexp{$MAPLOOP}, match={Acoustics}, replace={Acoust.}]
			\step[fieldsource=\regexp{$MAPLOOP}, match={Apparatus}, replace={App.}]
			\step[fieldsource=\regexp{$MAPLOOP}, match={Active}, replace={Act.}]
			\step[fieldsource=\regexp{$MAPLOOP}, match={Applications}, replace={Appl.}]
			\step[fieldsource=\regexp{$MAPLOOP}, match={Administration}, replace={Admin.}]
			\step[fieldsource=\regexp{$MAPLOOP}, match={Applied}, replace={Appl.}]
			\step[fieldsource=\regexp{$MAPLOOP}, match={Administrative}, replace={Administ.}]
			\step[fieldsource=\regexp{$MAPLOOP}, match={Approximate}, replace={Approx.}]
			\step[fieldsource=\regexp{$MAPLOOP}, match={Advanced}, replace={Adv.}]
			\step[fieldsource=\regexp{$MAPLOOP}, match={Advances}, replace={Adv.}]
			\step[fieldsource=\regexp{$MAPLOOP}, match={Archives}, replace={Arch.}]
			\step[fieldsource=\regexp{$MAPLOOP}, match={Archive}, replace={Arch.}]
			\step[fieldsource=\regexp{$MAPLOOP}, match={Aeronautics}, replace={Aeronaut.}]
			\step[fieldsource=\regexp{$MAPLOOP}, match={Artificial}, replace={Artif.}]
			\step[fieldsource=\regexp{$MAPLOOP}, match={Aerospace}, replace={Aerosp.}]
			\step[fieldsource=\regexp{$MAPLOOP}, match={Assembly}, replace={Assem.}]
			\step[fieldsource=\regexp{$MAPLOOP}, match={Affective}, replace={Affect.}]
			\step[fieldsource=\regexp{$MAPLOOP}, match={Association}, replace={Assoc.}]
			\step[fieldsource=\regexp{$MAPLOOP}, match={Africa}, replace={Afr.}]
			\step[fieldsource=\regexp{$MAPLOOP}, match={African}, replace={Afr.}]
			\step[fieldsource=\regexp{$MAPLOOP}, match={Astronomy}, replace={Astron.}]
			\step[fieldsource=\regexp{$MAPLOOP}, match={Aircraft}, replace={Aircr.}]
			\step[fieldsource=\regexp{$MAPLOOP}, match={Astronautics}, replace={Astronaut.}]
			\step[fieldsource=\regexp{$MAPLOOP}, match={Algebraic}, replace={Algebr.}]
			\step[fieldsource=\regexp{$MAPLOOP}, match={Astrophysics}, replace={Astrophys.}]
			\step[fieldsource=\regexp{$MAPLOOP}, match={American}, replace={Amer.}]
			\step[fieldsource=\regexp{$MAPLOOP}, match={Atmosphere}, replace={Atmos.}]
			\step[fieldsource=\regexp{$MAPLOOP}, match={Atomic}, replace={At.}]
			\step[fieldsource=\regexp{$MAPLOOP}, match={Atoms}, replace={At.}]
			\step[fieldsource=\regexp{$MAPLOOP}, match={Broadcasting}, replace={Broadcast.}]
			\step[fieldsource=\regexp{$MAPLOOP}, match={Australasian}, replace={Australas.}]
			\step[fieldsource=\regexp{$MAPLOOP}, match={Bulletin}, replace={Bull.}]
			\step[fieldsource=\regexp{$MAPLOOP}, match={Australia}, replace={Aust.}]
			\step[fieldsource=\regexp{$MAPLOOP}, match={Bureau}, replace={Bur.}]
			\step[fieldsource=\regexp{$MAPLOOP}, match={{Automatic }}, replace={Autom.}]
			\step[fieldsource=\regexp{$MAPLOOP}, match={Business}, replace={Bus.}]
			\step[fieldsource=\regexp{$MAPLOOP}, match={Automation}, replace={Automat.}]
			\step[fieldsource=\regexp{$MAPLOOP}, match={Canadian}, replace={Can.}]
			\step[fieldsource=\regexp{$MAPLOOP}, match={Automotive}, replace={Automot.}]
			\step[fieldsource=\regexp{$MAPLOOP}, match={Automobiles}, replace={Automob.}]
			\step[fieldsource=\regexp{$MAPLOOP}, match={Automobile}, replace={Automob.}]
			\step[fieldsource=\regexp{$MAPLOOP}, match={Ceramic}, replace={Ceram.}]
			\step[fieldsource=\regexp{$MAPLOOP}, match={Autonomous}, replace={Auton.}]
			\step[fieldsource=\regexp{$MAPLOOP}, match={Chemical}, replace={Chem.}]
			\step[fieldsource=\regexp{$MAPLOOP}, match={Behavioral}, replace={Behav.}]
			\step[fieldsource=\regexp{$MAPLOOP}, match={Behavior}, replace={Behav.}]
			\step[fieldsource=\regexp{$MAPLOOP}, match={Chinese}, replace={Chin.}]
			\step[fieldsource=\regexp{$MAPLOOP}, match={Belgian}, replace={Belg.}]
			\step[fieldsource=\regexp{$MAPLOOP}, match={Climatology}, replace={Climatol.}]
			\step[fieldsource=\regexp{$MAPLOOP}, match={Biochemical}, replace={Biochem.}]
			\step[fieldsource=\regexp{$MAPLOOP}, match={Clinical}, replace={Clin.}]
			\step[fieldsource=\regexp{$MAPLOOP}, match={Bioinformatics}, replace={Bioinf.}]
			\step[fieldsource=\regexp{$MAPLOOP}, match={Cognitive}, replace={Cogn.}]
			\step[fieldsource=\regexp{$MAPLOOP}, match={Biology, Biological}, replace={Biol.}]
			\step[fieldsource=\regexp{$MAPLOOP}, match={Colloquium}, replace={Colloq.}]
			\step[fieldsource=\regexp{$MAPLOOP}, match={Kolloquium}, replace={Kolloq.}]
			\step[fieldsource=\regexp{$MAPLOOP}, match={Biomedical}, replace={Biomed.}]
			\step[fieldsource=\regexp{$MAPLOOP}, match={Communications}, replace={Commun.}]
			\step[fieldsource=\regexp{$MAPLOOP}, match={Communication}, replace={Commun.}]
			\step[fieldsource=\regexp{$MAPLOOP}, match={Biophysics}, replace={Biophys.}]
			\step[fieldsource=\regexp{$MAPLOOP}, match={Compatibility}, replace={Compat.}]
			\step[fieldsource=\regexp{$MAPLOOP}, match={British}, replace={Brit.}]
			\step[fieldsource=\regexp{$MAPLOOP}, match={Components}, replace={Compon.}]
			\step[fieldsource=\regexp{$MAPLOOP}, match={Component}, replace={Compon.}]
			\step[fieldsource=\regexp{$MAPLOOP}, match={Computational}, replace={Comput.}]
			\step[fieldsource=\regexp{$MAPLOOP}, match={Delivery}, replace={Del.}]
			\step[fieldsource=\regexp{$MAPLOOP}, match={Computers}, replace={Comput.}]
			\step[fieldsource=\regexp{$MAPLOOP}, match={Computer}, replace={Comput.}]
			\step[fieldsource=\regexp{$MAPLOOP}, match={Department}, replace={Dept.}]
			\step[fieldsource=\regexp{$MAPLOOP}, match={Computing}, replace={Comput.}]
			\step[fieldsource=\regexp{$MAPLOOP}, match={Design}, replace={Des.}]
			\step[fieldsource=\regexp{$MAPLOOP}, match={Condensed}, replace={Condens.}]
			\step[fieldsource=\regexp{$MAPLOOP}, match={Detector}, replace={Detect.}]
			\step[fieldsource=\regexp{$MAPLOOP}, match={Conferences}, replace={Conf.}]
			\step[fieldsource=\regexp{$MAPLOOP}, match={Conference}, replace={Conf.}]
			\step[fieldsource=\regexp{$MAPLOOP}, match={Development}, replace={Develop.}]
			\step[fieldsource=\regexp{$MAPLOOP}, match={Congress}, replace={Congr.}]
			\step[fieldsource=\regexp{$MAPLOOP}, match={Differential}, replace={Differ.}]
			\step[fieldsource=\regexp{$MAPLOOP}, match={Consumer}, replace={Consum.}]
			\step[fieldsource=\regexp{$MAPLOOP}, match={Digest}, replace={Dig.}] 
			\step[fieldsource=\regexp{$MAPLOOP}, match={{ of the }}, replace={{ }}]
			\step[fieldsource=\regexp{$MAPLOOP}, match={{ of }}, replace={{ }}]
			\step[fieldsource=\regexp{$MAPLOOP}, match={{Of }}, replace={{ }}]
			\step[fieldsource=\regexp{$MAPLOOP}, match={{ on }}, replace={{ }}]
			\step[fieldsource=\regexp{$MAPLOOP}, match={{ On }}, replace={{ }}]
			\step[fieldsource=\regexp{$MAPLOOP}, match={{On }}, replace={{ }}]
			\step[fieldsource=\regexp{$MAPLOOP}, match={{ in }}, replace={{ }}]
			\step[fieldsource=\regexp{$MAPLOOP}, match={{ In }}, replace={{ }}]
			\step[fieldsource=\regexp{$MAPLOOP}, match={{In }}, replace={{ }}]
			\step[fieldsource=\regexp{$MAPLOOP}, match={{ the }}, replace={{ }}] 
			\step[fieldsource=\regexp{$MAPLOOP}, match={{ The }}, replace={{ }}] 
			\step[fieldsource=\regexp{$MAPLOOP}, match={{The }}, replace={}] 
			\step[fieldsource=\regexp{$MAPLOOP}, match={First}, replace={1st}]
			\step[fieldsource=\regexp{$MAPLOOP}, match={Second}, replace={2nd}]
			\step[fieldsource=\regexp{$MAPLOOP}, match={Third}, replace={3rd}]
			\step[fieldsource=\regexp{$MAPLOOP}, match={Fourth}, replace={4th}]
			\step[fieldsource=\regexp{$MAPLOOP}, match={Fifth}, replace={5th}]
			\step[fieldsource=\regexp{$MAPLOOP}, match={Sixth}, replace={6th}]
			\step[fieldsource=\regexp{$MAPLOOP}, match={Seventh}, replace={7th}]
			\step[fieldsource=\regexp{$MAPLOOP}, match={Eighth}, replace={8th}]
			\step[fieldsource=\regexp{$MAPLOOP}, match={Ninth}, replace={9th}]
			\step[fieldsource=\regexp{$MAPLOOP}, match={Tenth}, replace={10th}]
			\step[fieldsource=\regexp{$MAPLOOP}, match={Eleventh}, replace={11th}]
			\step[fieldsource=\regexp{$MAPLOOP}, match={Twelfth}, replace={12th}]
			\step[fieldsource=\regexp{$MAPLOOP}, match={Thirteenth}, replace={13th}]
			\step[fieldsource=\regexp{$MAPLOOP}, match={Fourteenth}, replace={14th}]
			\step[fieldsource=\regexp{$MAPLOOP}, match={Fifteenth}, replace={15th}]
			\step[fieldsource=\regexp{$MAPLOOP}, match={Sixteenth}, replace={16th}]
			\step[fieldsource=\regexp{$MAPLOOP}, match={Seventeenth}, replace={17th}]
			\step[fieldsource=\regexp{$MAPLOOP}, match={Eighteenth}, replace={18th}]
			\step[fieldsource=\regexp{$MAPLOOP}, match={Nineteenth}, replace={19th}]
			\step[fieldsource=\regexp{$MAPLOOP}, match={Twentieth}, replace={20th}]
		}
	}
}

%% file: macros.tex
\makeatletter
\newcounter{IEEE@bibentries}
\renewcommand\IEEEtriggeratref[1]{%
	\renewbibmacro{finentry}{%
		\stepcounter{IEEE@bibentries}%
		\ifthenelse{\equal{\value{IEEE@bibentries}}{#1}}
		{\finentry\@IEEEtriggercmd}
		{\finentry}%
	}%
}
\makeatother

\ifx \isaccepted \undefined
\newcommand\copyrighttext{%
	\footnotesize \centering This work has been submitted to the IEEE for possible publication.\\ Copyright may be transferred without notice, after which this version may no longer be accessible.}
\else
\newcommand\copyrighttext{%
	\footnotesize \parbox[t]{.11\textwidth}{\copyright{} \the\year~IEEE.} \parbox[t]{.89\textwidth}{Personal use of this material is permitted. Permission from IEEE must be obtained for all other uses, in any current or future media, including reprinting/republishing this material for advertising or promotional purposes, creating new collective works, for resale or redistribution to servers or lists, or reuse of any copyrighted component of this work in other works.}}
\fi
\newcommand\copyrightnotice{%
	\ifx \compileforpublish \undefined
	\else
	\begin{tikzpicture}[remember picture,overlay]
	\node[anchor=south,yshift=10.5pt] at (current page.south) {\parbox{\dimexpr\textwidth-\fboxsep-\fboxrule\relax}{\copyrighttext}};
	\end{tikzpicture}%
	\fi
}

\makeatletter
	\renewcommand\thefigure{\@arabic\c@figure}
	\renewcommand\fnum@figure{\figurename~\thefigure}
	\newcommand\captbc{\renewcommand\fnum@figure{\figurename~\thefigure~(continues on next page)}}
\makeatother

\makeatletter
	\renewcommand\thefigure{\@arabic\c@figure}
	\renewcommand\fnum@figure{\figurename~\thefigure}
	\newcommand\capcntd{\renewcommand\fnum@figure{\figurename~\thefigure~(continued)}}
\makeatother

\makeatletter
	\renewcommand\thetable{\@Roman\c@table}
	\renewcommand\fnum@table{\tablename~\thetable}
	\newcommand\tablecaptbc{\renewcommand\fnum@table{\tablename~\thetable~(continues on next page)}}
\makeatother

\makeatletter
	\renewcommand\thetable{\@Roman\c@table}
	\renewcommand\fnum@table{\tablename~\thetable}
	\newcommand\tablecapcntdtbc{\renewcommand\fnum@table{\tablename~\thetable~(continued, continues on next page)}}
\makeatother

\makeatletter
\renewcommand\thetable{\@Roman\c@table}
\renewcommand\fnum@table{\tablename~\thetable}
\newcommand\tablecapcntd{\renewcommand\fnum@table{\tablename~\thetable~(continued)}}
\makeatother

\makeatletter
	\tikzset{store number of columns in/.style={execute at end matrix={\xdef#1{\the\pgf@matrix@numberofcolumns}}},
			 store number of rows in/.style={execute at end matrix={\xdef#1{\the\pgfmatrixcurrentrow}}}
		 	}
\makeatother

\newcommand{\todo}[1]{\emph{\textcolor{red}{(TODO: #1)}}}
\renewcommand{\todo}[1]{}

\newcommand\ubar[1]{\stackunder[1.2pt]{$#1$}{\rule{.8ex}{.075ex}}}

\newcommand{\cf}{cf.}
\newcommand{\eg}{e.\,g.}
\newcommand{\ia}{i.\,a.}
\newcommand{\ie}{i.\,e.}

\definecolor{surveyRed}{named}{tuRed}%

\definecolor{surveyLightRed}{named}{tuRed}%
\colorlet{surveyLightRed}{tuRed!25}

\definecolor{surveyBlue}{named}{tuBlue}%

\definecolor{surveyLightBlue}{named}{tuBlue}%
\colorlet{surveyLightBlue}{tuBlue!25}

\pgfplotsset{
	show sum on right/.style={
		/pgfplots/scatter/@post marker code/.append code={%
			\pgfmathprintnumberto[fixed,assume math mode=true]{\pgfplotspointmeta}{\myval}
			\ifdim\myval pt<2 pt  
				\node[
				at={(normalized axis cs:%
					\pgfkeysvalueof{/data point/x},%
					\pgfkeysvalueof{/data point/y})%
				},
				anchor=west,inner sep=0,inner xsep=2pt,outer sep=0
				]
				{\myval};
			\else 
			\fi
		},
	},
}

\pgfplotsset{
	show sum on top/.style={
		/pgfplots/scatter/@post marker code/.append code={%
			\pgfmathprintnumberto[fixed,assume math mode=true]{\pgfplotspointmeta}{\myval}
			\ifdim\myval pt<2 pt  
				\node[
				at={(normalized axis cs:%
					\pgfkeysvalueof{/data point/x},%
					\pgfkeysvalueof{/data point/y})%
				},
				anchor=west,rotate=90,inner sep=0,inner xsep=2pt,outer sep=0
				]
				{\myval};
			\else 
			\fi
		},
	},
}						

\pgfplotsset{
	labelstyle/.style={
		visualization depends on={meta < 2 \as \valueisone},
		every node near coord/.append style={
			inner sep=0,
			outer sep=0, 
			fill= {\ifdim 1 pt=\valueisone pt \else white\fi}, 
			font=\footnotesize
		}
	}
}

\pgfplotsset{
	horizontalbar/.style={	    
		nodes near coords={%
			\pgfmathprintnumberto[fixed,assume math mode=true]{\pgfplotspointmeta}{\myval}%
			\pgfmathparse{\myval<2?:\myval}\pgfmathresult%
		},
		show sum on right,
		labelstyle		
	}
}	

\pgfplotsset{
	verticalbar/.style={	    
		nodes near coords={%
			\pgfmathprintnumberto[fixed,assume math mode=true]{\pgfplotspointmeta}{\myval}%
			\pgfmathparse{\myval<2?:\myval}\pgfmathresult%
		},
		show sum on top,
		labelstyle
	}
}

\tikzstyle{actuatorbar} = [draw=surveyRed, text=surveyRed,  pattern= north west lines, pattern color=surveyLightRed]
\tikzstyle{blowoutbar}  = [draw=surveyBlue,text=surveyBlue, pattern= north east lines, pattern color=surveyLightBlue]

%% file: abstract.tex
The advent of automated vehicles operating at SAE levels 4 and 5 poses high fault tolerance demands for all functions contributing to the driving task.
At the actuator level, fault-tolerant vehicle motion control, which exploits functional redundancies among the actuators, is one means to achieve the required degree of fault tolerance.
Therefore, we give a comprehensive overview of the state of the art in actuator fault-tolerant vehicle motion control with a focus on drive, brake, and steering degradations, as well as tire blowouts. 
This review shows that actuator fault-tolerant vehicle motion is a widely studied field; yet, the presented approaches differ with respect to many aspects.
To provide a starting point for future research, we survey the employed actuator topologies, the tolerated degradations, the presented control approaches, as well as the experiments conducted for validation.
Overall, and despite the large number of different approaches, the covered literature reveals the potential of increasing fault tolerance by fault-tolerant vehicle motion control. 
Thus, besides developing novel approaches or demonstrating real-time applicability, future research should aim at investigating limitations and enabling comparison of fault-tolerant motion control approaches in order to allow for a thorough safety argumentation. 

%% file: introduction.tex
\label{sec:introduction}

\input{literaturespecificcommands}

\IEEEPARstart{F}{ault}-tolerant vehicle systems, which have been the subject of intensive research for more than two decades, continue to gain importance with the advent of automated driving technology.
While the required safety level for the operation of automated vehicles according to SAE levels~4 and~5~\cite{sae_2018} is yet to be defined, it is indisputable that automated vehicles demand a high degree of fault tolerance throughout the automated driving system. %
On the part of vehicle actuators, redundant implementations are one way to compensate for the missing human fallback layer in SAE levels~4 and~5. 
However, redundancy is usually accompanied by increased costs caused by, \ia, additional weight and installation space complexity. 
These drawbacks become even more important for novel over-actuated topologies proposed in several automated vehicle prototypes, for instance those featuring wheel-individual drive, brake, and steering actuators.

Also in modern series production vehicles, there is a trend towards over-actuation in order to improve the vehicles' handling, ride comfort, and safety.
For instance, electronic stability control leverages wheel-individual brakes, and premium class vehicles feature chassis control systems that use torque vectoring and by-wire rear-axle steering.

The over-actuation offers additional opportunities for realizing fault-tolerant automated vehicles. 
Since drive, brake, and steering actuators generally impact both, the longitudinal and lateral vehicle motion, the resulting functional redundancies could be exploited either for avoiding physical redundancy at the actuator level or as a secondary fallback layer if the currently accepted single fault assumption for vehicle systems~\cite{isermann_2002} does not hold true for automated vehicles.

\stepcounter{footnote}

In either case,  the impact of the degraded actuators must be handled by fault-tolerant system layers superimposed on the actuator layer in order to obtain safe vehicle behavior.
In particular, vehicle motion control must be able to account for possible actuator degradations in order to be able to stabilize the vehicle on its given trajectory by using the remaining fully functional actuators%
\footnote{Following a hierarchical approach towards automated vehicle systems, \eg, as presented by \citet{matthaei_2015}, higher layers such as trajectory planning and decision making must adapt to the degradation as well.}.
Hence, we aim at giving a comprehensive overview of the current state of the art of actuator fault-tolerant vehicle motion control in this survey.

A few earlier literature reviews are available~\cite{
	zhang_2008,
	johansen_2013,
	wanner_2012a,
	vivas-lopez_2013,
	shyrokau_2014,
	kissai_2017,
	huang_2019b,
	shet_2020}, however, they cover only parts of the existing publications that address fault-tolerant motion control in the presence of actuator degradations. 
Cross-domain overviews are given by \citet{zhang_2008} as well as \citet{johansen_2013}.
In their \citeyear{zhang_2008} study, \citeauthor{zhang_2008} present a comprehensive review of reconfigurable control systems and include selected publications from the automotive domain. 
\citeauthor{johansen_2013}, in their overview from \citeyear{johansen_2013}, highlight the potential of control allocation towards fault tolerance, yet they give only one reference for a fault-tolerant control approach in the automotive domain.

Focusing on the automotive domain, \citet{wanner_2012a} review fault-tolerant vehicle design in general in their~\citeyear{wanner_2012a} study. 
With respect to fault-tolerant control in the presence of actuator degradations, the authors present general considerations as well as requirements and introduce a few selected publications on the topic.
\citet{vivas-lopez_2013} as well as \citet{shyrokau_2014} survey approaches towards global chassis control in their studies from~\citeyear{vivas-lopez_2013} and~\citeyear{shyrokau_2014}, respectively. 
Yet, the authors give only a few selected examples of fault-tolerant control approaches addressing actuator faults. 
The same applies to the more recent works of \citet{kissai_2017}, \citet{huang_2019b}, and \citet{shet_2020}, which are from~\citeyear{kissai_2017}, \citeyear{huang_2019b}, and~\citeyear{shet_2020}, respectively.
As \citeauthor{vivas-lopez_2013} and \citeauthor{shyrokau_2014}, \citeauthor{kissai_2017} present a survey on integrated vehicle motion control. 
Again, only one approach is mentioned that targets actuator fault-tolerant vehicle motion control. 
The work of \citeauthor{huang_2019b} focuses on steer-by-wire systems. 
While giving an extensive overview of steering-internal fault-tolerant control, it mentions only a few publications that present approaches to overcome steering degradations by means of other healthy actuators. 
Finally, also the recent survey of \citeauthor{shet_2020} contains only a small selection of publications outlining fault-tolerant vehicle motion control approaches.

This survey is meant to serve as a starting point for future research in the field of fault-tolerant vehicle motion control.
Hence, after refining the focus in \autoref{sec:focus}, the subsequent sections are structured by the different ways in which researchers and practitioners can approach fault-tolerant vehicle motion control.
On the one hand, they can look for approaches to handle selected actuator degradations; therefore, \autoref{sec:degradations} provides an overview of the different degradation types addressed in the literature. 
On the other hand, researchers and practitioners can search for fault-tolerant vehicle motion control approaches for a given set of drive, brake, and steering actuators.
Actuator topologies are hence reviewed in \autoref{sec:topologies}.
To compare novel control approaches to the state of the art, \autoref{sec:approaches} considers the different control targets, the employed control techniques, and the related control structures.
Last but not least, \autoref{sec:experiments} gives an overview of the experiments that are used to evaluate fault-tolerant vehicle motion control approaches.

%% file: literaturespecificcommands.tex
\newcommand{\increasingTableCitations}{
\nocite{ahmadi_2020}
\nocite{alipour_2014}
\nocite{almeida_2013}
\nocite{amato_2020}
\nocite{bian_2016}
\nocite{bosche_2009}
\nocite{boudali_2018}
\nocite{buente_2006}
\nocite{chen_2011}
\nocite{chen_2014b}
\nocite{chen_2017a}
\nocite{chen_2018}
\nocite{chen_2019}
\nocite{chen_2019a}
\nocite{chen_2020}
\nocite{chen_2020a}
\nocite{chu_2012}
\nocite{dominguez-garcia_2004}
\nocite{dumont_2006}
\nocite{feng_2013}
\nocite{fu_2020a}
\nocite{gaspar_2015}
\nocite{guo_2019a}
\nocite{guo_2020}
\nocite{guo_2020a}
\nocite{hac_2006}
\nocite{hac_2006a}
\nocite{haddad_2013}
\nocite{hiraoka_2004}
\nocite{hoedt_2013}
\nocite{hu_2015}
\nocite{hu_2016a}
\nocite{hu_2017a}
\nocite{hu_2018}
\nocite{hu_2019}
\nocite{inpark_2013}
\nocite{ito_2013}
\nocite{jeon_2013}
\nocite{jing_2015}
\nocite{jing_2018}
\nocite{jonasson_2007a}
\nocite{jonasson_2008}
\nocite{jonasson_2016}
\nocite{jonasson_2017}
\nocite{khelladi_2020}
\nocite{kim_2016}
\nocite{kirli_2017}
\nocite{kissai_2018}
\nocite{knobel_2009}
\nocite{kou_2010}
\nocite{kruger_2010}
\nocite{lee_2020}
\nocite{li_2009}
\nocite{li_2013}
\nocite{li_2014}
\nocite{li_2014a}
\nocite{li_2016}
\nocite{li_2016a}
\nocite{li_2021a}
\nocite{liu_2012}
\nocite{liu_2012a}
\nocite{liu_2013a}
\nocite{liu_2015}
\nocite{liu_2016a}
\nocite{liu_2017}
\nocite{liu_2020}
\nocite{lu_2014}
\nocite{lu_2016}
\nocite{luo_2016}
\nocite{luo_2019}
\nocite{marino_2007}
\nocite{mi_2015}
\nocite{mihaly_2017}
\nocite{miyazaki_2005}
\nocite{moseberg_2015}
\nocite{moseberg_2015a}
\nocite{moseberg_2016}
\nocite{mutoh_2007}
\nocite{mutoh_2009}
\nocite{nemeth_2012}
\nocite{nguyen_2017}
\nocite{pathak_2008}
\nocite{plumlee_2004}
\nocite{plumlee_2004a}
\nocite{polmans_2014}
\nocite{poussot-vassal_2008}
\nocite{poussot-vassal_2008a}
\nocite{ramanathanvenkita_2020}
\nocite{raveendran_2020}
\nocite{reinold_2010}
\nocite{ringdorfer_2013}
\nocite{sakthivel_2018}
\nocite{sename_2013}
\nocite{sho_2013}
\nocite{song_2003}
\nocite{song_2005}
\nocite{song_2011}
\nocite{stolte_2018}
\nocite{subroto_2020}
\nocite{sun_2014}
\nocite{temiz_2018}
\nocite{temiz_2019}
\nocite{temiz_2020}
\nocite{tian_2018}
\nocite{wada_2012}
\nocite{wada_2013}
\nocite{wan_2020}
\nocite{wang_2007}
\nocite{wang_2009}
\nocite{wang_2010a}
\nocite{wang_2011}
\nocite{wang_2011b}
\nocite{wang_2012}
\nocite{wang_2013}
\nocite{wang_2013b}
\nocite{wang_2014}
\nocite{wang_2014a}
\nocite{wang_2014b}
\nocite{wang_2014c}
\nocite{wang_2016a}
\nocite{wang_2016c}
\nocite{wang_2018}
\nocite{wang_2018b}
\nocite{wang_2019a}
\nocite{wang_2020}
\nocite{wang_2020c}
\nocite{wang_2020g}
\nocite{wanner_2012}
\nocite{wanner_2015}
\nocite{wanner_2015a}
\nocite{wong_2016}
\nocite{xin_2014}
\nocite{xin_2016}
\nocite{yang_2008}
\nocite{yang_2010}
\nocite{yim_2014}
\nocite{yin_2015}
\nocite{zhang_2014}
\nocite{zhang_2014a}
\nocite{zhang_2015}
\nocite{zhang_2016a}
\nocite{zhang_2016b}
\nocite{zhang_2017}
\nocite{zhang_2018a}
\nocite{zhang_2018b}
\nocite{zhang_2019}
\nocite{zhang_2020a}
\nocite{zhang_2020b}
\nocite{zong_2013}
\nocite{chen_2014a}
\nocite{fenyes_2020}
\nocite{guo_2012}
\nocite{haddad_2012}
\nocite{jing_2019}
\nocite{li_2013a}
\nocite{li_2020a}
\nocite{liu_2014}
\nocite{lu_2019}
\nocite{meng_2019}
\nocite{mo_2013}
\nocite{patwardhan_1994}
\nocite{patwardhan_1994a}
\nocite{patwardhan_1996}
\nocite{patwardhan_1997}
\nocite{wang_2015a}
\nocite{wang_2015b}
\nocite{wang_2016}
\nocite{wang_2016b}
\nocite{wang_2017}
\nocite{wang_2018a}
\nocite{yang_2019}
\nocite{yu_2014}
}

\def\numPapers{149}

\def\numDegBrakeDriveTorqueConst{15}
\def\numDegBrakeDriveTorqueZero{66}
\def\numDegBrakeDriveLockingSpinning{6}
\def\numDegBrakeDriveTorqueReducedRange{54}
\def\numDegSteeringReducedRange{20}
\def\numDegSteeringReducedDynamics{5}
\def\numDegSteeringFixed{35}
\def\numDegSteeringFree{23}
\newcommand{\citeDegBrakeDriveTorqueConst}{\cite{chen_2017a,chen_2019a,jing_2015,jonasson_2007a,jonasson_2008,ringdorfer_2013,stolte_2018,subroto_2020,wang_2013,wang_2014,wang_2014b,wang_2018,wanner_2012,wanner_2015,zhang_2020b}}
\newcommand{\citeDegBrakeDriveTorqueZero}{\cite{ahmadi_2020,chu_2012,feng_2013,fu_2020a,gaspar_2015,hac_2006,hac_2006a,hiraoka_2004,hoedt_2013,ito_2013,jeon_2013,jing_2018,jonasson_2008,kim_2016,kissai_2018,knobel_2009,kou_2010,kruger_2010,lee_2020,li_2014,li_2014a,li_2016,li_2016a,liu_2012,liu_2012a,liu_2013a,liu_2015,liu_2016a,liu_2017,lu_2014,luo_2016,luo_2019,mihaly_2017,miyazaki_2005,moseberg_2015,moseberg_2015a,moseberg_2016,mutoh_2007,mutoh_2009,nemeth_2012,nguyen_2017,pathak_2008,ramanathanvenkita_2020,sho_2013,stolte_2018,sun_2014,wada_2012,wada_2013,wan_2020,wang_2007,wang_2009,wang_2012,wang_2013,wang_2013b,wang_2014a,wang_2014c,wang_2018,wang_2020g,wong_2016,xin_2014,xin_2016,yang_2008,yang_2010,zhang_2014a,zhang_2020a,zong_2013}}
\newcommand{\citeDegBrakeDriveLockingSpinning}{\cite{hoedt_2013,moseberg_2015,moseberg_2015a,moseberg_2016,stolte_2018,wanner_2015a}}
\newcommand{\citeDegBrakeDriveTorqueReducedRange}{\cite{alipour_2014,almeida_2013,amato_2020,bian_2016,chen_2014b,chen_2018,chen_2020,chu_2012,guo_2019a,guo_2020,inpark_2013,jeon_2013,jing_2015,jing_2018,kim_2016,li_2013,liu_2020,nguyen_2017,poussot-vassal_2008,poussot-vassal_2008a,raveendran_2020,sakthivel_2018,sename_2013,song_2003,song_2005,stolte_2018,subroto_2020,temiz_2018,temiz_2019,temiz_2020,wang_2010a,wang_2011,wang_2011b,wang_2012,wang_2013,wang_2014,wang_2014a,wang_2014b,wang_2016c,wang_2018,wang_2019a,wong_2016,yang_2008,yang_2010,yin_2015,zhang_2014,zhang_2014a,zhang_2015,zhang_2016b,zhang_2018a,zhang_2018b,zhang_2019,zhang_2020a,zhang_2020b}}
\newcommand{\citeDegSteeringReducedRange}{\cite{bosche_2009,chen_2011,guo_2020,guo_2020a,haddad_2013,jing_2015,jing_2018,li_2013,sakthivel_2018,song_2011,stolte_2018,subroto_2020,temiz_2020,wada_2012,wada_2013,wang_2014a,wang_2014b,wang_2018,wang_2018b,wang_2020}}
\newcommand{\citeDegSteeringReducedDynamics}{\cite{stolte_2018,temiz_2019,wang_2014a,yang_2008,yang_2010}}
\newcommand{\citeDegSteeringFixed}{\cite{bosche_2009,boudali_2018,buente_2006,chen_2019,chen_2019a,dumont_2006,hiraoka_2004,hoedt_2013,jonasson_2008,khelladi_2020,kissai_2018,knobel_2009,li_2009,lu_2016,marino_2007,moseberg_2015,moseberg_2015a,moseberg_2016,nemeth_2012,pathak_2008,plumlee_2004,plumlee_2004a,reinold_2010,stolte_2018,tian_2018,wada_2012,wada_2013,wang_2007,wang_2014b,wang_2018,wang_2018b,wang_2020,wang_2020c,yim_2014,zhang_2016a}}
\newcommand{\citeDegSteeringFree}{\cite{chen_2020a,dominguez-garcia_2004,hu_2015,hu_2016a,hu_2017a,hu_2018,hu_2019,ito_2013,jonasson_2016,jonasson_2017,kirli_2017,li_2016,li_2016a,li_2021a,mi_2015,polmans_2014,tian_2018,wang_2014a,wang_2014b,wang_2016a,yang_2008,yang_2010,zhang_2017}}

\def\numCtrlTargetPose{1}
\def\numCtrlTargetPath{27}
\def\numCtrlTargetVeD{65}
\def\numCtrlTargetLaD{48}
\def\numCtrlTargetLoD{6}
\def\numCtrlTargetYA{2}
\def\numCtrlTargetElse{0}
\newcommand{\citeCtrlTargetPose}{stolte_2018}
\newcommand{\citeCtrlTargetPath}{ahmadi_2020,boudali_2018,chen_2011,chen_2018,chen_2019,chen_2019a,chen_2020a,dumont_2006,guo_2020,guo_2020a,haddad_2013,hu_2016a,hu_2019,khelladi_2020,luo_2019,pathak_2008,song_2011,wang_2020,wang_2020c,yang_2008,yang_2010,zhang_2014,zhang_2014a,zhang_2015,zhang_2018a,zhang_2019,zhang_2020b}
\newcommand{\citeCtrlTargetVeD}{alipour_2014,almeida_2013,amato_2020,bian_2016,buente_2006,chen_2017a,chen_2020,feng_2013,gaspar_2015,guo_2019a,hac_2006,hac_2006a,hoedt_2013,inpark_2013,jonasson_2007a,jonasson_2008,kim_2016,kirli_2017,knobel_2009,kou_2010,kruger_2010,li_2013,li_2014,li_2014a,li_2016,li_2016a,liu_2012,liu_2013a,liu_2015,liu_2016a,liu_2020,luo_2016,mihaly_2017,moseberg_2015,moseberg_2015a,moseberg_2016,nguyen_2017,reinold_2010,subroto_2020,sun_2014,temiz_2018,temiz_2019,temiz_2020,wan_2020,wang_2007,wang_2009,wang_2010a,wang_2011,wang_2011b,wang_2012,wang_2013,wang_2014,wang_2014a,wang_2014b,wang_2018,wang_2018b,wang_2019a,wanner_2012,wanner_2015,wanner_2015a,zhang_2016a,zhang_2016b,zhang_2018b,zhang_2020a,zong_2013}
\newcommand{\citeCtrlTargetLaD}{bosche_2009,chen_2014b,dominguez-garcia_2004,fu_2020a,hiraoka_2004,hu_2015,hu_2017a,hu_2018,ito_2013,jeon_2013,jing_2015,jing_2018,jonasson_2016,jonasson_2017,kissai_2018,lee_2020,li_2009,li_2021a,liu_2012a,liu_2017,lu_2014,lu_2016,marino_2007,mi_2015,nemeth_2012,plumlee_2004,plumlee_2004a,polmans_2014,poussot-vassal_2008,poussot-vassal_2008a,ramanathanvenkita_2020,ringdorfer_2013,sakthivel_2018,sename_2013,sho_2013,tian_2018,wada_2012,wada_2013,wang_2014c,wang_2016a,wang_2016c,wang_2020g,wong_2016,xin_2014,xin_2016,yim_2014,yin_2015,zhang_2017}
\newcommand{\citeCtrlTargetLoD}{chu_2012,mutoh_2007,mutoh_2009,song_2003,song_2005,wang_2013b}
\newcommand{\citeCtrlTargetYA}{miyazaki_2005,raveendran_2020}
\newcommand{\citeCtrlTargetElse}{}

\def\numCtrlStructureHierarchical{66}
\def\numCtrlStructureParallel{30}
\def\numCtrlStructureNotStructured{65}
\def\numCtrlStructureAlternatives{6}

\newcommand{\citeCtrlStructureHierarchical}{ahmadi_2020,almeida_2013,amato_2020,boudali_2018,buente_2006,chen_2014b,chen_2019,chen_2019a,chen_2020,chen_2020a,dumont_2006,feng_2013,fu_2020a,gaspar_2015,guo_2020,hoedt_2013,hu_2015,inpark_2013,jeon_2013,jing_2018,jonasson_2007a,jonasson_2008,khelladi_2020,kissai_2018,knobel_2009,kou_2010,kruger_2010,li_2009,li_2014,liu_2013a,liu_2015,liu_2017,liu_2020,lu_2016,moseberg_2015,moseberg_2015a,moseberg_2016,ringdorfer_2013,sho_2013,subroto_2020,temiz_2018,temiz_2019,temiz_2020,wada_2012,wada_2013,wan_2020,wang_2007,wang_2009,wang_2010a,wang_2011b,wang_2012,wang_2014a,wang_2014c,wang_2016c,wang_2019a,wang_2020g,wanner_2012,wanner_2015,wanner_2015a,wong_2016,xin_2014,xin_2016,yim_2014,zhang_2018b,zhang_2020b,zong_2013}
\newcommand{\citeCtrlStructureParallel}{almeida_2013,boudali_2018,chen_2014b,fu_2020a,hu_2016a,hu_2017a,hu_2018,ito_2013,jonasson_2016,jonasson_2017,khelladi_2020,kirli_2017,lee_2020,liu_2012,moseberg_2015,moseberg_2015a,moseberg_2016,nguyen_2017,pathak_2008,wang_2011,wang_2018,wang_2018b,wang_2019a,wang_2020,wang_2020c,xin_2014,xin_2016,yang_2008,yang_2010,zhang_2018b}
\newcommand{\citeCtrlStructureSingle}{alipour_2014,bian_2016,bosche_2009,chen_2011,chen_2017a,chen_2018,chu_2012,dominguez-garcia_2004,guo_2019a,guo_2020a,hac_2006,hac_2006a,haddad_2013,hiraoka_2004,hu_2019,jing_2015,kim_2016,li_2013,li_2014a,li_2016,li_2016a,li_2021a,liu_2012a,liu_2016a,lu_2014,luo_2016,luo_2019,marino_2007,mi_2015,mihaly_2017,miyazaki_2005,mutoh_2007,mutoh_2009,nemeth_2012,plumlee_2004,plumlee_2004a,polmans_2014,poussot-vassal_2008,poussot-vassal_2008a,ramanathanvenkita_2020,raveendran_2020,reinold_2010,sakthivel_2018,sename_2013,song_2003,song_2005,song_2011,stolte_2018,sun_2014,tian_2018,wang_2013,wang_2013b,wang_2014,wang_2014b,wang_2016a,yin_2015,zhang_2014,zhang_2014a,zhang_2015,zhang_2016a,zhang_2016b,zhang_2017,zhang_2018a,zhang_2019,zhang_2020a}
\newcommand{\citeCtrlAlternativeApproaches}{bian_2016,ito_2013,miyazaki_2005,wang_2014c,wanner_2012,wanner_2015}

\newcommand{\citeSteerByDrive}{\cite{chen_2020a,hu_2015,hu_2016a,hu_2017a,hu_2018,hu_2019,kirli_2017,li_2021a,mi_2015,polmans_2014,tian_2018,wang_2016a}}
\newcommand{\citeSteerByBrake}{\cite{dominguez-garcia_2004,jonasson_2016,jonasson_2017,zhang_2017}}

\newcommand{\citeCtrlLawPCA}{chen_2014b,feng_2013,khelladi_2020,knobel_2009,moseberg_2015,moseberg_2015a,moseberg_2016,subroto_2020,wanner_2012,wanner_2015,wanner_2015a,yim_2014}

\newcommand{\citeCtrlLawAvCA}{amato_2020,ito_2013,jing_2018,jonasson_2007a,liu_2017,sho_2013,xin_2014,xin_2016}

\newcommand{\citeCtrlLawFzCA}{ahmadi_2020,gaspar_2015,hac_2006a,li_2009,li_2014a,ramanathanvenkita_2020,ringdorfer_2013,wang_2013b}

\newcommand{\citeCtrlLawLtwo}{wang_2014b}

\newcommand{\citeExpMiL}{ahmadi_2020,alipour_2014,almeida_2013,amato_2020,bian_2016,bosche_2009,boudali_2018,buente_2006,chen_2011,chen_2014b,chen_2017a,chen_2018,chen_2019,chen_2019a,chen_2020,chen_2020a,dominguez-garcia_2004,dumont_2006,feng_2013,fu_2020a,gaspar_2015,guo_2019a,guo_2020,guo_2020a,hac_2006,hac_2006a,haddad_2013,hiraoka_2004,hoedt_2013,hu_2015,hu_2016a,hu_2017a,hu_2018,hu_2019,inpark_2013,ito_2013,jeon_2013,jing_2015,jing_2018,jonasson_2007a,jonasson_2008,jonasson_2016,jonasson_2017,khelladi_2020,kirli_2017,kissai_2018,knobel_2009,kou_2010,lee_2020,li_2009,li_2013,li_2014,li_2014a,li_2016,li_2016a,liu_2012,liu_2012a,liu_2013a,liu_2016a,liu_2017,liu_2020,lu_2014,lu_2016,luo_2016,luo_2019,marino_2007,mi_2015,mihaly_2017,miyazaki_2005,moseberg_2015,moseberg_2015a,moseberg_2016,mutoh_2007,mutoh_2009,nemeth_2012,nguyen_2017,pathak_2008,plumlee_2004,plumlee_2004a,polmans_2014,poussot-vassal_2008,poussot-vassal_2008a,reinold_2010,ringdorfer_2013,sakthivel_2018,sename_2013,sho_2013,song_2003,song_2005,song_2011,stolte_2018,subroto_2020,sun_2014,temiz_2018,temiz_2019,temiz_2020,tian_2018,wada_2012,wada_2013,wan_2020,wang_2007,wang_2009,wang_2010a,wang_2011,wang_2011b,wang_2012,wang_2013b,wang_2014a,wang_2014b,wang_2014c,wang_2016a,wang_2016c,wang_2018,wang_2018b,wang_2019a,wang_2020,wang_2020c,wang_2020g,wanner_2012,wanner_2015,wong_2016,xin_2014,xin_2016,yang_2008,yang_2010,yim_2014,yin_2015,zhang_2014,zhang_2014a,zhang_2015,zhang_2016a,zhang_2016b,zhang_2017,zhang_2018a,zhang_2018b,zhang_2019,zhang_2020a,zong_2013}
\newcommand{\citeExpHiL}{kim_2016,raveendran_2020,zhang_2020b}
\newcommand{\citeExpViL}{chu_2012,kruger_2010,li_2021a,liu_2015,ramanathanvenkita_2020,wang_2013,wang_2014,wanner_2015a}

\newcommand{\citeSimToolCarSim}{alipour_2014,almeida_2013,amato_2020,chen_2014b,chen_2017a,chen_2018,chen_2019,chen_2019a,chen_2020a,fu_2020a,gaspar_2015,guo_2019a,haddad_2013,hiraoka_2004,hu_2016a,hu_2017a,hu_2018,hu_2019,inpark_2013,jeon_2013,jing_2015,jing_2018,kim_2016,kou_2010,lee_2020,liu_2016a,liu_2017,liu_2020,luo_2016,luo_2019,mi_2015,nemeth_2012,nguyen_2017,sho_2013,tian_2018,wan_2020,wang_2007,wang_2009,wang_2010a,wang_2011,wang_2011b,wang_2012,wang_2013b,wang_2014a,wang_2014b,wang_2014c,wang_2016a,wang_2016c,wang_2020g,wong_2016,yim_2014,yin_2015,zhang_2016b,zhang_2017,zhang_2018a,zhang_2018b,zhang_2019,zhang_2020a,zhang_2020b}
\newcommand{\citeSimToolCarMaker}{khelladi_2020,raveendran_2020}
\newcommand{\citeSimToolDyna}{ringdorfer_2013}
\newcommand{\citeSimToolAdams}{guo_2020,guo_2020a}
\newcommand{\citeSimToolAVL}{chen_2020}
\newcommand{\citeSimToolVeDYNA}{wang_2018,wang_2018b,wang_2019a,wang_2020,wang_2020c}

\newcommand{\citeSimToolAMESim}{kissai_2018}

\newcommand{\citeTrajDLC}{chen_2014b,chen_2020a,fu_2020a,guo_2019a,guo_2020a,hu_2018,kim_2016,kirli_2017,kissai_2018,kou_2010,kruger_2010,li_2016,li_2016a,li_2021a,marino_2007,nguyen_2017,plumlee_2004a,poussot-vassal_2008,poussot-vassal_2008a,ramanathanvenkita_2020,stolte_2018,tian_2018,wang_2007,wang_2009,wang_2014b,wang_2014c,wang_2016a,wang_2018,wang_2018b,wang_2019a,wong_2016,yim_2014,yin_2015,zhang_2020a,zhang_2020b}
\newcommand{\citeTrajSLC}{alipour_2014,almeida_2013,boudali_2018,chen_2019,chen_2019a,guo_2019a,hiraoka_2004,hu_2015,hu_2016a,hu_2017a,hu_2019,jing_2015,jing_2018,knobel_2009,li_2016,li_2016a,liu_2020,lu_2016,mi_2015,pathak_2008,plumlee_2004,sakthivel_2018,subroto_2020,temiz_2018,temiz_2019,temiz_2020,wang_2010a,wang_2011,wang_2011b,wang_2012,wang_2014b,wang_2018,wang_2018b,wang_2019a,wang_2020,wang_2020c,wang_2020g,zhang_2016b,zhang_2018b,zhang_2019}
\newcommand{\citeTrajJTurn}{bian_2016,hu_2016a,hu_2019,jing_2018,li_2009,li_2014,li_2016,li_2016a,li_2021a,liu_2013a,subroto_2020,tian_2018,wang_2010a,wang_2011,wang_2011b,wang_2012,wang_2014b,wang_2016a,wang_2018,wang_2018b,wang_2019a,wang_2020c,zhang_2020b,zong_2013}
\newcommand{\citeTrajCirc}{ahmadi_2020,buente_2006,chen_2011,chen_2019,chen_2019a,chen_2020a,hiraoka_2004,hu_2018,ito_2013,jonasson_2007a,jonasson_2008,jonasson_2016,jonasson_2017,lee_2020,li_2021a,liu_2016a,lu_2016,moseberg_2016,mutoh_2007,ringdorfer_2013,sho_2013,song_2011,subroto_2020,wang_2013,wang_2014,wanner_2012,wanner_2015,xin_2014,xin_2016,zhang_2017}
\newcommand{\citeTrajCurved}{ahmadi_2020,bian_2016,buente_2006,chen_2011,chen_2019,chen_2019a,chen_2020a,hiraoka_2004,hu_2016a,hu_2018,hu_2019,ito_2013,jing_2018,jonasson_2007a,jonasson_2008,jonasson_2016,jonasson_2017,lee_2020,li_2009,li_2014,li_2016,li_2016a,liu_2013a,liu_2016a,lu_2016,moseberg_2016,mutoh_2007,ringdorfer_2013,sho_2013,song_2011,subroto_2020,tian_2018,wang_2010a,wang_2011,wang_2011b,wang_2012,wang_2013,wang_2014,wang_2014b,wang_2016a,wang_2018,wang_2018b,wang_2019a,wang_2020c,wanner_2012,wanner_2015,xin_2014,xin_2016,zhang_2017,zhang_2020b,zong_2013}
\newcommand{\citeTrajStraight}{almeida_2013,amato_2020,chen_2011,chen_2017a,chen_2018,feng_2013,fu_2020a,hac_2006,hac_2006a,inpark_2013,jeon_2013,jonasson_2007a,lee_2020,li_2014,li_2014a,liu_2012,liu_2012a,liu_2016a,liu_2017,liu_2020,lu_2014,luo_2016,luo_2019,miyazaki_2005,moseberg_2015,moseberg_2015a,mutoh_2007,mutoh_2009,pathak_2008,raveendran_2020,reinold_2010,ringdorfer_2013,sho_2013,song_2003,song_2005,song_2011,sun_2014,wan_2020,wang_2007,wang_2009,wang_2011,wang_2012,wang_2013,wang_2013b,wang_2014a,wang_2016c,wang_2018,wang_2020,wang_2020g,wanner_2015a,yang_2008,yang_2010,zhang_2014,zhang_2014a,zhang_2015,zhang_2016b,zhang_2018a,zhang_2018b,zhang_2019,zong_2013}
\newcommand{\citeTrajOther}{bosche_2009,chen_2018,chen_2020,chu_2012,dominguez-garcia_2004,dumont_2006,gaspar_2015,guo_2020,haddad_2013,hoedt_2013,khelladi_2020,li_2013,liu_2015,mihaly_2017,nemeth_2012,polmans_2014,wada_2012,wada_2013,wang_2020g,wong_2016,zhang_2016a,zhang_2017}

\newcommand{\publicationsPerYearTableData}{(1990,0)
(1991,0)
(1992,0)
(1993,0)
(1994,0)
(1995,0)
(1996,0)
(1997,0)
(1998,0)
(1999,0)
(2000,0)
(2001,0)
(2002,0)
(2003,1)
(2004,4)
(2005,2)
(2006,4)
(2007,4)
(2008,5)
(2009,5)
(2010,5)
(2011,4)
(2012,7)
(2013,16)
(2014,15)
(2015,11)
(2016,16)
(2017,8)
(2018,13)
(2019,8)
(2020,20)
(2021,1)
}

\newcommand{\citeTirePapers}{\cite{chen_2014a,fenyes_2020,guo_2012,haddad_2012,jing_2019,li_2013a,li_2020a,liu_2014,lu_2019,meng_2019,mo_2013,patwardhan_1994,patwardhan_1994a,patwardhan_1996,patwardhan_1997,wang_2015a,wang_2015b,wang_2016,wang_2016b,wang_2017,wang_2018a,yang_2019,yu_2014}}
\def\numPapersTire{23}

\def\numCtrlTargetTirePose{0}
\def\numCtrlTargetTirePath{16}
\def\numCtrlTargetTireVeD{1}
\def\numCtrlTargetTireLaD{6}
\def\numCtrlTargetTireLoD{0}
\def\numCtrlTargetTireYA{0}
\def\numCtrlTargetTireElse{0}
\newcommand{\citeCtrlTargetTirePose}{}
\newcommand{\citeCtrlTargetTirePath}{fenyes_2020,guo_2012,haddad_2012,jing_2019,li_2020a,patwardhan_1994,patwardhan_1994a,patwardhan_1996,patwardhan_1997,wang_2015a,wang_2015b,wang_2016,wang_2016b,wang_2017,wang_2018a,yang_2019}
\newcommand{\citeCtrlTargetTireVeD}{yu_2014}
\newcommand{\citeCtrlTargetTireLaD}{chen_2014a,li_2013a,liu_2014,lu_2019,meng_2019,mo_2013}
\newcommand{\citeCtrlTargetTireLoD}{}
\newcommand{\citeCtrlTargetTireYA}{}
\newcommand{\citeCtrlTargetTireElse}{}

\def\numCtrlStructureTireHierarchical{7}
\def\numCtrlStructureTireParallel{5}
\def\numCtrlStructureTireNotStructured{12}
\def\numCtrlStructureTireAlternatives{1}

\newcommand{\citeCtrlStructureHierarchicalTire}{jing_2019,liu_2014,meng_2019,mo_2013,patwardhan_1994a,yang_2019,yu_2014}
\newcommand{\citeCtrlStructureParallelTire}{haddad_2012,wang_2016,wang_2016b,wang_2017,yang_2019}
\newcommand{\citeCtrlStructureSingleTire}{chen_2014a,fenyes_2020,guo_2012,li_2013a,li_2020a,lu_2019,patwardhan_1994,patwardhan_1996,patwardhan_1997,wang_2015a,wang_2015b,wang_2018a}
\newcommand{\citeCtrlAlternativeApproachesTire}{yu_2014}

\newcommand{\citeCtrlLawTirePCA}{yang_2019}

\newcommand{\citeCtrlLawTireAvCA}{}

\newcommand{\citeCtrlLawTireFzCA}{}

\newcommand{\citeCtrlLawTireTBCA}{meng_2019,mo_2013}

\newcommand{\citeCtrlLawTireLtwo}{fenyes_2020}

\newcommand{\citeExpMiLTire}{fenyes_2020,guo_2012,haddad_2012,li_2013a,li_2020a,liu_2014,lu_2019,meng_2019,mo_2013,patwardhan_1994,patwardhan_1994a,wang_2015a,wang_2015b,wang_2016,wang_2016b,wang_2017,wang_2018a,yang_2019,yu_2014}
\newcommand{\citeExpHiLTire}{jing_2019}
\newcommand{\citeExpViLTire}{chen_2014a,patwardhan_1996,patwardhan_1997}

\newcommand{\citeSimToolCarSimTire}{haddad_2012,li_2020a,liu_2014,lu_2019,meng_2019,yu_2014}
\newcommand{\citeSimToolCarMakerTire}{fenyes_2020}
\newcommand{\citeSimToolDynaTire}{}
\newcommand{\citeSimToolAdamsTire}{}
\newcommand{\citeSimToolAVLTire}{}
\newcommand{\citeSimToolVeDYNATire}{guo_2012,jing_2019,wang_2015a,wang_2015b,wang_2016,wang_2016b,wang_2017,wang_2018a}

\newcommand{\citeSimToolAMESimTire}{}

\newcommand{\citeTrajDLCTire}{}
\newcommand{\citeTrajSLCTire}{haddad_2012}
\newcommand{\citeTrajJTurnTire}{patwardhan_1994}
\newcommand{\citeTrajCircTire}{guo_2012,li_2013a,liu_2014,meng_2019}
\newcommand{\citeTrajCurvedTire}{guo_2012,li_2013a,liu_2014,meng_2019,patwardhan_1994}
\newcommand{\citeTrajStraightTire}{chen_2014a,guo_2012,jing_2019,li_2013a,li_2020a,liu_2014,lu_2019,meng_2019,mo_2013,wang_2015a,wang_2015b,wang_2016,wang_2016b,wang_2017,wang_2018a,yu_2014}
\newcommand{\citeTrajOtherTire}{fenyes_2020,patwardhan_1994a,patwardhan_1996,patwardhan_1997,wang_2016,yang_2019}

\newcommand{\tirePublicationsPerYearTableData}{(1990,0)
(1991,0)
(1992,0)
(1993,0)
(1994,2)
(1995,0)
(1996,1)
(1997,1)
(1998,0)
(1999,0)
(2000,0)
(2001,0)
(2002,0)
(2003,0)
(2004,0)
(2005,0)
(2006,0)
(2007,0)
(2008,0)
(2009,0)
(2010,0)
(2011,0)
(2012,2)
(2013,2)
(2014,3)
(2015,2)
(2016,2)
(2017,1)
(2018,1)
(2019,4)
(2020,2)
(2021,0)
}

\newcommand{\citeSuspesionPapers}{\cite{ahmad_2020,chamseddine_2008,dawei_2020,dawei_2021,du_2020,fergani_2014,gaspar_2008,gaspar_2009,gaspar_2010,gaspar_2012,jing_2015a,jing_2015b,li_2011,li_2012,li_2021,lin_2019,liu_2016,liu_2020a,moradi_2014,moradi_2015,morato_2018,morato_2020,pan_2020,papagiannis_2019,poussot-vassal_2008,reinold_2010,retianza_2019,sename_2013,sun_2018,sun_2020,tudon-martinez_2013,wang_2015,wang_2019c,wang_2020e,wu_2021,xiong_2020,yu_2019,zhang_2020c,zhang_2021}}
\def\numSuspensionPapers{39}

\def\numDriveAWI{107}
\def\numDriveAW{2}
\def\numDriveFA{3}
\def\numDriveRA{2}
\def\numDriveFAI{1}
\def\numDriveRAI{8}
\def\numDriveFARAI{1}
\def\numDriveUnknown{50}

\def\numBrakeAWI{50}
\def\numBrakeAW{2}
\def\numBrakeFA{0}
\def\numBrakeRA{0}
\def\numBrakeFAI{0}
\def\numBrakeRAI{2}
\def\numBrakeFARAI{0}
\def\numBrakeUnknown{120}

\def\numSteeringAWI{22}
\def\numSteeringAW{28}
\def\numSteeringFA{76}
\def\numSteeringRA{1}
\def\numSteeringFAI{3}
\def\numSteeringRAI{0}
\def\numSteeringFARAI{0}
\def\numSteeringUnknown{44}

\def\numTopologyDriveAWIBrakeAWISteeringAWI{9}
\def\numTopologyDriveAWIBrakeAWISteeringAW{2}
\def\numTopologyDriveAWIBrakeAWISteeringUnknown{4}
\def\numTopologyDriveAWIBrakeUnknownSteeringAWI{12}
\def\numTopologyDriveAWIBrakeUnknownSteeringAW{14}
\def\numTopologyDriveAWIBrakeUnknownSteeringFA{39}
\def\numTopologyDriveAWIBrakeUnknownSteeringFAI{2}
\def\numTopologyDriveAWIBrakeUnknownSteeringUnknown{23}
\def\numTopologyDriveAWBrakeUnknownSteeringUnknown{2}
\def\numTopologyDriveFABrakeUnknownSteeringAW{0}
\def\numTopologyDriveFABrakeUnknownSteeringFA{0}
\def\numTopologyDriveFABrakeUnknownSteeringRA{0}
\def\numTopologyDriveRABrakeAWSteeringUnknown{2}
\def\numTopologyDriveFAIBrakeUnknownSteeringFA{1}
\def\numTopologyDriveRAIBrakeAWISteeringUnknown{1}
\def\numTopologyDriveRAIBrakeUnknownSteeringFA{4}
\def\numTopologyDriveRAIBrakeUnknownSteeringUnknown{2}
\def\numTopologyDriveFARAIBrakeUnknownSteeringAW{1}
\def\numTopologyDriveUnknownBrakeAWISteeringAWI{1}
\def\numTopologyDriveUnknownBrakeAWISteeringAW{4}
\def\numTopologyDriveUnknownBrakeAWISteeringFA{12}
\def\numTopologyDriveUnknownBrakeAWISteeringFAI{1}
\def\numTopologyDriveUnknownBrakeAWISteeringUnknown{6}
\def\numTopologyDriveUnknownBrakeRAISteeringFA{2}
\def\numTopologyDriveUnknownBrakeUnknownSteeringAW{5}
\def\numTopologyDriveUnknownBrakeUnknownSteeringFA{0}

\newcommand{\topologyTableData}{
(\numTopologyDriveUnknownBrakeUnknownSteeringFA,topoHHC)
(\numTopologyDriveUnknownBrakeUnknownSteeringAW,topoHHB)
(\numTopologyDriveUnknownBrakeRAISteeringFA,topoHFC)
(\numTopologyDriveUnknownBrakeAWISteeringUnknown,topoHAH)
(\numTopologyDriveUnknownBrakeAWISteeringFAI,topoHAE)
(\numTopologyDriveUnknownBrakeAWISteeringFA,topoHAC)
(\numTopologyDriveUnknownBrakeAWISteeringAW,topoHAB)
(\numTopologyDriveUnknownBrakeAWISteeringAWI,topoHAA)
(\numTopologyDriveFARAIBrakeUnknownSteeringAW,topoGHB)
(\numTopologyDriveRAIBrakeUnknownSteeringUnknown,topoFHH)
(\numTopologyDriveRAIBrakeUnknownSteeringFA,topoFHC)
(\numTopologyDriveRAIBrakeAWISteeringUnknown,topoFAH)
(\numTopologyDriveFAIBrakeUnknownSteeringFA,topoEHC)
(\numTopologyDriveRABrakeAWSteeringUnknown,topoDBH)
(\numTopologyDriveFABrakeUnknownSteeringRA,topoCHD)
(\numTopologyDriveFABrakeUnknownSteeringFA,topoCHC)
(\numTopologyDriveFABrakeUnknownSteeringAW,topoCHB)
(\numTopologyDriveAWBrakeUnknownSteeringUnknown,topoBHH)
(\numTopologyDriveAWIBrakeUnknownSteeringUnknown,topoAHH)
(\numTopologyDriveAWIBrakeUnknownSteeringFAI,topoAHE)
(\numTopologyDriveAWIBrakeUnknownSteeringFA,topoAHC)
(\numTopologyDriveAWIBrakeUnknownSteeringAW,topoAHB)
(\numTopologyDriveAWIBrakeUnknownSteeringAWI,topoAHA)
(\numTopologyDriveAWIBrakeAWISteeringUnknown,topoAAH)
(\numTopologyDriveAWIBrakeAWISteeringAW,topoAAB)
(\numTopologyDriveAWIBrakeAWISteeringAWI,topoAAA)
}

\def\numTopologyTireDriveAWIBrakeAWISteeringAWI{0}
\def\numTopologyTireDriveAWIBrakeAWISteeringAW{0}
\def\numTopologyTireDriveAWIBrakeAWISteeringUnknown{0}
\def\numTopologyTireDriveAWIBrakeUnknownSteeringAWI{0}
\def\numTopologyTireDriveAWIBrakeUnknownSteeringAW{0}
\def\numTopologyTireDriveAWIBrakeUnknownSteeringFA{2}
\def\numTopologyTireDriveAWIBrakeUnknownSteeringFAI{0}
\def\numTopologyTireDriveAWIBrakeUnknownSteeringUnknown{0}
\def\numTopologyTireDriveAWBrakeUnknownSteeringUnknown{0}
\def\numTopologyTireDriveFABrakeUnknownSteeringAW{1}
\def\numTopologyTireDriveFABrakeUnknownSteeringFA{1}
\def\numTopologyTireDriveFABrakeUnknownSteeringRA{1}
\def\numTopologyTireDriveRABrakeAWSteeringUnknown{0}
\def\numTopologyTireDriveFAIBrakeUnknownSteeringFA{0}
\def\numTopologyTireDriveRAIBrakeAWISteeringUnknown{0}
\def\numTopologyTireDriveRAIBrakeUnknownSteeringFA{1}
\def\numTopologyTireDriveRAIBrakeUnknownSteeringUnknown{0}
\def\numTopologyTireDriveFARAIBrakeUnknownSteeringAW{0}
\def\numTopologyTireDriveUnknownBrakeAWISteeringAWI{0}
\def\numTopologyTireDriveUnknownBrakeAWISteeringAW{0}
\def\numTopologyTireDriveUnknownBrakeAWISteeringFA{6}
\def\numTopologyTireDriveUnknownBrakeAWISteeringFAI{0}
\def\numTopologyTireDriveUnknownBrakeAWISteeringUnknown{4}
\def\numTopologyTireDriveUnknownBrakeRAISteeringFA{0}
\def\numTopologyTireDriveUnknownBrakeUnknownSteeringAW{1}
\def\numTopologyTireDriveUnknownBrakeUnknownSteeringFA{8}

\newcommand{\topologyTableDataTire}{
(\numTopologyTireDriveUnknownBrakeUnknownSteeringFA,topoHHC)
(\numTopologyTireDriveUnknownBrakeUnknownSteeringAW,topoHHB)
(\numTopologyTireDriveUnknownBrakeRAISteeringFA,topoHFC)
(\numTopologyTireDriveUnknownBrakeAWISteeringUnknown,topoHAH)
(\numTopologyTireDriveUnknownBrakeAWISteeringFAI,topoHAE)
(\numTopologyTireDriveUnknownBrakeAWISteeringFA,topoHAC)
(\numTopologyTireDriveUnknownBrakeAWISteeringAW,topoHAB)
(\numTopologyTireDriveUnknownBrakeAWISteeringAWI,topoHAA)
(\numTopologyTireDriveFARAIBrakeUnknownSteeringAW,topoGHB)
(\numTopologyTireDriveRAIBrakeUnknownSteeringUnknown,topoFHH)
(\numTopologyTireDriveRAIBrakeUnknownSteeringFA,topoFHC)
(\numTopologyTireDriveRAIBrakeAWISteeringUnknown,topoFAH)
(\numTopologyTireDriveFAIBrakeUnknownSteeringFA,topoEHC)
(\numTopologyTireDriveRABrakeAWSteeringUnknown,topoDBH)
(\numTopologyTireDriveFABrakeUnknownSteeringRA,topoCHD)
(\numTopologyTireDriveFABrakeUnknownSteeringFA,topoCHC)
(\numTopologyTireDriveFABrakeUnknownSteeringAW,topoCHB)
(\numTopologyTireDriveAWBrakeUnknownSteeringUnknown,topoBHH)
(\numTopologyTireDriveAWIBrakeUnknownSteeringUnknown,topoAHH)
(\numTopologyTireDriveAWIBrakeUnknownSteeringFAI,topoAHE)
(\numTopologyTireDriveAWIBrakeUnknownSteeringFA,topoAHC)
(\numTopologyTireDriveAWIBrakeUnknownSteeringAW,topoAHB)
(\numTopologyTireDriveAWIBrakeUnknownSteeringAWI,topoAHA)
(\numTopologyTireDriveAWIBrakeAWISteeringUnknown,topoAAH)
(\numTopologyTireDriveAWIBrakeAWISteeringAW,topoAAB)
(\numTopologyTireDriveAWIBrakeAWISteeringAWI,topoAAA)
}

\newcommand{\citeDriveAWIBrakeAWISteeringAWI}{jonasson_2007a,jonasson_2008,moseberg_2015,moseberg_2015a,moseberg_2016,stolte_2018,wang_2007,wang_2009,wang_2014a}
\newcommand{\citeDriveAWIBrakeAWISteeringAW}{knobel_2009,plumlee_2004a}
\newcommand{\citeDriveAWIBrakeAWISteeringUnknown}{jeon_2013,kim_2016,wanner_2012,wanner_2015}
\newcommand{\citeDriveAWIBrakeUnknownSteeringAWI}{bian_2016,hoedt_2013,li_2013,li_2014,li_2016,li_2016a,liu_2013a,liu_2015,reinold_2010,temiz_2020,zhang_2014a,zong_2013}
\newcommand{\citeDriveAWIBrakeUnknownSteeringAW}{buente_2006,dumont_2006,haddad_2013,liu_2012a,pathak_2008,subroto_2020,temiz_2018,temiz_2019,yang_2008,yang_2010,zhang_2014,zhang_2015,zhang_2016a,zhang_2016b}
\newcommand{\citeDriveAWIBrakeUnknownSteeringFA}{chen_2019,chen_2019a,chu_2012,gaspar_2015,guo_2019a,guo_2020,hu_2015,hu_2016a,hu_2017a,hu_2018,hu_2019,jing_2015,jing_2018,kirli_2017,li_2021a,lu_2016,luo_2016,luo_2019,mi_2015,mihaly_2017,ramanathanvenkita_2020,sakthivel_2018,tian_2018,wan_2020,wang_2014b,wang_2016a,wang_2016c,wang_2018,wang_2018b,wang_2019a,wang_2020,wang_2020c,wanner_2015a,wong_2016,zhang_2018a,zhang_2018b,zhang_2019,zhang_2020a,zhang_2020b,meng_2019,yang_2019}
\newcommand{\citeDriveAWIBrakeUnknownSteeringFAI}{hiraoka_2004,yin_2015}
\newcommand{\citeDriveAWIBrakeUnknownSteeringUnknown}{alipour_2014,almeida_2013,amato_2020,chen_2014b,chen_2020,fu_2020a,lee_2020,li_2014a,liu_2012,liu_2017,liu_2020,nguyen_2017,polmans_2014,sun_2014,wang_2010a,wang_2011,wang_2011b,wang_2012,wang_2013,wang_2014,wang_2020g,xin_2014,xin_2016}
\newcommand{\citeDriveAWBrakeUnknownSteeringUnknown}{mutoh_2007,mutoh_2009}
\newcommand{\citeDriveFABrakeUnknownSteeringAW}{patwardhan_1994a}
\newcommand{\citeDriveFABrakeUnknownSteeringFA}{patwardhan_1994a}
\newcommand{\citeDriveFABrakeUnknownSteeringRA}{patwardhan_1994a}
\newcommand{\citeDriveRABrakeAWSteeringUnknown}{song_2003,song_2005}
\newcommand{\citeDriveFAIBrakeUnknownSteeringFA}{chen_2020a}
\newcommand{\citeDriveRAIBrakeAWISteeringUnknown}{ringdorfer_2013}
\newcommand{\citeDriveRAIBrakeUnknownSteeringFA}{ito_2013,miyazaki_2005,wada_2012,wada_2013,fenyes_2020}
\newcommand{\citeDriveRAIBrakeUnknownSteeringUnknown}{chen_2017a,chen_2018}
\newcommand{\citeDriveFARAIBrakeUnknownSteeringAW}{ahmadi_2020}
\newcommand{\citeDriveUnknownBrakeAWISteeringAWI}{feng_2013}
\newcommand{\citeDriveUnknownBrakeAWISteeringAW}{kissai_2018,kou_2010,kruger_2010,lu_2014}
\newcommand{\citeDriveUnknownBrakeAWISteeringFA}{boudali_2018,dominguez-garcia_2004,hac_2006,hac_2006a,jonasson_2016,khelladi_2020,li_2009,nemeth_2012,plumlee_2004,raveendran_2020,sename_2013,zhang_2017,jing_2019,patwardhan_1994,wang_2015a,wang_2016,wang_2016b,wang_2017}
\newcommand{\citeDriveUnknownBrakeAWISteeringFAI}{yim_2014}
\newcommand{\citeDriveUnknownBrakeAWISteeringUnknown}{inpark_2013,jonasson_2017,liu_2016a,sho_2013,wang_2013b,wang_2014c,chen_2014a,liu_2014,mo_2013,yu_2014}
\newcommand{\citeDriveUnknownBrakeRAISteeringFA}{poussot-vassal_2008,poussot-vassal_2008a}
\newcommand{\citeDriveUnknownBrakeUnknownSteeringAW}{bosche_2009,chen_2011,guo_2020a,marino_2007,song_2011,haddad_2012}
\newcommand{\citeDriveUnknownBrakeUnknownSteeringFA}{guo_2012,li_2013a,li_2020a,lu_2019,patwardhan_1996,patwardhan_1997,wang_2015b,wang_2018a}

\def\topologyXMax{45}

\pgfplotsset{topologyYticks/.style={
ytick={topoAHC,topoAHH,topoHAC,topoAHB,topoAHA,topoHAH,topoAAA,topoHHC,topoHHB,topoFHC,topoAAH,topoHAB,topoAAB,topoAHE,topoBHH,topoDBH,topoFHH,topoHFC,topoCHB,topoCHC,topoCHD,topoEHC,topoFAH,topoGHB,topoHAA,topoHAE},
symbolic y coords={topoAHC,topoAHH,topoHAC,topoAHB,topoAHA,topoHAH,topoAAA,topoHHC,topoHHB,topoFHC,topoAAH,topoHAB,topoAAB,topoAHE,topoBHH,topoDBH,topoFHH,topoHFC,topoCHB,topoCHC,topoCHD,topoEHC,topoFAH,topoGHB,topoHAA,topoHAE}
}}

\newcommand{\upmostTopologyTick}{topoAHC}

\pgfplotsset{topologyYtickLabels/.style={yticklabels={
\tikz[baseline]{\node[labelNodeStyleDrive] at(0,0)(){AW-I};}$\vert$\tikz[baseline]{\node[labelNodeStyleBrake] at(0,0)(){---};}$\vert$\tikz[baseline]{\node[labelNodeStyleBrake] at(0,0)(){FA};},
\tikz[baseline]{\node[labelNodeStyleDrive] at(0,0)(){AW-I};}$\vert$\tikz[baseline]{\node[labelNodeStyleBrake] at(0,0)(){---};}$\vert$\tikz[baseline]{\node[labelNodeStyleBrake] at(0,0)(){---};},
\tikz[baseline]{\node[labelNodeStyleDrive] at(0,0)(){---};}$\vert$\tikz[baseline]{\node[labelNodeStyleBrake] at(0,0)(){AW-I};}$\vert$\tikz[baseline]{\node[labelNodeStyleBrake] at(0,0)(){FA};},
\tikz[baseline]{\node[labelNodeStyleDrive] at(0,0)(){AW-I};}$\vert$\tikz[baseline]{\node[labelNodeStyleBrake] at(0,0)(){---};}$\vert$\tikz[baseline]{\node[labelNodeStyleBrake] at(0,0)(){AW};},
\tikz[baseline]{\node[labelNodeStyleDrive] at(0,0)(){AW-I};}$\vert$\tikz[baseline]{\node[labelNodeStyleBrake] at(0,0)(){---};}$\vert$\tikz[baseline]{\node[labelNodeStyleBrake] at(0,0)(){AW-I};},
\tikz[baseline]{\node[labelNodeStyleDrive] at(0,0)(){---};}$\vert$\tikz[baseline]{\node[labelNodeStyleBrake] at(0,0)(){AW-I};}$\vert$\tikz[baseline]{\node[labelNodeStyleBrake] at(0,0)(){---};},
\tikz[baseline]{\node[labelNodeStyleDrive] at(0,0)(){AW-I};}$\vert$\tikz[baseline]{\node[labelNodeStyleBrake] at(0,0)(){AW-I};}$\vert$\tikz[baseline]{\node[labelNodeStyleBrake] at(0,0)(){AW-I};},
\tikz[baseline]{\node[labelNodeStyleDrive] at(0,0)(){---};}$\vert$\tikz[baseline]{\node[labelNodeStyleBrake] at(0,0)(){---};}$\vert$\tikz[baseline]{\node[labelNodeStyleBrake] at(0,0)(){FA};},
\tikz[baseline]{\node[labelNodeStyleDrive] at(0,0)(){---};}$\vert$\tikz[baseline]{\node[labelNodeStyleBrake] at(0,0)(){---};}$\vert$\tikz[baseline]{\node[labelNodeStyleBrake] at(0,0)(){AW};},
\tikz[baseline]{\node[labelNodeStyleDrive] at(0,0)(){RA-I};}$\vert$\tikz[baseline]{\node[labelNodeStyleBrake] at(0,0)(){---};}$\vert$\tikz[baseline]{\node[labelNodeStyleBrake] at(0,0)(){FA};},
\tikz[baseline]{\node[labelNodeStyleDrive] at(0,0)(){AW-I};}$\vert$\tikz[baseline]{\node[labelNodeStyleBrake] at(0,0)(){AW-I};}$\vert$\tikz[baseline]{\node[labelNodeStyleBrake] at(0,0)(){---};},
\tikz[baseline]{\node[labelNodeStyleDrive] at(0,0)(){---};}$\vert$\tikz[baseline]{\node[labelNodeStyleBrake] at(0,0)(){AW-I};}$\vert$\tikz[baseline]{\node[labelNodeStyleBrake] at(0,0)(){AW};},
\tikz[baseline]{\node[labelNodeStyleDrive] at(0,0)(){AW-I};}$\vert$\tikz[baseline]{\node[labelNodeStyleBrake] at(0,0)(){AW-I};}$\vert$\tikz[baseline]{\node[labelNodeStyleBrake] at(0,0)(){AW};},
\tikz[baseline]{\node[labelNodeStyleDrive] at(0,0)(){AW-I};}$\vert$\tikz[baseline]{\node[labelNodeStyleBrake] at(0,0)(){---};}$\vert$\tikz[baseline]{\node[labelNodeStyleBrake] at(0,0)(){FA-I};},
\tikz[baseline]{\node[labelNodeStyleDrive] at(0,0)(){AW};}$\vert$\tikz[baseline]{\node[labelNodeStyleBrake] at(0,0)(){---};}$\vert$\tikz[baseline]{\node[labelNodeStyleBrake] at(0,0)(){---};},
\tikz[baseline]{\node[labelNodeStyleDrive] at(0,0)(){RA};}$\vert$\tikz[baseline]{\node[labelNodeStyleBrake] at(0,0)(){AW};}$\vert$\tikz[baseline]{\node[labelNodeStyleBrake] at(0,0)(){---};},
\tikz[baseline]{\node[labelNodeStyleDrive] at(0,0)(){RA-I};}$\vert$\tikz[baseline]{\node[labelNodeStyleBrake] at(0,0)(){---};}$\vert$\tikz[baseline]{\node[labelNodeStyleBrake] at(0,0)(){---};},
\tikz[baseline]{\node[labelNodeStyleDrive] at(0,0)(){---};}$\vert$\tikz[baseline]{\node[labelNodeStyleBrake] at(0,0)(){RA-I};}$\vert$\tikz[baseline]{\node[labelNodeStyleBrake] at(0,0)(){FA};},
\tikz[baseline]{\node[labelNodeStyleDrive] at(0,0)(){FA};}$\vert$\tikz[baseline]{\node[labelNodeStyleBrake] at(0,0)(){---};}$\vert$\tikz[baseline]{\node[labelNodeStyleBrake] at(0,0)(){AW};},
\tikz[baseline]{\node[labelNodeStyleDrive] at(0,0)(){FA};}$\vert$\tikz[baseline]{\node[labelNodeStyleBrake] at(0,0)(){---};}$\vert$\tikz[baseline]{\node[labelNodeStyleBrake] at(0,0)(){FA};},
\tikz[baseline]{\node[labelNodeStyleDrive] at(0,0)(){FA};}$\vert$\tikz[baseline]{\node[labelNodeStyleBrake] at(0,0)(){---};}$\vert$\tikz[baseline]{\node[labelNodeStyleBrake] at(0,0)(){RA};},
\tikz[baseline]{\node[labelNodeStyleDrive] at(0,0)(){FA-I};}$\vert$\tikz[baseline]{\node[labelNodeStyleBrake] at(0,0)(){---};}$\vert$\tikz[baseline]{\node[labelNodeStyleBrake] at(0,0)(){FA};},
\tikz[baseline]{\node[labelNodeStyleDrive] at(0,0)(){RA-I};}$\vert$\tikz[baseline]{\node[labelNodeStyleBrake] at(0,0)(){AW-I};}$\vert$\tikz[baseline]{\node[labelNodeStyleBrake] at(0,0)(){---};},
\tikz[baseline]{\node[labelNodeStyleDrive] at(0,0)(){FA+RA-I};}$\vert$\tikz[baseline]{\node[labelNodeStyleBrake] at(0,0)(){---};}$\vert$\tikz[baseline]{\node[labelNodeStyleBrake] at(0,0)(){AW};},
\tikz[baseline]{\node[labelNodeStyleDrive] at(0,0)(){---};}$\vert$\tikz[baseline]{\node[labelNodeStyleBrake] at(0,0)(){AW-I};}$\vert$\tikz[baseline]{\node[labelNodeStyleBrake] at(0,0)(){AW-I};},
\tikz[baseline]{\node[labelNodeStyleDrive] at(0,0)(){---};}$\vert$\tikz[baseline]{\node[labelNodeStyleBrake] at(0,0)(){AW-I};}$\vert$\tikz[baseline]{\node[labelNodeStyleBrake] at(0,0)(){FA-I};}
}}}

\def\numCtrlLawOCA{42}
\newcommand{\citeCtrlLawOCA}{almeida_2013,bian_2016,boudali_2018,buente_2006,chen_2020,chu_2012,fu_2020a,guo_2020,hac_2006,inpark_2013,jonasson_2008,kissai_2018,kou_2010,kruger_2010,li_2014,liu_2012,liu_2013a,liu_2015,liu_2016a,liu_2020,lu_2016,plumlee_2004,reinold_2010,sun_2014,wada_2012,wada_2013,wan_2020,wang_2007,wang_2009,wang_2010a,wang_2011b,wang_2012,wang_2014a,wang_2014c,wang_2016c,wang_2020g,wanner_2012,wanner_2015,wong_2016,zhang_2018b,zhang_2020b,zong_2013}
\def\numCtrlLawTireOCA{2}
\newcommand{\citeCtrlLawTireOCA}{jing_2019,yu_2014}
\def\numCtrlLawSM{39}
\newcommand{\citeCtrlLawSM}{alipour_2014,almeida_2013,bian_2016,chen_2019,chen_2019a,chen_2020a,dumont_2006,feng_2013,fu_2020a,guo_2019a,hu_2017a,hu_2018,hu_2019,inpark_2013,jeon_2013,kim_2016,li_2016,li_2016a,liu_2020,lu_2014,lu_2016,raveendran_2020,ringdorfer_2013,sho_2013,song_2003,song_2005,subroto_2020,tian_2018,wan_2020,wang_2007,wang_2009,wang_2010a,wang_2012,wang_2013,wang_2014c,yim_2014,zhang_2018a,zhang_2018b,zhang_2019}
\def\numCtrlLawTireSM{2}
\newcommand{\citeCtrlLawTireSM}{mo_2013,yu_2014}
\def\numCtrlLawPID{34}
\newcommand{\citeCtrlLawPID}{ahmadi_2020,almeida_2013,amato_2020,buente_2006,jonasson_2007a,jonasson_2008,jonasson_2016,jonasson_2017,kirli_2017,lee_2020,li_2009,li_2021a,liu_2020,marino_2007,miyazaki_2005,moseberg_2015,moseberg_2015a,moseberg_2016,nguyen_2017,pathak_2008,temiz_2018,temiz_2019,temiz_2020,wang_2014a,wang_2016c,wang_2018,wang_2019a,wang_2020g,wanner_2012,wanner_2015,wanner_2015a,wong_2016,zhang_2017,zhang_2018b}
\def\numCtrlLawTirePID{4}
\newcommand{\citeCtrlLawTirePID}{haddad_2012,li_2013a,lu_2019,wang_2016}
\def\numCtrlLawFF{20}
\newcommand{\citeCtrlLawFF}{boudali_2018,chen_2014b,fu_2020a,ito_2013,jonasson_2016,jonasson_2017,khelladi_2020,kirli_2017,li_2009,liu_2017,moseberg_2015,moseberg_2015a,moseberg_2016,nguyen_2017,wang_2011,wang_2018,wang_2018b,wang_2019a,wang_2020,wang_2020c}
\def\numCtrlLawTireFF{6}
\newcommand{\citeCtrlLawTireFF}{patwardhan_1994a,patwardhan_1996,patwardhan_1997,wang_2016,wang_2016b,wang_2017}
\def\numCtrlLawACA{21}
\newcommand{\citeCtrlLawACA}{chen_2014b,chen_2019,chen_2019a,feng_2013,hoedt_2013,hu_2015,jeon_2013,khelladi_2020,knobel_2009,moseberg_2015,moseberg_2015a,moseberg_2016,subroto_2020,temiz_2018,temiz_2019,temiz_2020,wang_2019a,wanner_2012,wanner_2015,wanner_2015a,yim_2014}
\def\numCtrlLawTireACA{1}
\newcommand{\citeCtrlLawTireACA}{yang_2019}
\def\numCtrlLawHinf{16}
\newcommand{\citeCtrlLawHinf}{chen_2017a,chen_2018,guo_2020,guo_2020a,jing_2015,jing_2018,kissai_2018,mi_2015,poussot-vassal_2008,poussot-vassal_2008a,sename_2013,wada_2012,wada_2013,wang_2014b,wang_2016a,wang_2020}
\def\numCtrlLawTireHinf{3}
\newcommand{\citeCtrlLawTireHinf}{fenyes_2020,jing_2019,wang_2015a}
\def\numCtrlLawLT{16}
\newcommand{\citeCtrlLawLT}{bosche_2009,chen_2011,gaspar_2015,hu_2016a,hu_2017a,hu_2018,mihaly_2017,nemeth_2012,song_2011,wang_2011,wang_2011b,yang_2008,yang_2010,zhang_2014,zhang_2014a,zhang_2015}
\def\numCtrlLawTireLT{3}
\newcommand{\citeCtrlLawTireLT}{haddad_2012,li_2020a,wang_2016b}
\def\numCtrlLawDCA{16}
\newcommand{\citeCtrlLawDCA}{ahmadi_2020,amato_2020,gaspar_2015,hac_2006a,ito_2013,jing_2018,jonasson_2007a,li_2009,li_2014a,liu_2017,ramanathanvenkita_2020,ringdorfer_2013,sho_2013,wang_2013b,xin_2014,xin_2016}
\def\numCtrlLawTireDCA{2}
\newcommand{\citeCtrlLawTireDCA}{meng_2019,mo_2013}
\def\numCtrlLawRB{12}
\newcommand{\citeCtrlLawRB}{chen_2020,ito_2013,lee_2020,liu_2012,liu_2012a,miyazaki_2005,mutoh_2007,mutoh_2009,pathak_2008,wang_2014c,xin_2014,xin_2016}
\def\numCtrlLawTireRB{2}
\newcommand{\citeCtrlLawTireRB}{liu_2014,yu_2014}
\def\numCtrlLawLQR{10}
\newcommand{\citeCtrlLawLQR}{chen_2014b,dumont_2006,hiraoka_2004,hu_2015,plumlee_2004a,polmans_2014,yang_2008,yang_2010,zhang_2016a,zhang_2020a}
\def\numCtrlLawTireLQR{3}
\newcommand{\citeCtrlLawTireLQR}{liu_2014,patwardhan_1994,patwardhan_1994a}
\def\numCtrlLawMPC{9}
\newcommand{\citeCtrlLawMPC}{chen_2020a,kou_2010,li_2014,liu_2013a,liu_2015,stolte_2018,yin_2015,zhang_2020b,zong_2013}
\def\numCtrlLawTireMPC{4}
\newcommand{\citeCtrlLawTireMPC}{guo_2012,wang_2015b,wang_2018a,yang_2019}
\def\numCtrlLawSF{10}
\newcommand{\citeCtrlLawSF}{boudali_2018,dominguez-garcia_2004,hu_2016a,hu_2017a,hu_2018,khelladi_2020,wang_2014,wang_2018b,wang_2020c,zhang_2016b}
\def\numCtrlLawTireSF{3}
\newcommand{\citeCtrlLawTireSF}{meng_2019,wang_2017,yang_2019}
\def\numCtrlLawFuzzy{3}
\newcommand{\citeCtrlLawFuzzy}{sakthivel_2018,xin_2014,xin_2016}
\def\numCtrlLawTireFuzzy{2}
\newcommand{\citeCtrlLawTireFuzzy}{chen_2014a,lu_2019}
\def\numCtrlLawFB{4}
\newcommand{\citeCtrlLawFB}{haddad_2013,hoedt_2013,knobel_2009,kruger_2010}
\def\numCtrlLawTireFB{0}
\newcommand{\citeCtrlLawTireFB}{}
\def\numCtrlLawMFA{3}
\newcommand{\citeCtrlLawMFA}{li_2013,luo_2016,luo_2019}
\def\numCtrlLawTireMFA{0}
\newcommand{\citeCtrlLawTireMFA}{}

\pgfplotsset{controlLawYtickLabels/.style={
symbolic y coords={law1,law2,law3,law4,law5,law6,law7,law8,law9,law10,law11,law12,law13,law14,law15},
ytick={law1,law2,law3,law4,law5,law6,law7,law8,law9,law10,law11,law12,law13,law14,law15},
yticklabels={OCA,SM,PID,FF,ACA,$\mathcal{H}_\infty$,LT,DCA,RB,LQR,MPC,SF,Fuzzy,FB,MFA}
}}

\newcommand{\controlLawTable}{
(\numCtrlLawOCA,law1)
(\numCtrlLawSM,law2)
(\numCtrlLawPID,law3)
(\numCtrlLawFF,law4)
(\numCtrlLawACA,law5)
(\numCtrlLawHinf,law6)
(\numCtrlLawLT,law7)
(\numCtrlLawDCA,law8)
(\numCtrlLawRB,law9)
(\numCtrlLawLQR,law10)
(\numCtrlLawMPC,law11)
(\numCtrlLawSF,law12)
(\numCtrlLawFuzzy,law13)
(\numCtrlLawFB,law14)
(\numCtrlLawMFA,law15)
}
\newcommand{\controlLawTableTire}{
(\numCtrlLawTireOCA,law1)
(\numCtrlLawTireSM,law2)
(\numCtrlLawTirePID,law3)
(\numCtrlLawTireFF,law4)
(\numCtrlLawTireACA,law5)
(\numCtrlLawTireHinf,law6)
(\numCtrlLawTireLT,law7)
(\numCtrlLawTireDCA,law8)
(\numCtrlLawTireRB,law9)
(\numCtrlLawTireLQR,law10)
(\numCtrlLawTireMPC,law11)
(\numCtrlLawTireSF,law12)
(\numCtrlLawTireFuzzy,law13)
(\numCtrlLawTireFB,law14)
(\numCtrlLawTireMFA,law15)
}
\newcommand{\controlLawOverviewLegend}{OCA:~Optimal control allocation; SM:~Sliding mode; PID:~Proportional, integral, derivative or alike; FF:~Feed forward; ACA:~Analytical control allocation; $\mathcal{H}_\infty$:~$\mathcal{H}_\infty$ control; LT:~Lyapunov theory-based; DCA:~Direct control allocation; RB:~Rule-based reconfiguration; LQR:~Linear quadratic regulator; MPC:~Model predictive control; SF:~State feedback control; Fuzzy:~Fuzzy control; FB:~Flatness-based; MFA:~Model-free adaptive control}

\newcommand{\getControlLawWidth}{
\setlength{\ControlLawWidthTemp}{\widthof{\footnotesize PD\&DCA}}
\ifthenelse{\lengthtest{\ControlLawWidthTemp>\ControlLawWidth}}{\setlength{\ControlLawWidth}{\ControlLawWidthTemp}}{}
\setlength{\ControlLawWidthTemp}{\widthof{\footnotesize SM}}
\ifthenelse{\lengthtest{\ControlLawWidthTemp>\ControlLawWidth}}{\setlength{\ControlLawWidth}{\ControlLawWidthTemp}}{}
\setlength{\ControlLawWidthTemp}{\widthof{\footnotesize (PI+SM)\&OCA}}
\ifthenelse{\lengthtest{\ControlLawWidthTemp>\ControlLawWidth}}{\setlength{\ControlLawWidth}{\ControlLawWidthTemp}}{}
\setlength{\ControlLawWidthTemp}{\widthof{\footnotesize PI\&DCA}}
\ifthenelse{\lengthtest{\ControlLawWidthTemp>\ControlLawWidth}}{\setlength{\ControlLawWidth}{\ControlLawWidthTemp}}{}
\setlength{\ControlLawWidthTemp}{\widthof{\footnotesize SM/OCA}}
\ifthenelse{\lengthtest{\ControlLawWidthTemp>\ControlLawWidth}}{\setlength{\ControlLawWidth}{\ControlLawWidthTemp}}{}
\setlength{\ControlLawWidthTemp}{\widthof{\footnotesize LT}}
\ifthenelse{\lengthtest{\ControlLawWidthTemp>\ControlLawWidth}}{\setlength{\ControlLawWidth}{\ControlLawWidthTemp}}{}
\setlength{\ControlLawWidthTemp}{\widthof{\footnotesize (FF+SF)\&OCA}}
\ifthenelse{\lengthtest{\ControlLawWidthTemp>\ControlLawWidth}}{\setlength{\ControlLawWidth}{\ControlLawWidthTemp}}{}
\setlength{\ControlLawWidthTemp}{\widthof{\footnotesize PD\&OCA}}
\ifthenelse{\lengthtest{\ControlLawWidthTemp>\ControlLawWidth}}{\setlength{\ControlLawWidth}{\ControlLawWidthTemp}}{}
\setlength{\ControlLawWidthTemp}{\widthof{\footnotesize LT}}
\ifthenelse{\lengthtest{\ControlLawWidthTemp>\ControlLawWidth}}{\setlength{\ControlLawWidth}{\ControlLawWidthTemp}}{}
\setlength{\ControlLawWidthTemp}{\widthof{\footnotesize (LQR+FF)\&ACA}}
\ifthenelse{\lengthtest{\ControlLawWidthTemp>\ControlLawWidth}}{\setlength{\ControlLawWidth}{\ControlLawWidthTemp}}{}
\setlength{\ControlLawWidthTemp}{\widthof{\footnotesize $\mathcal{H}_\infty$}}
\ifthenelse{\lengthtest{\ControlLawWidthTemp>\ControlLawWidth}}{\setlength{\ControlLawWidth}{\ControlLawWidthTemp}}{}
\setlength{\ControlLawWidthTemp}{\widthof{\footnotesize $\mathcal{H}_\infty$}}
\ifthenelse{\lengthtest{\ControlLawWidthTemp>\ControlLawWidth}}{\setlength{\ControlLawWidth}{\ControlLawWidthTemp}}{}
\setlength{\ControlLawWidthTemp}{\widthof{\footnotesize SM\&ACA}}
\ifthenelse{\lengthtest{\ControlLawWidthTemp>\ControlLawWidth}}{\setlength{\ControlLawWidth}{\ControlLawWidthTemp}}{}
\setlength{\ControlLawWidthTemp}{\widthof{\footnotesize SM\&ACA}}
\ifthenelse{\lengthtest{\ControlLawWidthTemp>\ControlLawWidth}}{\setlength{\ControlLawWidth}{\ControlLawWidthTemp}}{}
\setlength{\ControlLawWidthTemp}{\widthof{\footnotesize RB\&OCA}}
\ifthenelse{\lengthtest{\ControlLawWidthTemp>\ControlLawWidth}}{\setlength{\ControlLawWidth}{\ControlLawWidthTemp}}{}
\setlength{\ControlLawWidthTemp}{\widthof{\footnotesize MPC\&SM}}
\ifthenelse{\lengthtest{\ControlLawWidthTemp>\ControlLawWidth}}{\setlength{\ControlLawWidth}{\ControlLawWidthTemp}}{}
\setlength{\ControlLawWidthTemp}{\widthof{\footnotesize OCA}}
\ifthenelse{\lengthtest{\ControlLawWidthTemp>\ControlLawWidth}}{\setlength{\ControlLawWidth}{\ControlLawWidthTemp}}{}
\setlength{\ControlLawWidthTemp}{\widthof{\footnotesize SF}}
\ifthenelse{\lengthtest{\ControlLawWidthTemp>\ControlLawWidth}}{\setlength{\ControlLawWidth}{\ControlLawWidthTemp}}{}
\setlength{\ControlLawWidthTemp}{\widthof{\footnotesize SM\&LQR}}
\ifthenelse{\lengthtest{\ControlLawWidthTemp>\ControlLawWidth}}{\setlength{\ControlLawWidth}{\ControlLawWidthTemp}}{}
\setlength{\ControlLawWidthTemp}{\widthof{\footnotesize SM\&ACA}}
\ifthenelse{\lengthtest{\ControlLawWidthTemp>\ControlLawWidth}}{\setlength{\ControlLawWidth}{\ControlLawWidthTemp}}{}
\setlength{\ControlLawWidthTemp}{\widthof{\footnotesize (FF+SM)\&OCA}}
\ifthenelse{\lengthtest{\ControlLawWidthTemp>\ControlLawWidth}}{\setlength{\ControlLawWidth}{\ControlLawWidthTemp}}{}
\setlength{\ControlLawWidthTemp}{\widthof{\footnotesize LT\&DCA}}
\ifthenelse{\lengthtest{\ControlLawWidthTemp>\ControlLawWidth}}{\setlength{\ControlLawWidth}{\ControlLawWidthTemp}}{}
\setlength{\ControlLawWidthTemp}{\widthof{\footnotesize SM}}
\ifthenelse{\lengthtest{\ControlLawWidthTemp>\ControlLawWidth}}{\setlength{\ControlLawWidth}{\ControlLawWidthTemp}}{}
\setlength{\ControlLawWidthTemp}{\widthof{\footnotesize $\mathcal{H}_\infty$\&OCA}}
\ifthenelse{\lengthtest{\ControlLawWidthTemp>\ControlLawWidth}}{\setlength{\ControlLawWidth}{\ControlLawWidthTemp}}{}
\setlength{\ControlLawWidthTemp}{\widthof{\footnotesize $\mathcal{H}_\infty$}}
\ifthenelse{\lengthtest{\ControlLawWidthTemp>\ControlLawWidth}}{\setlength{\ControlLawWidth}{\ControlLawWidthTemp}}{}
\setlength{\ControlLawWidthTemp}{\widthof{\footnotesize OCA}}
\ifthenelse{\lengthtest{\ControlLawWidthTemp>\ControlLawWidth}}{\setlength{\ControlLawWidth}{\ControlLawWidthTemp}}{}
\setlength{\ControlLawWidthTemp}{\widthof{\footnotesize DCA}}
\ifthenelse{\lengthtest{\ControlLawWidthTemp>\ControlLawWidth}}{\setlength{\ControlLawWidth}{\ControlLawWidthTemp}}{}
\setlength{\ControlLawWidthTemp}{\widthof{\footnotesize FB}}
\ifthenelse{\lengthtest{\ControlLawWidthTemp>\ControlLawWidth}}{\setlength{\ControlLawWidth}{\ControlLawWidthTemp}}{}
\setlength{\ControlLawWidthTemp}{\widthof{\footnotesize LQR}}
\ifthenelse{\lengthtest{\ControlLawWidthTemp>\ControlLawWidth}}{\setlength{\ControlLawWidth}{\ControlLawWidthTemp}}{}
\setlength{\ControlLawWidthTemp}{\widthof{\footnotesize FB\&ACA}}
\ifthenelse{\lengthtest{\ControlLawWidthTemp>\ControlLawWidth}}{\setlength{\ControlLawWidth}{\ControlLawWidthTemp}}{}
\setlength{\ControlLawWidthTemp}{\widthof{\footnotesize LQR\&ACA}}
\ifthenelse{\lengthtest{\ControlLawWidthTemp>\ControlLawWidth}}{\setlength{\ControlLawWidth}{\ControlLawWidthTemp}}{}
\setlength{\ControlLawWidthTemp}{\widthof{\footnotesize SF+LT}}
\ifthenelse{\lengthtest{\ControlLawWidthTemp>\ControlLawWidth}}{\setlength{\ControlLawWidth}{\ControlLawWidthTemp}}{}
\setlength{\ControlLawWidthTemp}{\widthof{\footnotesize SF+LT+SM}}
\ifthenelse{\lengthtest{\ControlLawWidthTemp>\ControlLawWidth}}{\setlength{\ControlLawWidth}{\ControlLawWidthTemp}}{}
\setlength{\ControlLawWidthTemp}{\widthof{\footnotesize SF+LT+SM}}
\ifthenelse{\lengthtest{\ControlLawWidthTemp>\ControlLawWidth}}{\setlength{\ControlLawWidth}{\ControlLawWidthTemp}}{}
\setlength{\ControlLawWidthTemp}{\widthof{\footnotesize SM}}
\ifthenelse{\lengthtest{\ControlLawWidthTemp>\ControlLawWidth}}{\setlength{\ControlLawWidth}{\ControlLawWidthTemp}}{}
\setlength{\ControlLawWidthTemp}{\widthof{\footnotesize SM\&OCA}}
\ifthenelse{\lengthtest{\ControlLawWidthTemp>\ControlLawWidth}}{\setlength{\ControlLawWidth}{\ControlLawWidthTemp}}{}
\setlength{\ControlLawWidthTemp}{\widthof{\footnotesize (RB+FF)/DCA}}
\ifthenelse{\lengthtest{\ControlLawWidthTemp>\ControlLawWidth}}{\setlength{\ControlLawWidth}{\ControlLawWidthTemp}}{}
\setlength{\ControlLawWidthTemp}{\widthof{\footnotesize SM\&ACA}}
\ifthenelse{\lengthtest{\ControlLawWidthTemp>\ControlLawWidth}}{\setlength{\ControlLawWidth}{\ControlLawWidthTemp}}{}
\setlength{\ControlLawWidthTemp}{\widthof{\footnotesize $\mathcal{H}_\infty$}}
\ifthenelse{\lengthtest{\ControlLawWidthTemp>\ControlLawWidth}}{\setlength{\ControlLawWidth}{\ControlLawWidthTemp}}{}
\setlength{\ControlLawWidthTemp}{\widthof{\footnotesize $\mathcal{H}_\infty$\&DCA}}
\ifthenelse{\lengthtest{\ControlLawWidthTemp>\ControlLawWidth}}{\setlength{\ControlLawWidth}{\ControlLawWidthTemp}}{}
\setlength{\ControlLawWidthTemp}{\widthof{\footnotesize PI\&DCA}}
\ifthenelse{\lengthtest{\ControlLawWidthTemp>\ControlLawWidth}}{\setlength{\ControlLawWidth}{\ControlLawWidthTemp}}{}
\setlength{\ControlLawWidthTemp}{\widthof{\footnotesize PI\&OCA}}
\ifthenelse{\lengthtest{\ControlLawWidthTemp>\ControlLawWidth}}{\setlength{\ControlLawWidth}{\ControlLawWidthTemp}}{}
\setlength{\ControlLawWidthTemp}{\widthof{\footnotesize FF+PID}}
\ifthenelse{\lengthtest{\ControlLawWidthTemp>\ControlLawWidth}}{\setlength{\ControlLawWidth}{\ControlLawWidthTemp}}{}
\setlength{\ControlLawWidthTemp}{\widthof{\footnotesize FF+PID}}
\ifthenelse{\lengthtest{\ControlLawWidthTemp>\ControlLawWidth}}{\setlength{\ControlLawWidth}{\ControlLawWidthTemp}}{}
\setlength{\ControlLawWidthTemp}{\widthof{\footnotesize (FF+SF)\&ACA}}
\ifthenelse{\lengthtest{\ControlLawWidthTemp>\ControlLawWidth}}{\setlength{\ControlLawWidth}{\ControlLawWidthTemp}}{}
\setlength{\ControlLawWidthTemp}{\widthof{\footnotesize SM}}
\ifthenelse{\lengthtest{\ControlLawWidthTemp>\ControlLawWidth}}{\setlength{\ControlLawWidth}{\ControlLawWidthTemp}}{}
\setlength{\ControlLawWidthTemp}{\widthof{\footnotesize FF+PID}}
\ifthenelse{\lengthtest{\ControlLawWidthTemp>\ControlLawWidth}}{\setlength{\ControlLawWidth}{\ControlLawWidthTemp}}{}
\setlength{\ControlLawWidthTemp}{\widthof{\footnotesize $\mathcal{H}_\infty$\&OCA}}
\ifthenelse{\lengthtest{\ControlLawWidthTemp>\ControlLawWidth}}{\setlength{\ControlLawWidth}{\ControlLawWidthTemp}}{}
\setlength{\ControlLawWidthTemp}{\widthof{\footnotesize FB\&ACA}}
\ifthenelse{\lengthtest{\ControlLawWidthTemp>\ControlLawWidth}}{\setlength{\ControlLawWidth}{\ControlLawWidthTemp}}{}
\setlength{\ControlLawWidthTemp}{\widthof{\footnotesize MPC\&OCA}}
\ifthenelse{\lengthtest{\ControlLawWidthTemp>\ControlLawWidth}}{\setlength{\ControlLawWidth}{\ControlLawWidthTemp}}{}
\setlength{\ControlLawWidthTemp}{\widthof{\footnotesize FB\&OCA}}
\ifthenelse{\lengthtest{\ControlLawWidthTemp>\ControlLawWidth}}{\setlength{\ControlLawWidth}{\ControlLawWidthTemp}}{}
\setlength{\ControlLawWidthTemp}{\widthof{\footnotesize RB+P}}
\ifthenelse{\lengthtest{\ControlLawWidthTemp>\ControlLawWidth}}{\setlength{\ControlLawWidth}{\ControlLawWidthTemp}}{}
\setlength{\ControlLawWidthTemp}{\widthof{\footnotesize FF\&PI\&DCA}}
\ifthenelse{\lengthtest{\ControlLawWidthTemp>\ControlLawWidth}}{\setlength{\ControlLawWidth}{\ControlLawWidthTemp}}{}
\setlength{\ControlLawWidthTemp}{\widthof{\footnotesize MFA}}
\ifthenelse{\lengthtest{\ControlLawWidthTemp>\ControlLawWidth}}{\setlength{\ControlLawWidth}{\ControlLawWidthTemp}}{}
\setlength{\ControlLawWidthTemp}{\widthof{\footnotesize MPC\&OCA}}
\ifthenelse{\lengthtest{\ControlLawWidthTemp>\ControlLawWidth}}{\setlength{\ControlLawWidth}{\ControlLawWidthTemp}}{}
\setlength{\ControlLawWidthTemp}{\widthof{\footnotesize DCA}}
\ifthenelse{\lengthtest{\ControlLawWidthTemp>\ControlLawWidth}}{\setlength{\ControlLawWidth}{\ControlLawWidthTemp}}{}
\setlength{\ControlLawWidthTemp}{\widthof{\footnotesize SM}}
\ifthenelse{\lengthtest{\ControlLawWidthTemp>\ControlLawWidth}}{\setlength{\ControlLawWidth}{\ControlLawWidthTemp}}{}
\setlength{\ControlLawWidthTemp}{\widthof{\footnotesize SM}}
\ifthenelse{\lengthtest{\ControlLawWidthTemp>\ControlLawWidth}}{\setlength{\ControlLawWidth}{\ControlLawWidthTemp}}{}
\setlength{\ControlLawWidthTemp}{\widthof{\footnotesize PID}}
\ifthenelse{\lengthtest{\ControlLawWidthTemp>\ControlLawWidth}}{\setlength{\ControlLawWidth}{\ControlLawWidthTemp}}{}
\setlength{\ControlLawWidthTemp}{\widthof{\footnotesize RB+OCA}}
\ifthenelse{\lengthtest{\ControlLawWidthTemp>\ControlLawWidth}}{\setlength{\ControlLawWidth}{\ControlLawWidthTemp}}{}
\setlength{\ControlLawWidthTemp}{\widthof{\footnotesize RB}}
\ifthenelse{\lengthtest{\ControlLawWidthTemp>\ControlLawWidth}}{\setlength{\ControlLawWidth}{\ControlLawWidthTemp}}{}
\setlength{\ControlLawWidthTemp}{\widthof{\footnotesize MPC\&OCA}}
\ifthenelse{\lengthtest{\ControlLawWidthTemp>\ControlLawWidth}}{\setlength{\ControlLawWidth}{\ControlLawWidthTemp}}{}
\setlength{\ControlLawWidthTemp}{\widthof{\footnotesize MPC\&OCA}}
\ifthenelse{\lengthtest{\ControlLawWidthTemp>\ControlLawWidth}}{\setlength{\ControlLawWidth}{\ControlLawWidthTemp}}{}
\setlength{\ControlLawWidthTemp}{\widthof{\footnotesize OCA}}
\ifthenelse{\lengthtest{\ControlLawWidthTemp>\ControlLawWidth}}{\setlength{\ControlLawWidth}{\ControlLawWidthTemp}}{}
\setlength{\ControlLawWidthTemp}{\widthof{\footnotesize FF\&DCA}}
\ifthenelse{\lengthtest{\ControlLawWidthTemp>\ControlLawWidth}}{\setlength{\ControlLawWidth}{\ControlLawWidthTemp}}{}
\setlength{\ControlLawWidthTemp}{\widthof{\footnotesize SM\&PI\&OCA}}
\ifthenelse{\lengthtest{\ControlLawWidthTemp>\ControlLawWidth}}{\setlength{\ControlLawWidth}{\ControlLawWidthTemp}}{}
\setlength{\ControlLawWidthTemp}{\widthof{\footnotesize SM}}
\ifthenelse{\lengthtest{\ControlLawWidthTemp>\ControlLawWidth}}{\setlength{\ControlLawWidth}{\ControlLawWidthTemp}}{}
\setlength{\ControlLawWidthTemp}{\widthof{\footnotesize SM\&OCA}}
\ifthenelse{\lengthtest{\ControlLawWidthTemp>\ControlLawWidth}}{\setlength{\ControlLawWidth}{\ControlLawWidthTemp}}{}
\setlength{\ControlLawWidthTemp}{\widthof{\footnotesize MFA}}
\ifthenelse{\lengthtest{\ControlLawWidthTemp>\ControlLawWidth}}{\setlength{\ControlLawWidth}{\ControlLawWidthTemp}}{}
\setlength{\ControlLawWidthTemp}{\widthof{\footnotesize MFA}}
\ifthenelse{\lengthtest{\ControlLawWidthTemp>\ControlLawWidth}}{\setlength{\ControlLawWidth}{\ControlLawWidthTemp}}{}
\setlength{\ControlLawWidthTemp}{\widthof{\footnotesize PI}}
\ifthenelse{\lengthtest{\ControlLawWidthTemp>\ControlLawWidth}}{\setlength{\ControlLawWidth}{\ControlLawWidthTemp}}{}
\setlength{\ControlLawWidthTemp}{\widthof{\footnotesize $\mathcal{H}_\infty$}}
\ifthenelse{\lengthtest{\ControlLawWidthTemp>\ControlLawWidth}}{\setlength{\ControlLawWidth}{\ControlLawWidthTemp}}{}
\setlength{\ControlLawWidthTemp}{\widthof{\footnotesize LT}}
\ifthenelse{\lengthtest{\ControlLawWidthTemp>\ControlLawWidth}}{\setlength{\ControlLawWidth}{\ControlLawWidthTemp}}{}
\setlength{\ControlLawWidthTemp}{\widthof{\footnotesize RB/P}}
\ifthenelse{\lengthtest{\ControlLawWidthTemp>\ControlLawWidth}}{\setlength{\ControlLawWidth}{\ControlLawWidthTemp}}{}
\setlength{\ControlLawWidthTemp}{\widthof{\footnotesize (FF+P(D))\&ACA}}
\ifthenelse{\lengthtest{\ControlLawWidthTemp>\ControlLawWidth}}{\setlength{\ControlLawWidth}{\ControlLawWidthTemp}}{}
\setlength{\ControlLawWidthTemp}{\widthof{\footnotesize (FF+P(D))\&ACA}}
\ifthenelse{\lengthtest{\ControlLawWidthTemp>\ControlLawWidth}}{\setlength{\ControlLawWidth}{\ControlLawWidthTemp}}{}
\setlength{\ControlLawWidthTemp}{\widthof{\footnotesize (FF+P(D))\&ACA}}
\ifthenelse{\lengthtest{\ControlLawWidthTemp>\ControlLawWidth}}{\setlength{\ControlLawWidth}{\ControlLawWidthTemp}}{}
\setlength{\ControlLawWidthTemp}{\widthof{\footnotesize RB}}
\ifthenelse{\lengthtest{\ControlLawWidthTemp>\ControlLawWidth}}{\setlength{\ControlLawWidth}{\ControlLawWidthTemp}}{}
\setlength{\ControlLawWidthTemp}{\widthof{\footnotesize RB}}
\ifthenelse{\lengthtest{\ControlLawWidthTemp>\ControlLawWidth}}{\setlength{\ControlLawWidth}{\ControlLawWidthTemp}}{}
\setlength{\ControlLawWidthTemp}{\widthof{\footnotesize LT}}
\ifthenelse{\lengthtest{\ControlLawWidthTemp>\ControlLawWidth}}{\setlength{\ControlLawWidth}{\ControlLawWidthTemp}}{}
\setlength{\ControlLawWidthTemp}{\widthof{\footnotesize FF+P}}
\ifthenelse{\lengthtest{\ControlLawWidthTemp>\ControlLawWidth}}{\setlength{\ControlLawWidth}{\ControlLawWidthTemp}}{}
\setlength{\ControlLawWidthTemp}{\widthof{\footnotesize PD+RB}}
\ifthenelse{\lengthtest{\ControlLawWidthTemp>\ControlLawWidth}}{\setlength{\ControlLawWidth}{\ControlLawWidthTemp}}{}
\setlength{\ControlLawWidthTemp}{\widthof{\footnotesize OCA}}
\ifthenelse{\lengthtest{\ControlLawWidthTemp>\ControlLawWidth}}{\setlength{\ControlLawWidth}{\ControlLawWidthTemp}}{}
\setlength{\ControlLawWidthTemp}{\widthof{\footnotesize LQR}}
\ifthenelse{\lengthtest{\ControlLawWidthTemp>\ControlLawWidth}}{\setlength{\ControlLawWidth}{\ControlLawWidthTemp}}{}
\setlength{\ControlLawWidthTemp}{\widthof{\footnotesize LQR}}
\ifthenelse{\lengthtest{\ControlLawWidthTemp>\ControlLawWidth}}{\setlength{\ControlLawWidth}{\ControlLawWidthTemp}}{}
\setlength{\ControlLawWidthTemp}{\widthof{\footnotesize $\mathcal{H}_\infty$}}
\ifthenelse{\lengthtest{\ControlLawWidthTemp>\ControlLawWidth}}{\setlength{\ControlLawWidth}{\ControlLawWidthTemp}}{}
\setlength{\ControlLawWidthTemp}{\widthof{\footnotesize $\mathcal{H}_\infty$}}
\ifthenelse{\lengthtest{\ControlLawWidthTemp>\ControlLawWidth}}{\setlength{\ControlLawWidth}{\ControlLawWidthTemp}}{}
\setlength{\ControlLawWidthTemp}{\widthof{\footnotesize DCA}}
\ifthenelse{\lengthtest{\ControlLawWidthTemp>\ControlLawWidth}}{\setlength{\ControlLawWidth}{\ControlLawWidthTemp}}{}
\setlength{\ControlLawWidthTemp}{\widthof{\footnotesize SM}}
\ifthenelse{\lengthtest{\ControlLawWidthTemp>\ControlLawWidth}}{\setlength{\ControlLawWidth}{\ControlLawWidthTemp}}{}
\setlength{\ControlLawWidthTemp}{\widthof{\footnotesize OCA}}
\ifthenelse{\lengthtest{\ControlLawWidthTemp>\ControlLawWidth}}{\setlength{\ControlLawWidth}{\ControlLawWidthTemp}}{}
\setlength{\ControlLawWidthTemp}{\widthof{\footnotesize SM\&DCA}}
\ifthenelse{\lengthtest{\ControlLawWidthTemp>\ControlLawWidth}}{\setlength{\ControlLawWidth}{\ControlLawWidthTemp}}{}
\setlength{\ControlLawWidthTemp}{\widthof{\footnotesize Fuzzy}}
\ifthenelse{\lengthtest{\ControlLawWidthTemp>\ControlLawWidth}}{\setlength{\ControlLawWidth}{\ControlLawWidthTemp}}{}
\setlength{\ControlLawWidthTemp}{\widthof{\footnotesize $\mathcal{H}_\infty$}}
\ifthenelse{\lengthtest{\ControlLawWidthTemp>\ControlLawWidth}}{\setlength{\ControlLawWidth}{\ControlLawWidthTemp}}{}
\setlength{\ControlLawWidthTemp}{\widthof{\footnotesize SM\&DCA}}
\ifthenelse{\lengthtest{\ControlLawWidthTemp>\ControlLawWidth}}{\setlength{\ControlLawWidth}{\ControlLawWidthTemp}}{}
\setlength{\ControlLawWidthTemp}{\widthof{\footnotesize SM}}
\ifthenelse{\lengthtest{\ControlLawWidthTemp>\ControlLawWidth}}{\setlength{\ControlLawWidth}{\ControlLawWidthTemp}}{}
\setlength{\ControlLawWidthTemp}{\widthof{\footnotesize SM}}
\ifthenelse{\lengthtest{\ControlLawWidthTemp>\ControlLawWidth}}{\setlength{\ControlLawWidth}{\ControlLawWidthTemp}}{}
\setlength{\ControlLawWidthTemp}{\widthof{\footnotesize LT}}
\ifthenelse{\lengthtest{\ControlLawWidthTemp>\ControlLawWidth}}{\setlength{\ControlLawWidth}{\ControlLawWidthTemp}}{}
\setlength{\ControlLawWidthTemp}{\widthof{\footnotesize MPC}}
\ifthenelse{\lengthtest{\ControlLawWidthTemp>\ControlLawWidth}}{\setlength{\ControlLawWidth}{\ControlLawWidthTemp}}{}
\setlength{\ControlLawWidthTemp}{\widthof{\footnotesize SM\&ACA}}
\ifthenelse{\lengthtest{\ControlLawWidthTemp>\ControlLawWidth}}{\setlength{\ControlLawWidth}{\ControlLawWidthTemp}}{}
\setlength{\ControlLawWidthTemp}{\widthof{\footnotesize OCA}}
\ifthenelse{\lengthtest{\ControlLawWidthTemp>\ControlLawWidth}}{\setlength{\ControlLawWidth}{\ControlLawWidthTemp}}{}
\setlength{\ControlLawWidthTemp}{\widthof{\footnotesize PI\&ACA}}
\ifthenelse{\lengthtest{\ControlLawWidthTemp>\ControlLawWidth}}{\setlength{\ControlLawWidth}{\ControlLawWidthTemp}}{}
\setlength{\ControlLawWidthTemp}{\widthof{\footnotesize PI\&ACA}}
\ifthenelse{\lengthtest{\ControlLawWidthTemp>\ControlLawWidth}}{\setlength{\ControlLawWidth}{\ControlLawWidthTemp}}{}
\setlength{\ControlLawWidthTemp}{\widthof{\footnotesize PI\&ACA}}
\ifthenelse{\lengthtest{\ControlLawWidthTemp>\ControlLawWidth}}{\setlength{\ControlLawWidth}{\ControlLawWidthTemp}}{}
\setlength{\ControlLawWidthTemp}{\widthof{\footnotesize SM}}
\ifthenelse{\lengthtest{\ControlLawWidthTemp>\ControlLawWidth}}{\setlength{\ControlLawWidth}{\ControlLawWidthTemp}}{}
\setlength{\ControlLawWidthTemp}{\widthof{\footnotesize $\mathcal{H}_\infty$\&OCA}}
\ifthenelse{\lengthtest{\ControlLawWidthTemp>\ControlLawWidth}}{\setlength{\ControlLawWidth}{\ControlLawWidthTemp}}{}
\setlength{\ControlLawWidthTemp}{\widthof{\footnotesize $\mathcal{H}_\infty$\&OCA}}
\ifthenelse{\lengthtest{\ControlLawWidthTemp>\ControlLawWidth}}{\setlength{\ControlLawWidth}{\ControlLawWidthTemp}}{}
\setlength{\ControlLawWidthTemp}{\widthof{\footnotesize SM\&OCA}}
\ifthenelse{\lengthtest{\ControlLawWidthTemp>\ControlLawWidth}}{\setlength{\ControlLawWidth}{\ControlLawWidthTemp}}{}
\setlength{\ControlLawWidthTemp}{\widthof{\footnotesize SM\&OCA}}
\ifthenelse{\lengthtest{\ControlLawWidthTemp>\ControlLawWidth}}{\setlength{\ControlLawWidth}{\ControlLawWidthTemp}}{}
\setlength{\ControlLawWidthTemp}{\widthof{\footnotesize SM\&OCA}}
\ifthenelse{\lengthtest{\ControlLawWidthTemp>\ControlLawWidth}}{\setlength{\ControlLawWidth}{\ControlLawWidthTemp}}{}
\setlength{\ControlLawWidthTemp}{\widthof{\footnotesize SM\&OCA}}
\ifthenelse{\lengthtest{\ControlLawWidthTemp>\ControlLawWidth}}{\setlength{\ControlLawWidth}{\ControlLawWidthTemp}}{}
\setlength{\ControlLawWidthTemp}{\widthof{\footnotesize FF+LT}}
\ifthenelse{\lengthtest{\ControlLawWidthTemp>\ControlLawWidth}}{\setlength{\ControlLawWidth}{\ControlLawWidthTemp}}{}
\setlength{\ControlLawWidthTemp}{\widthof{\footnotesize LT\&OCA}}
\ifthenelse{\lengthtest{\ControlLawWidthTemp>\ControlLawWidth}}{\setlength{\ControlLawWidth}{\ControlLawWidthTemp}}{}
\setlength{\ControlLawWidthTemp}{\widthof{\footnotesize SM\&OCA}}
\ifthenelse{\lengthtest{\ControlLawWidthTemp>\ControlLawWidth}}{\setlength{\ControlLawWidth}{\ControlLawWidthTemp}}{}
\setlength{\ControlLawWidthTemp}{\widthof{\footnotesize SM}}
\ifthenelse{\lengthtest{\ControlLawWidthTemp>\ControlLawWidth}}{\setlength{\ControlLawWidth}{\ControlLawWidthTemp}}{}
\setlength{\ControlLawWidthTemp}{\widthof{\footnotesize DCA}}
\ifthenelse{\lengthtest{\ControlLawWidthTemp>\ControlLawWidth}}{\setlength{\ControlLawWidth}{\ControlLawWidthTemp}}{}
\setlength{\ControlLawWidthTemp}{\widthof{\footnotesize SF}}
\ifthenelse{\lengthtest{\ControlLawWidthTemp>\ControlLawWidth}}{\setlength{\ControlLawWidth}{\ControlLawWidthTemp}}{}
\setlength{\ControlLawWidthTemp}{\widthof{\footnotesize PI\&OCA}}
\ifthenelse{\lengthtest{\ControlLawWidthTemp>\ControlLawWidth}}{\setlength{\ControlLawWidth}{\ControlLawWidthTemp}}{}
\setlength{\ControlLawWidthTemp}{\widthof{\footnotesize $\mathcal{H}_\infty$}}
\ifthenelse{\lengthtest{\ControlLawWidthTemp>\ControlLawWidth}}{\setlength{\ControlLawWidth}{\ControlLawWidthTemp}}{}
\setlength{\ControlLawWidthTemp}{\widthof{\footnotesize (SM\&OCA)/RB}}
\ifthenelse{\lengthtest{\ControlLawWidthTemp>\ControlLawWidth}}{\setlength{\ControlLawWidth}{\ControlLawWidthTemp}}{}
\setlength{\ControlLawWidthTemp}{\widthof{\footnotesize $\mathcal{H}_\infty$}}
\ifthenelse{\lengthtest{\ControlLawWidthTemp>\ControlLawWidth}}{\setlength{\ControlLawWidth}{\ControlLawWidthTemp}}{}
\setlength{\ControlLawWidthTemp}{\widthof{\footnotesize PI\&OCA}}
\ifthenelse{\lengthtest{\ControlLawWidthTemp>\ControlLawWidth}}{\setlength{\ControlLawWidth}{\ControlLawWidthTemp}}{}
\setlength{\ControlLawWidthTemp}{\widthof{\footnotesize FF+PI}}
\ifthenelse{\lengthtest{\ControlLawWidthTemp>\ControlLawWidth}}{\setlength{\ControlLawWidth}{\ControlLawWidthTemp}}{}
\setlength{\ControlLawWidthTemp}{\widthof{\footnotesize FF+SF}}
\ifthenelse{\lengthtest{\ControlLawWidthTemp>\ControlLawWidth}}{\setlength{\ControlLawWidth}{\ControlLawWidthTemp}}{}
\setlength{\ControlLawWidthTemp}{\widthof{\footnotesize (FF+P)\&ACA}}
\ifthenelse{\lengthtest{\ControlLawWidthTemp>\ControlLawWidth}}{\setlength{\ControlLawWidth}{\ControlLawWidthTemp}}{}
\setlength{\ControlLawWidthTemp}{\widthof{\footnotesize FF+$\mathcal{H}_\infty$}}
\ifthenelse{\lengthtest{\ControlLawWidthTemp>\ControlLawWidth}}{\setlength{\ControlLawWidth}{\ControlLawWidthTemp}}{}
\setlength{\ControlLawWidthTemp}{\widthof{\footnotesize FF+SF}}
\ifthenelse{\lengthtest{\ControlLawWidthTemp>\ControlLawWidth}}{\setlength{\ControlLawWidth}{\ControlLawWidthTemp}}{}
\setlength{\ControlLawWidthTemp}{\widthof{\footnotesize PID\&OCA}}
\ifthenelse{\lengthtest{\ControlLawWidthTemp>\ControlLawWidth}}{\setlength{\ControlLawWidth}{\ControlLawWidthTemp}}{}
\setlength{\ControlLawWidthTemp}{\widthof{\footnotesize PI\&(ACA/OCA)}}
\ifthenelse{\lengthtest{\ControlLawWidthTemp>\ControlLawWidth}}{\setlength{\ControlLawWidth}{\ControlLawWidthTemp}}{}
\setlength{\ControlLawWidthTemp}{\widthof{\footnotesize PI\&(ACA/OCA)}}
\ifthenelse{\lengthtest{\ControlLawWidthTemp>\ControlLawWidth}}{\setlength{\ControlLawWidth}{\ControlLawWidthTemp}}{}
\setlength{\ControlLawWidthTemp}{\widthof{\footnotesize PI\&ACA}}
\ifthenelse{\lengthtest{\ControlLawWidthTemp>\ControlLawWidth}}{\setlength{\ControlLawWidth}{\ControlLawWidthTemp}}{}
\setlength{\ControlLawWidthTemp}{\widthof{\footnotesize P\&OCA}}
\ifthenelse{\lengthtest{\ControlLawWidthTemp>\ControlLawWidth}}{\setlength{\ControlLawWidth}{\ControlLawWidthTemp}}{}
\setlength{\ControlLawWidthTemp}{\widthof{\footnotesize Fuzzy\&(RB+DCA)}}
\ifthenelse{\lengthtest{\ControlLawWidthTemp>\ControlLawWidth}}{\setlength{\ControlLawWidth}{\ControlLawWidthTemp}}{}
\setlength{\ControlLawWidthTemp}{\widthof{\footnotesize Fuzzy\&(RB+DCA)}}
\ifthenelse{\lengthtest{\ControlLawWidthTemp>\ControlLawWidth}}{\setlength{\ControlLawWidth}{\ControlLawWidthTemp}}{}
\setlength{\ControlLawWidthTemp}{\widthof{\footnotesize LQR+LT}}
\ifthenelse{\lengthtest{\ControlLawWidthTemp>\ControlLawWidth}}{\setlength{\ControlLawWidth}{\ControlLawWidthTemp}}{}
\setlength{\ControlLawWidthTemp}{\widthof{\footnotesize LQR+LT}}
\ifthenelse{\lengthtest{\ControlLawWidthTemp>\ControlLawWidth}}{\setlength{\ControlLawWidth}{\ControlLawWidthTemp}}{}
\setlength{\ControlLawWidthTemp}{\widthof{\footnotesize SM\&ACA}}
\ifthenelse{\lengthtest{\ControlLawWidthTemp>\ControlLawWidth}}{\setlength{\ControlLawWidth}{\ControlLawWidthTemp}}{}
\setlength{\ControlLawWidthTemp}{\widthof{\footnotesize MPC}}
\ifthenelse{\lengthtest{\ControlLawWidthTemp>\ControlLawWidth}}{\setlength{\ControlLawWidth}{\ControlLawWidthTemp}}{}
\setlength{\ControlLawWidthTemp}{\widthof{\footnotesize LT}}
\ifthenelse{\lengthtest{\ControlLawWidthTemp>\ControlLawWidth}}{\setlength{\ControlLawWidth}{\ControlLawWidthTemp}}{}
\setlength{\ControlLawWidthTemp}{\widthof{\footnotesize LT}}
\ifthenelse{\lengthtest{\ControlLawWidthTemp>\ControlLawWidth}}{\setlength{\ControlLawWidth}{\ControlLawWidthTemp}}{}
\setlength{\ControlLawWidthTemp}{\widthof{\footnotesize LT}}
\ifthenelse{\lengthtest{\ControlLawWidthTemp>\ControlLawWidth}}{\setlength{\ControlLawWidth}{\ControlLawWidthTemp}}{}
\setlength{\ControlLawWidthTemp}{\widthof{\footnotesize LQR}}
\ifthenelse{\lengthtest{\ControlLawWidthTemp>\ControlLawWidth}}{\setlength{\ControlLawWidth}{\ControlLawWidthTemp}}{}
\setlength{\ControlLawWidthTemp}{\widthof{\footnotesize SF}}
\ifthenelse{\lengthtest{\ControlLawWidthTemp>\ControlLawWidth}}{\setlength{\ControlLawWidth}{\ControlLawWidthTemp}}{}
\setlength{\ControlLawWidthTemp}{\widthof{\footnotesize PI}}
\ifthenelse{\lengthtest{\ControlLawWidthTemp>\ControlLawWidth}}{\setlength{\ControlLawWidth}{\ControlLawWidthTemp}}{}
\setlength{\ControlLawWidthTemp}{\widthof{\footnotesize SM}}
\ifthenelse{\lengthtest{\ControlLawWidthTemp>\ControlLawWidth}}{\setlength{\ControlLawWidth}{\ControlLawWidthTemp}}{}
\setlength{\ControlLawWidthTemp}{\widthof{\footnotesize (SM+PI)\&OCA}}
\ifthenelse{\lengthtest{\ControlLawWidthTemp>\ControlLawWidth}}{\setlength{\ControlLawWidth}{\ControlLawWidthTemp}}{}
\setlength{\ControlLawWidthTemp}{\widthof{\footnotesize SM}}
\ifthenelse{\lengthtest{\ControlLawWidthTemp>\ControlLawWidth}}{\setlength{\ControlLawWidth}{\ControlLawWidthTemp}}{}
\setlength{\ControlLawWidthTemp}{\widthof{\footnotesize LQR}}
\ifthenelse{\lengthtest{\ControlLawWidthTemp>\ControlLawWidth}}{\setlength{\ControlLawWidth}{\ControlLawWidthTemp}}{}
\setlength{\ControlLawWidthTemp}{\widthof{\footnotesize MPC\&OCA}}
\ifthenelse{\lengthtest{\ControlLawWidthTemp>\ControlLawWidth}}{\setlength{\ControlLawWidth}{\ControlLawWidthTemp}}{}
\setlength{\ControlLawWidthTemp}{\widthof{\footnotesize MPC\&OCA}}
\ifthenelse{\lengthtest{\ControlLawWidthTemp>\ControlLawWidth}}{\setlength{\ControlLawWidth}{\ControlLawWidthTemp}}{}
}

\newcommand{\getControlLawWidthTire}{
\setlength{\ControlLawWidthTemp}{\widthof{\footnotesize Fuzzy}}
\ifthenelse{\lengthtest{\ControlLawWidthTemp> \ControlLawWidth}}{\setlength{\ControlLawWidth}{\ControlLawWidthTemp}}{}
\setlength{\ControlLawWidthTemp}{\widthof{\footnotesize $\mathcal{H}_\infty$}}
\ifthenelse{\lengthtest{\ControlLawWidthTemp> \ControlLawWidth}}{\setlength{\ControlLawWidth}{\ControlLawWidthTemp}}{}
\setlength{\ControlLawWidthTemp}{\widthof{\footnotesize MPC}}
\ifthenelse{\lengthtest{\ControlLawWidthTemp> \ControlLawWidth}}{\setlength{\ControlLawWidth}{\ControlLawWidthTemp}}{}
\setlength{\ControlLawWidthTemp}{\widthof{\footnotesize PID+LT}}
\ifthenelse{\lengthtest{\ControlLawWidthTemp> \ControlLawWidth}}{\setlength{\ControlLawWidth}{\ControlLawWidthTemp}}{}
\setlength{\ControlLawWidthTemp}{\widthof{\footnotesize $\mathcal{H}_\infty$\&OCA}}
\ifthenelse{\lengthtest{\ControlLawWidthTemp> \ControlLawWidth}}{\setlength{\ControlLawWidth}{\ControlLawWidthTemp}}{}
\setlength{\ControlLawWidthTemp}{\widthof{\footnotesize PID}}
\ifthenelse{\lengthtest{\ControlLawWidthTemp> \ControlLawWidth}}{\setlength{\ControlLawWidth}{\ControlLawWidthTemp}}{}
\setlength{\ControlLawWidthTemp}{\widthof{\footnotesize LT}}
\ifthenelse{\lengthtest{\ControlLawWidthTemp> \ControlLawWidth}}{\setlength{\ControlLawWidth}{\ControlLawWidthTemp}}{}
\setlength{\ControlLawWidthTemp}{\widthof{\footnotesize LQR\&RB}}
\ifthenelse{\lengthtest{\ControlLawWidthTemp> \ControlLawWidth}}{\setlength{\ControlLawWidth}{\ControlLawWidthTemp}}{}
\setlength{\ControlLawWidthTemp}{\widthof{\footnotesize Fuzzy PID}}
\ifthenelse{\lengthtest{\ControlLawWidthTemp> \ControlLawWidth}}{\setlength{\ControlLawWidth}{\ControlLawWidthTemp}}{}
\setlength{\ControlLawWidthTemp}{\widthof{\footnotesize SF\&DCA}}
\ifthenelse{\lengthtest{\ControlLawWidthTemp> \ControlLawWidth}}{\setlength{\ControlLawWidth}{\ControlLawWidthTemp}}{}
\setlength{\ControlLawWidthTemp}{\widthof{\footnotesize SM\&DCA}}
\ifthenelse{\lengthtest{\ControlLawWidthTemp> \ControlLawWidth}}{\setlength{\ControlLawWidth}{\ControlLawWidthTemp}}{}
\setlength{\ControlLawWidthTemp}{\widthof{\footnotesize LQR}}
\ifthenelse{\lengthtest{\ControlLawWidthTemp> \ControlLawWidth}}{\setlength{\ControlLawWidth}{\ControlLawWidthTemp}}{}
\setlength{\ControlLawWidthTemp}{\widthof{\footnotesize FF\&LQR}}
\ifthenelse{\lengthtest{\ControlLawWidthTemp> \ControlLawWidth}}{\setlength{\ControlLawWidth}{\ControlLawWidthTemp}}{}
\setlength{\ControlLawWidthTemp}{\widthof{\footnotesize FF}}
\ifthenelse{\lengthtest{\ControlLawWidthTemp> \ControlLawWidth}}{\setlength{\ControlLawWidth}{\ControlLawWidthTemp}}{}
\setlength{\ControlLawWidthTemp}{\widthof{\footnotesize FF}}
\ifthenelse{\lengthtest{\ControlLawWidthTemp> \ControlLawWidth}}{\setlength{\ControlLawWidth}{\ControlLawWidthTemp}}{}
\setlength{\ControlLawWidthTemp}{\widthof{\footnotesize $\mathcal{H}_\infty$}}
\ifthenelse{\lengthtest{\ControlLawWidthTemp> \ControlLawWidth}}{\setlength{\ControlLawWidth}{\ControlLawWidthTemp}}{}
\setlength{\ControlLawWidthTemp}{\widthof{\footnotesize MPC}}
\ifthenelse{\lengthtest{\ControlLawWidthTemp> \ControlLawWidth}}{\setlength{\ControlLawWidth}{\ControlLawWidthTemp}}{}
\setlength{\ControlLawWidthTemp}{\widthof{\footnotesize FF+PID}}
\ifthenelse{\lengthtest{\ControlLawWidthTemp> \ControlLawWidth}}{\setlength{\ControlLawWidth}{\ControlLawWidthTemp}}{}
\setlength{\ControlLawWidthTemp}{\widthof{\footnotesize FF+LT}}
\ifthenelse{\lengthtest{\ControlLawWidthTemp> \ControlLawWidth}}{\setlength{\ControlLawWidth}{\ControlLawWidthTemp}}{}
\setlength{\ControlLawWidthTemp}{\widthof{\footnotesize FF+SF}}
\ifthenelse{\lengthtest{\ControlLawWidthTemp> \ControlLawWidth}}{\setlength{\ControlLawWidth}{\ControlLawWidthTemp}}{}
\setlength{\ControlLawWidthTemp}{\widthof{\footnotesize MPC}}
\ifthenelse{\lengthtest{\ControlLawWidthTemp> \ControlLawWidth}}{\setlength{\ControlLawWidth}{\ControlLawWidthTemp}}{}
\setlength{\ControlLawWidthTemp}{\widthof{\footnotesize (MPC+SF)\&ACA}}
\ifthenelse{\lengthtest{\ControlLawWidthTemp> \ControlLawWidth}}{\setlength{\ControlLawWidth}{\ControlLawWidthTemp}}{}
\setlength{\ControlLawWidthTemp}{\widthof{\footnotesize (SM\&OCA)/RB}}
\ifthenelse{\lengthtest{\ControlLawWidthTemp> \ControlLawWidth}}{\setlength{\ControlLawWidth}{\ControlLawWidthTemp}}{}
}

%% file: focus.tex
\label{sec:focus}

Within in the field of automated driving, we concentrate on road vehicles. 
Fault-tolerant control approaches for off-road vehicles, \eg, \cite{proetzsch_2005,djeziri_2013}, omnidirectional robots~\cite{rotondo_2015,loureiro_2014,hacene_2019,lounici_2020}, as well as other mobile robots~\cite{jin_2019,zhang_2019a,doran_2020} are not further discussed as their specific application domain poses different requirements from a safety perspective compared to road vehicles. 
The same applies to fault-tolerant motion control of special purpose vehicles and vehicles with more than four wheels, for instance~\cite{bera_2012,nah_2013,liang_2017,volz_2018,prasad_2019,lu_2020,sathishkumar_2020}.

\increasingTableCitations
Specifically, we focus on drive, brake, and steering actuators as these are present in every road vehicle.
Moreover, tires are arguably the most critical mechanical components with regard to motion safety as they transfer all forces generated by actuators. 
Transfer characteristics as well as vehicle dynamics change significantly after tire blowouts and therefore impact vehicle motion critically.  
Thus, we include publications that target vehicle motion control after tire blowouts. %
In contrast, active suspension systems as optional components are not further considered%
\footnote{Fault-tolerant control approaches for active suspension systems are presented, \eg, in~\citeSuspesionPapers. \todo{Suspension paper durchfräsen}}.
As a consequence, we purely highlight planar vehicle motion aspects, although vertical motion is partially considered in the covered literature, too. 

All publications surveyed in Sections \ref{sec:degradations}--\ref{sec:experiments} introduce approaches to exploit the over-actuation of a vehicle.
Therefore, publications showing fault tolerance at the actuator level are not considered. 
Moreover, we focus on publications that outline the specific control approach for reaching fault tolerance.
Publications that only show results are neglected, \eg,~\cite{hayama_2008,hayama_2010,sondermann-woelke_2011,raste_2015a,gauger_2016,sandmann_2017,yue_2019,wang_2020f}.
The same applies to publications that only state that an approach is potentially fault-tolerant without demonstrating the implementation or presenting experimental results.

\begin{figure}
	\centering
	\input{publicationsPerYearBarPlot.tex}
	\caption{
		Overview of the publication years of the literature covered in the core of this survey. %
		(\protect\tikz[baseline=-\the\dimexpr\fontdimen22\textfont2\relax ]{\protect\node[anchor=base,actuatorbar] (temp) {};} publications addressing actuator degradations; 
		 \protect\tikz[baseline=-\the\dimexpr\fontdimen22\textfont2\relax ]{\protect\node[anchor=base,blowoutbar] (temp) {};} publications addressing tire degradations).
	}
	\label{fig:publicationsperyear}
\end{figure}
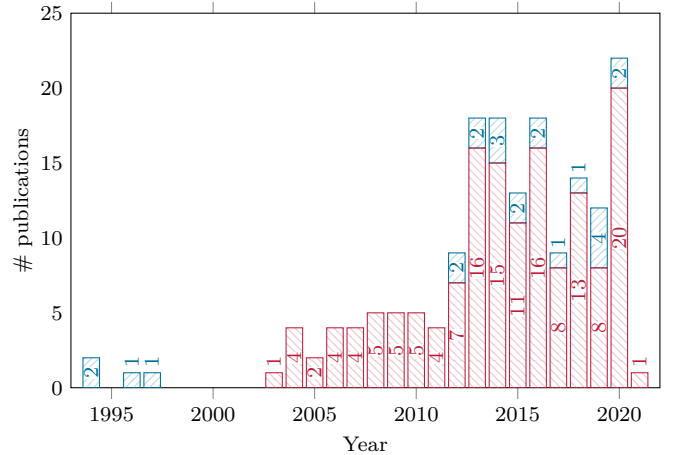

Some publications that appear relevant based on their titles and abstracts unfortunately cannot be included in this survey because they are only available to us in Chinese~\cite{liu_2013,guo_2007,wang_2007a,huang_2009,wang_2010,guo_2011,chu_2012a,wang_2013a,zheng_2013,zheng_2012a,zheng_2012,liu_2015a,yu_2016,tao_2018}, Japanese~\cite{ito_2012}, or not at all~\cite{wallmark_2007,mutoh_2006,kawakami_2001,li_2017,liu_2021}\todo{\cite{liu_2021} entfernen, wenn verfügbar}%
\footnote{For \cite{chu_2012a} and \cite{ito_2012}, there are similar publications of the same authors in English, which are part of this survey, see \cite{chu_2012} and \cite{ito_2013}, respectively. 
	For \cite{wallmark_2007}, which is cited in \cite{jonasson_2009}, very similar content is found in~\cite{jonasson_2008} according to the authors~\cite{wallmark_2019}. 
	For \cite{mutoh_2006}, \ia, mentioned in~\cite{xin_2014,xin_2016}, a paper with similar title is available~\cite{mutoh_2007}.
	\cite{kawakami_2001} is cited in~\cite{luo_2019}.}.

Within this focus, a total of \the\numexpr\numPapers+\numPapersTire\relax~publications forms the core of this survey.
Hence, without claiming to be exhaustive, the survey provides a representative picture of the current state of the art in fault-tolerant vehicle motion control research.
An overview of the contained publications is presented in \autoref{tab:literature}, which summarizes employed actuator topology, covered degradations, control approaches, as well as conducted experiments for each publication.

The literature covered in this survey indicates that actuator fault-tolerant vehicle motion control has been the subject of research for roughly a quarter of a century as illustrated in \autoref{fig:publicationsperyear}.
After the potential of fault-tolerant vehicle motion control had already been recognized earlier (1989 at the latest, \cf~\cite{fruechte_1989}), there were a few publications in the mid-1990s.
However, it was not until the beginning of the 21st century that fault-tolerant motion control came more into the attention of researchers, with research intensifying significantly over the past decade.

\todo{Noch mal mit Marcus Nolte diskutieren, ob Platooning Anwendungen doch rein sollen, \citet{hao_2020,spooner_1995,spooner_1997,wang_2006a}}

\input{literaturereview.tex}

%% file: publicationsPerYearBarPlot.tex
\newlength{\numberheight}
\setlength{\numberheight}{\heightof{\footnotesize 1}}
\setlength{\numberheight}{1.1\numberheight}
\begin{tikzpicture}
	\footnotesize
	\begin{axis}[
	    ybar stacked,
	    width =\linewidth,
	    scale only axis, 
	    x=.27cm,
	    y=\numberheight,
	    bar width=0.22cm,
	    xlabel={\footnotesize Year},
	    ylabel={\footnotesize \# publications},
		xtick align=outside,
		xticklabel style={/pgf/number format/1000 sep=,font=\footnotesize },%
	    yticklabel style={font=\footnotesize,align=right,inner sep=0pt,xshift=-0.1cm},%
	    ytick style={draw=none},
	    nodes near coords,
		xmin = 1993,
		xmax = 2022,
		ymin = 0, 
		ymax = 25,
		verticalbar
	    ]
	    \addplot+[ybar, actuatorbar,every node near coord/.append style={rotate=90}] plot coordinates {\publicationsPerYearTableData};
	    \addplot+[ybar, blowoutbar,every node near coord/.append style={rotate=90}] plot coordinates {\tirePublicationsPerYearTableData};
	\end{axis}
\end{tikzpicture}

%% file: literaturereview.tex
\tikzstyle{headervertical}=[anchor=south,inner sep =0, outer sep =0, minimum height = \DegradationHeight, yshift=-1mm]

\newlength{\BulletYShift}
\newlength{\BulletYShiftCorr}
\setlength{\BulletYShift}{\heightof{\footnotesize (a)}}
\setlength{\BulletYShiftCorr}{\depthof{\footnotesize (a)}}
\addtolength{\BulletYShift}{-\BulletYShiftCorr}
\setlength{\BulletYShift}{.5\BulletYShift}
\tikzstyle{drawcircle}=[circle, minimum size=1mm, fill=black, yshift=\BulletYShift,anchor=center]

\newcommand{\referenceLegend}{	DLC: 	Double lane change; 
								SLC: 	Single lane change; 
								Crv.:	Curved; 
								Str.:	Straight; 
								Oth.:	Other.
							}
\newcommand{\topologyLegend}{	%
								AW:~All-wheel; 
								FA:~Front axle; 
								RA:~Rear axle; 
								-I:~Wheel-individual; 
								($\cdot$):~Not actively used. 
							}
\newcommand{\controlTargetLegend}{	LaD:~Lateral dynamics;
									LoD:~Longitudinal dynamics;
									Path:~Path tracking; 
									Pose:~Temporal sequence of poses; 
									VeD:~Vehicle dynamcis; 
									YA:~Yaw angle.
								}
\newcommand{\degradationsLegend}{	Dr.:~Drive;
									Br.:~Brake;
									Steer.:~Steering;
									T.:~Tire.
									TU:~Unintended wheel torque; 
									T0:~Zero wheel torque; 
									L$\vert$S:~Locking or spinning wheel; 
									TR:~Reduced wheel torque range; 
									SR:~Reduced steering angle range; 
									SD:~Reduced steering dynamics; 
									SU:~Unintended steering angle; 
									S0:~Zero steering torque;
									BO:~Tire blowout.
								}

\newcommand{\bottomtext}{
	\begin{minipage}{\TableWidth}
		\footnotesize
		\begin{tabular}{p{\LegendWidth}p{\TableWidth-\LegendWidth-4.0\tabcolsep-\columnseparator}}
			\emph{Actuator topologies}: &\topologyLegend\\
			\emph{Degradations}: & \degradationsLegend\\
			\emph{Control targets}: &\controlTargetLegend\\
			\emph{Control techniques}: & \controlLawLegend \&:~Hierarchical control structure; +:~Parallel control structure; /:~Alternative control techniques.\\
		\emph{Experiments}: &
				HiL:	Hardware in the Loop; 
				MiL:	Model in the Loop; 
				ViL:	Vehicle in the Loop;\\		
		\emph{Reference}: &\referenceLegend			 
		\end{tabular}
	\end{minipage}
}

\newlength{\SourceWidth}
\setlength{\SourceWidth}{\widthof{\footnotesize[100]}}%
\newlength{\TopologyBrakeWidth}
\setlength{\TopologyBrakeWidth}{\widthof{\footnotesize(AW-I)}}%
\newlength{\TopologyDriveWidth}
\setlength{\TopologyDriveWidth}{\widthof{\footnotesize FA+RA-I}}%
\newlength{\TopologySteeringWidth}
\setlength{\TopologySteeringWidth}{\widthof{\footnotesize AW/FA/RA}}
\newlength{\DegradationWidth}
\setlength{\DegradationWidth}{\widthof{\rotatebox{90}{\footnotesize L$\vert$S}}}%
\newlength{\DegradationHeight}
\setlength{\DegradationHeight}{\heightof{\rotatebox{90}{BO}}}
\newlength{\ControlLawWidth}
\newlength{\ControlLawWidthTemp}
\setlength{\ControlLawWidth}{\widthof{\footnotesize SM)}}%
\setlength{\ControlLawWidthTemp}{\widthof{\footnotesize SM)}}%
\getControlLawWidth
\newlength{\ControlTargetWidth}
\setlength{\ControlTargetWidth}{\widthof{\footnotesize Target}}%
\newlength{\ExperimentWidth}
\setlength{\ExperimentWidth}{\widthof{\footnotesize MiL\&HiL}}%
\newlength{\ExperimentHeight}
\setlength{\ExperimentHeight}{\widthof{\footnotesize DLC}}%
\addtolength{\ExperimentHeight}{\totalheightof{\footnotesize Reference}}
\newlength{\ManeuverWidth}							
\setlength{\ManeuverWidth}{\widthof{\footnotesize Reference}}%
\setlength{\ManeuverWidth}{\ManeuverWidth*\ratio{1pt}{5pt}}%

\newlength{\tabcolsepStandard}
\setlength{\tabcolsepStandard}{\tabcolsep}
\setlength{\tabcolsep}{0.5\tabcolsep}
\newlength{\columnseparator}
\setlength{\columnseparator}{0.00cm}

\newlength{\AuthorWidth}
\setlength{\AuthorWidth}{\textwidth}
\addtolength{\AuthorWidth}{-\SourceWidth}
\addtolength{\AuthorWidth}{-\TopologyBrakeWidth}
\addtolength{\AuthorWidth}{-\TopologyDriveWidth}
\addtolength{\AuthorWidth}{-\TopologySteeringWidth}
\addtolength{\AuthorWidth}{-9.0\DegradationWidth}
\addtolength{\AuthorWidth}{-\ControlLawWidth}
\addtolength{\AuthorWidth}{-\ControlTargetWidth}
\addtolength{\AuthorWidth}{-\ExperimentWidth}
\addtolength{\AuthorWidth}{-5.0\ManeuverWidth}
\addtolength{\AuthorWidth}{-22.0\columnseparator}
\addtolength{\AuthorWidth}{-\columnseparator}
\newlength{\TableWidth}
\setlength{\TableWidth}{\textwidth}
\newlength{\LegendWidth}
\setlength{\LegendWidth}{\widthof{\footnotesize Control techniques:}}

\newcommand{\postProcessEmptyCells}{}
\newcommand{\controlLawLegend}{}

\newcommand{\drawliteratureoverview}[3]{%
	\begin{tikzpicture}
		\footnotesize
		
		\pgfdeclarelayer{foreground}
		\pgfdeclarelayer{background}
		\pgfsetlayers{background,foreground}
		
		\begin{pgfonlayer}{foreground}
			\tikzstyle{column 1}=[
				text width = \AuthorWidth,
				anchor=base east,
				align=right
			]	
			\tikzstyle{row 1}=[
				text depth=.5ex
			]	
			\tikzstyle{row 1 column 2}=[
				minimum width = \SourceWidth
			]	
			\tikzstyle{row 1 column 3}=[
				minimum width = \TopologyDriveWidth
			]						
			\tikzstyle{row 1 column 4}=[
				minimum width = \TopologyBrakeWidth
			]						
			\tikzstyle{row 1 column 5}=[
				minimum width = \TopologySteeringWidth
			]
			\tikzstyle{row 1 column 6}=[
				minimum width = \DegradationWidth
			]
			\tikzstyle{row 1 column 7}=[
				minimum width = \DegradationWidth
			]
			\tikzstyle{row 1 column 8}=[
				minimum width = \DegradationWidth
			]
			\tikzstyle{row 1 column 9}=[
				minimum width = \DegradationWidth
			]
			\tikzstyle{row 1 column 10}=[
				minimum width = \DegradationWidth
			]
			\tikzstyle{row 1 column 11}=[
				minimum width = \DegradationWidth
			]
			\tikzstyle{row 1 column 12}=[
				minimum width = \DegradationWidth
			]
			\tikzstyle{row 1 column 13}=[
				minimum width = \DegradationWidth
			]
			\tikzstyle{row 1 column 14}=[
				minimum width = \DegradationWidth
			]
			\tikzstyle{row 1 column 15}=[
				minimum width = \ControlTargetWidth
			]
			\tikzstyle{row 1 column 16}=[
				minimum width = \ControlLawWidth
			]
			\tikzstyle{row 1 column 17}=[
				minimum width = \ExperimentWidth
			]
			\tikzstyle{row 1 column 18}=[
				minimum width = \ManeuverWidth
			]
			\tikzstyle{row 1 column 19}=[
				minimum width = \ManeuverWidth
			]
			\tikzstyle{row 1 column 20}=[
				minimum width = \ManeuverWidth
			]
			\tikzstyle{row 1 column 21}=[
				minimum width = \ManeuverWidth
			]
			\tikzstyle{row 1 column 22}=[
				minimum width = \ManeuverWidth
			]
			\matrix (M) [	
				inner sep=0cm,
				matrix of nodes,
				nodes in empty cells,
				column sep={\columnseparator},
				row sep = 0.01em,
				nodes={	
					inner sep = 0, 
					outer sep = 0,
					anchor=base,
				},
				store number of columns in=\colcount,
				store number of rows in=\rowcount
			]{	
				\input{#1}\\
			};%
		\end{pgfonlayer}
		
		\begin{pgfonlayer}{background}
			\pgfmathtruncatemacro\rowcount{\rowcount-1}

			\foreach \i in {6,...,14,18,19,20,...,\colcount}
			{
				\draw (M-2-\i|-M-2-1.east)--(M-\rowcount-\i|-M-\rowcount-1.east);
			}
			\foreach \i in {2,...,\rowcount}
			{
				\draw (M-\i-1.east)--(M-\i-1-|M-\i-2.west);
				\draw (M-\i-1.east-|M-\i-2.east)--(M-\i-1-|M-\i-3.west);
				\draw (M-\i-1.east-|M-\i-3.east)--(M-\i-1-|M-\i-4.west);
				\draw (M-\i-1.east-|M-\i-4.east)--(M-\i-1-|M-\i-5.west);
				\draw (M-\i-1.east-|M-\i-5.east)--(M-\i-1-|M-\i-15.west);
				\draw (M-\i-1.east-|M-\i-15.east)--(M-\i-1-|M-\i-16.west);
				\draw (M-\i-1.east-|M-\i-16.east)--(M-\i-1-|M-\i-17.west);
				\draw (M-\i-1.east-|M-\i-17.east)--(M-\i-1-|M-\i-\colcount.center);
			}
			\draw (M-2-1.east-|M-2-\colcount.west)-|(M-2-1.south-|M-2-\colcount.south);
			\draw (M-\rowcount-1.east-|M-\rowcount-\colcount.west)-|(M-\rowcount-1.north-|M-\rowcount-\colcount.north);

			\draw ($(M-1-3.north west)+(0.075cm,0.025cm)$)--($(M-1-5.north east)+(-0.075cm,0.025cm)$)			node [midway, above=0.1cm, anchor=base,inner sep=0,outer sep=0] (topo){Actuator topology};
			\draw ($(M-1-6.north west)+(0.075cm,0.025cm)$)--($(M-1-9.north east)+(-0.075cm,0.025cm)$)			node [midway, above=0.1cm, anchor=base,inner sep=0,outer sep=0] (drive){Dr.\&Br.};		
			\draw ($(M-1-10.north west)+(0.075cm,0.025cm)$)--($(M-1-13.north east)+(-0.075cm,0.025cm)$)			node [midway, above=0.1cm, anchor=base,inner sep=0,outer sep=0] (steering){Steer.};	
			\draw ($(M-1-14.north west)+(0.075cm,0.025cm)$)--($(M-1-14.north east)+(-0.075cm,0.025cm)$)			node [midway, above=0.1cm, anchor=base,inner sep=0,outer sep=0] (tire){T.};	
			\draw ($(drive.north west)+(0.075cm,0.025cm)$)--($(M-1-14.east|-drive.north)+(-0.075cm,0.025cm)$)  	node [midway, above=0.1cm, anchor=base,inner sep=0,outer sep=0] {Degradation};		
			\draw ($(M-1-15.north west)+(0.075cm,0.025cm)$)--($(M-1-16.north east)+(-0.075cm,0.025cm)$) 		node [midway, above=0.1cm, anchor=base,inner sep=0,outer sep=0] {Control};
			\draw ($(M-1-18.north west)+(0.075cm,+0.025cm)$)--($(M-1-18.north-|M-1-\colcount.east)+(-0.075cm,0.025cm)$)node [midway, above=0.1cm, anchor=base,inner sep=0,outer sep=0] (reference){Reference};		
			\draw ($(M-1-17.west|-reference.north)+(0.075cm,+0.025cm)$)--($(reference.north east)+(-0.075cm,0.025cm)$)node [midway, above=0.1cm, anchor=base,inner sep=0,outer sep=0] (maneuver){Experiment};		
			\draw [line width=\heavyrulewidth]($(maneuver.north-|M-1-\colcount.east)+(0,0.025cm)$)--($(M-1-1.west|-maneuver.north)+(0,0.025cm)$);
			\draw [line width=\lightrulewidth]($(M-1-1.south west)+(0,-0.025cm)$)--($(M-1-\colcount.east|-M-1-1.south)+(0,-0.025cm)$);
			\draw [line width=\lightrulewidth](M-\rowcount-1.south west)--(M-\rowcount-1.south-|M-1-\colcount.east);
			\input{#2}
			\node at (M-\rowcount-1.south west) [anchor=north west, inner sep = 0cm,outer sep =0](key){\bottomtext} ;
			\draw [line width=\heavyrulewidth](key.south west)--(key.south-|M-1-\colcount.east);
			
			\postProcessEmptyCells
		
		\end{pgfonlayer}
	\end{tikzpicture}
}

\newcommand{\drawliteratureoverviewtire}[3]{%
	\renewcommand{\degradationsLegend}{
		RP:		Reduced Pressure;
		BO:		Blowout.
	}	
	\setlength{\DegradationWidth}{\widthof{Degradation}}%
	\addtolength{\DegradationWidth}{-0.5\DegradationWidth}
	\setlength{\TopologyWidthWide}{\widthof{\footnotesize AW/FA/RA}}%
	\setlength{\AuthorWidth}{\textwidth}
	\addtolength{\AuthorWidth}{-\SourceWidth}
	\addtolength{\AuthorWidth}{-2.0\TopologyWidth}
	\addtolength{\AuthorWidth}{-1.0\TopologyWidthWide}
	\addtolength{\AuthorWidth}{-2.0\DegradationWidth}
	\setlength{\ControlLawWidth}{\widthof{\footnotesize AW}}
	\setlength{\ControlLawWidthTemp}{\widthof{\footnotesize AW}}
	\getControlLawWidthTire
	\addtolength{\AuthorWidth}{-\ControlLawWidth}
	\addtolength{\AuthorWidth}{-\ControlTargetWidth}
	\addtolength{\AuthorWidth}{-\ExperimentWidth}
	\addtolength{\AuthorWidth}{-5.0\ManeuverWidth}
	\addtolength{\AuthorWidth}{-15.0\columnseparator}
	\addtolength{\AuthorWidth}{-\columnseparator}
	\begin{tikzpicture}
		\footnotesize
		
		\pgfdeclarelayer{foreground}
		\pgfdeclarelayer{background}
		\pgfsetlayers{background,foreground}
		
		\begin{pgfonlayer}{foreground}
			\tikzstyle{column 1}=[
				text width = \AuthorWidth,
				anchor=base east,
				align=right
			]	
			\tikzstyle{row 1 column 2}=[
				minimum width = \SourceWidth
			]	
			\tikzstyle{row 1 column 3}=[
				minimum width = \TopologyWidth
			]						
			\tikzstyle{row 1 column 4}=[
				minimum width = \TopologyWidth
			]						
			\tikzstyle{row 1 column 5}=[
				minimum width = \TopologyWidthWide
			]
			\tikzstyle{row 1 column 6}=[
				minimum width = \DegradationWidth
			]
			\tikzstyle{row 1 column 7}=[
				minimum width = \DegradationWidth
			]
			\tikzstyle{row 1 column 8}=[
				minimum width = \ControlTargetWidth
			]
			\tikzstyle{row 1 column 9}=[
				minimum width = \ControlLawWidth
			]
			\tikzstyle{row 1 column 10}=[
				minimum width = \ExperimentWidth
			]
			\tikzstyle{row 1 column 11}=[
				minimum width = \ManeuverWidth
			]
			\tikzstyle{row 1 column 12}=[
				minimum width = \ManeuverWidth
			]
			\tikzstyle{row 1 column 13}=[
				minimum width = \ManeuverWidth
			]
			\tikzstyle{row 1 column 14}=[
				minimum width = \ManeuverWidth
			]
			\tikzstyle{row 1 column 15}=[
				minimum width = \ManeuverWidth
			]

			\matrix (M) [	
			inner sep=0cm,
			matrix of nodes,
			nodes in empty cells,
			column sep={\columnseparator},
			row sep = 0.01em,
			nodes={	
				inner sep = 0, 
				outer sep = 0,
				anchor=base
			},
			store number of columns in=\colcount,
			store number of rows in=\rowcount
			]{	
				\input{#1}\\
			};%
		\end{pgfonlayer}
		
		\begin{pgfonlayer}{background}
			\pgfmathtruncatemacro\rowcount{\rowcount-1}

			\foreach \i in {6,7,11,12,...,\colcount}
			{
				\draw (M-2-\i|-M-2-1.east)--(M-\rowcount-\i|-M-\rowcount-1.east);
			}
			\foreach \i in {2,...,\rowcount}
			{
				\draw (M-\i-1.east)				--(M-\i-1-|M-\i-2.west);
				\draw (M-\i-1.east-|M-\i-2.east)--(M-\i-1-|M-\i-3.west);
				\draw (M-\i-1.east-|M-\i-3.east)--(M-\i-1-|M-\i-4.west);
				\draw (M-\i-1.east-|M-\i-4.east)--(M-\i-1-|M-\i-5.west);
				\draw (M-\i-1.east-|M-\i-5.east)--(M-\i-1-|M-\i-8.west);
				\draw (M-\i-1.east-|M-\i-8.east)--(M-\i-1-|M-\i-9.west);
				\draw (M-\i-1.east-|M-\i-9.east)--(M-\i-1-|M-\i-10.west);
				\draw (M-\i-1.east-|M-\i-10.east)--(M-\i-1-|M-\i-\colcount.center);
			}
			\draw (M-2-1.east-|M-2-\colcount.west)-|(M-2-1.south-|M-2-\colcount.south);
			\draw (M-\rowcount-1.east-|M-\rowcount-\colcount.west)-|(M-\rowcount-1.north-|M-\rowcount-\colcount.north);

			\draw ($(M-1-3.north west) +(0.075cm,0.025cm)$)--($(M-1-5.north east)+(-0.075cm,0.025cm)$)			node [midway, above=0.1cm, anchor=base,inner sep=0,outer sep=0] (topo){Actuator topology};
			\draw ($(M-1-6.north west)+(0.075cm,0.025cm)$)--($(M-1-7.north east)+(-0.075cm,0.025cm)$) 			node [midway, above=0.1cm, anchor=base,inner sep=0,outer sep=0] {Degradation};
			\draw ($(M-1-8.north west)+(0.075cm,0.025cm)$)--($(M-1-9.north east)+(-0.075cm,0.025cm)$)  			node [midway, above=0.1cm, anchor=base,inner sep=0,outer sep=0] {Control};		
			\draw ($(M-1-11.north west)+(0.075cm,+0.025cm)$)--($(M-1-11.north-|M-1-15.east)+(-0.075cm,0.025cm)$)node [midway, above=0.1cm, anchor=base,inner sep=0,outer sep=0] (maneuver){Reference};		
			\draw [line width=\heavyrulewidth]($(maneuver.north-|M-1-\colcount.east)+(0,0.025cm)$)--($(M-1-1.west|-maneuver.north)+(0,0.025cm)$);
			\draw [line width=\lightrulewidth](M-1-1.south west)--(M-1-\colcount.east|-M-1-1.south);
			\draw [line width=\lightrulewidth](M-\rowcount-1.south west)--(M-\rowcount-1.south-|M-1-\colcount.east);
			\input{#2}
			\node at (M-\rowcount-1.south west) [anchor=north west, inner sep = 0cm,outer sep =0](key){\bottomtext} ;
			\draw [line width=\heavyrulewidth](key.south west)--(key.south-|M-1-\colcount.east);

			\postProcessEmptyCells
		
		\end{pgfonlayer}
	\end{tikzpicture}
}

\newcommand{\tablecaption}{Overview of fault-tolerant control approaches targeting steering, drive, brake, and tire degradations}
\begin{table*}
	\centering
	\tablecaptbc
	\caption{\tablecaption}
	\vspace{-1em}
	\label{tab:literature}		
	\begin{tikzpicture}
		\footnotesize
		
		\pgfdeclarelayer{foreground}
		\pgfdeclarelayer{background}
		\pgfsetlayers{background,foreground}
		
		\begin{pgfonlayer}{foreground}
			\tikzstyle{column 1}=[
				text width = \AuthorWidth,
				anchor=base east,
				align=right
			]	
			\tikzstyle{row 1}=[
				text depth=.5ex
			]	
			\tikzstyle{row 1 column 2}=[
				minimum width = \SourceWidth
			]	
			\tikzstyle{row 1 column 3}=[
				minimum width = \TopologyDriveWidth
			]						
			\tikzstyle{row 1 column 4}=[
				minimum width = \TopologyBrakeWidth
			]						
			\tikzstyle{row 1 column 5}=[
				minimum width = \TopologySteeringWidth
			]
			\tikzstyle{row 1 column 6}=[
				minimum width = \DegradationWidth
			]
			\tikzstyle{row 1 column 7}=[
				minimum width = \DegradationWidth
			]
			\tikzstyle{row 1 column 8}=[
				minimum width = \DegradationWidth
			]
			\tikzstyle{row 1 column 9}=[
				minimum width = \DegradationWidth
			]
			\tikzstyle{row 1 column 10}=[
				minimum width = \DegradationWidth
			]
			\tikzstyle{row 1 column 11}=[
				minimum width = \DegradationWidth
			]
			\tikzstyle{row 1 column 12}=[
				minimum width = \DegradationWidth
			]
			\tikzstyle{row 1 column 13}=[
				minimum width = \DegradationWidth
			]
			\tikzstyle{row 1 column 14}=[
				minimum width = \DegradationWidth
			]
			\tikzstyle{row 1 column 15}=[
				minimum width = \ControlTargetWidth
			]
			\tikzstyle{row 1 column 16}=[
				minimum width = \ControlLawWidth
			]
			\tikzstyle{row 1 column 17}=[
				minimum width = \ExperimentWidth
			]
			\tikzstyle{row 1 column 18}=[
				minimum width = \ManeuverWidth
			]
			\tikzstyle{row 1 column 19}=[
				minimum width = \ManeuverWidth
			]
			\tikzstyle{row 1 column 20}=[
				minimum width = \ManeuverWidth
			]
			\tikzstyle{row 1 column 21}=[
				minimum width = \ManeuverWidth
			]
			\tikzstyle{row 1 column 22}=[
				minimum width = \ManeuverWidth
			]
			\matrix (M) [	
				inner sep=0cm,
				matrix of nodes,
				nodes in empty cells,
				column sep={\columnseparator},
				row sep = 0.01em,
				nodes={	
					inner sep = 0, 
					outer sep = 0,
					anchor=base,
				},
				store number of columns in=\colcount,
				store number of rows in=\rowcount
			]{	
				\input{SteeringDriveBrake1.tab}\\
			};%
		\end{pgfonlayer}
		
		\begin{pgfonlayer}{background}
			\pgfmathtruncatemacro\rowcount{\rowcount-1}

			\foreach \i in {6,...,14,18,19,20,...,\colcount}
			{
				\draw (M-2-\i|-M-2-1.east)--(M-\rowcount-\i|-M-\rowcount-1.east);
			}
			\foreach \i in {2,...,\rowcount}
			{
				\draw (M-\i-1.east)--(M-\i-1-|M-\i-2.west);
				\draw (M-\i-1.east-|M-\i-2.east)--(M-\i-1-|M-\i-3.west);
				\draw (M-\i-1.east-|M-\i-3.east)--(M-\i-1-|M-\i-4.west);
				\draw (M-\i-1.east-|M-\i-4.east)--(M-\i-1-|M-\i-5.west);
				\draw (M-\i-1.east-|M-\i-5.east)--(M-\i-1-|M-\i-15.west);
				\draw (M-\i-1.east-|M-\i-15.east)--(M-\i-1-|M-\i-16.west);
				\draw (M-\i-1.east-|M-\i-16.east)--(M-\i-1-|M-\i-17.west);
				\draw (M-\i-1.east-|M-\i-17.east)--(M-\i-1-|M-\i-\colcount.center);
			}
			\draw (M-2-1.east-|M-2-\colcount.west)-|(M-2-1.south-|M-2-\colcount.south);
			\draw (M-\rowcount-1.east-|M-\rowcount-\colcount.west)-|(M-\rowcount-1.north-|M-\rowcount-\colcount.north);

			\draw ($(M-1-3.north west)+(0.075cm,0.025cm)$)--($(M-1-5.north east)+(-0.075cm,0.025cm)$)			node [midway, above=0.1cm, anchor=base,inner sep=0,outer sep=0] (topo){Actuator topology};
			\draw ($(M-1-6.north west)+(0.075cm,0.025cm)$)--($(M-1-9.north east)+(-0.075cm,0.025cm)$)			node [midway, above=0.1cm, anchor=base,inner sep=0,outer sep=0] (drive){Dr.\&Br.};		
			\draw ($(M-1-10.north west)+(0.075cm,0.025cm)$)--($(M-1-13.north east)+(-0.075cm,0.025cm)$)			node [midway, above=0.1cm, anchor=base,inner sep=0,outer sep=0] (steering){Steer.};	
			\draw ($(M-1-14.north west)+(0.075cm,0.025cm)$)--($(M-1-14.north east)+(-0.075cm,0.025cm)$)			node [midway, above=0.1cm, anchor=base,inner sep=0,outer sep=0] (tire){T.};	
			\draw ($(drive.north west)+(0.075cm,0.025cm)$)--($(M-1-14.east|-drive.north)+(-0.075cm,0.025cm)$)  	node [midway, above=0.1cm, anchor=base,inner sep=0,outer sep=0] {Degradation};		
			\draw ($(M-1-15.north west)+(0.075cm,0.025cm)$)--($(M-1-16.north east)+(-0.075cm,0.025cm)$) 		node [midway, above=0.1cm, anchor=base,inner sep=0,outer sep=0] {Control};
			\draw ($(M-1-18.north west)+(0.075cm,+0.025cm)$)--($(M-1-18.north-|M-1-\colcount.east)+(-0.075cm,0.025cm)$)node [midway, above=0.1cm, anchor=base,inner sep=0,outer sep=0] (reference){Reference};		
			\draw ($(M-1-17.west|-reference.north)+(0.075cm,+0.025cm)$)--($(reference.north east)+(-0.075cm,0.025cm)$)node [midway, above=0.1cm, anchor=base,inner sep=0,outer sep=0] (maneuver){Experiment};		
			\draw [line width=\heavyrulewidth]($(maneuver.north-|M-1-\colcount.east)+(0,0.025cm)$)--($(M-1-1.west|-maneuver.north)+(0,0.025cm)$);
			\draw [line width=\lightrulewidth]($(M-1-1.south west)+(0,-0.025cm)$)--($(M-1-\colcount.east|-M-1-1.south)+(0,-0.025cm)$);
			\draw [line width=\lightrulewidth](M-\rowcount-1.south west)--(M-\rowcount-1.south-|M-1-\colcount.east);
			\input{SteeringDriveBrake1}
			\node at (M-\rowcount-1.south west) [anchor=north west, inner sep = 0cm,outer sep =0](key){\bottomtext} ;
			\draw [line width=\heavyrulewidth](key.south west)--(key.south-|M-1-\colcount.east);
			
			\postProcessEmptyCells
		
		\end{pgfonlayer}
	\end{tikzpicture}

\end{table*}

\begin{table*}
	\centering
	\addtocounter{table}{-1}
	\tablecapcntdtbc%
	\caption{\tablecaption}
	\vspace{-1em}
	\begin{tikzpicture}
		\footnotesize
		
		\pgfdeclarelayer{foreground}
		\pgfdeclarelayer{background}
		\pgfsetlayers{background,foreground}
		
		\begin{pgfonlayer}{foreground}
			\tikzstyle{column 1}=[
				text width = \AuthorWidth,
				anchor=base east,
				align=right
			]	
			\tikzstyle{row 1}=[
				text depth=.5ex
			]	
			\tikzstyle{row 1 column 2}=[
				minimum width = \SourceWidth
			]	
			\tikzstyle{row 1 column 3}=[
				minimum width = \TopologyDriveWidth
			]						
			\tikzstyle{row 1 column 4}=[
				minimum width = \TopologyBrakeWidth
			]						
			\tikzstyle{row 1 column 5}=[
				minimum width = \TopologySteeringWidth
			]
			\tikzstyle{row 1 column 6}=[
				minimum width = \DegradationWidth
			]
			\tikzstyle{row 1 column 7}=[
				minimum width = \DegradationWidth
			]
			\tikzstyle{row 1 column 8}=[
				minimum width = \DegradationWidth
			]
			\tikzstyle{row 1 column 9}=[
				minimum width = \DegradationWidth
			]
			\tikzstyle{row 1 column 10}=[
				minimum width = \DegradationWidth
			]
			\tikzstyle{row 1 column 11}=[
				minimum width = \DegradationWidth
			]
			\tikzstyle{row 1 column 12}=[
				minimum width = \DegradationWidth
			]
			\tikzstyle{row 1 column 13}=[
				minimum width = \DegradationWidth
			]
			\tikzstyle{row 1 column 14}=[
				minimum width = \DegradationWidth
			]
			\tikzstyle{row 1 column 15}=[
				minimum width = \ControlTargetWidth
			]
			\tikzstyle{row 1 column 16}=[
				minimum width = \ControlLawWidth
			]
			\tikzstyle{row 1 column 17}=[
				minimum width = \ExperimentWidth
			]
			\tikzstyle{row 1 column 18}=[
				minimum width = \ManeuverWidth
			]
			\tikzstyle{row 1 column 19}=[
				minimum width = \ManeuverWidth
			]
			\tikzstyle{row 1 column 20}=[
				minimum width = \ManeuverWidth
			]
			\tikzstyle{row 1 column 21}=[
				minimum width = \ManeuverWidth
			]
			\tikzstyle{row 1 column 22}=[
				minimum width = \ManeuverWidth
			]
			\matrix (M) [	
				inner sep=0cm,
				matrix of nodes,
				nodes in empty cells,
				column sep={\columnseparator},
				row sep = 0.01em,
				nodes={	
					inner sep = 0, 
					outer sep = 0,
					anchor=base,
				},
				store number of columns in=\colcount,
				store number of rows in=\rowcount
			]{	
				\input{SteeringDriveBrake2.tab}\\
			};%
		\end{pgfonlayer}
		
		\begin{pgfonlayer}{background}
			\pgfmathtruncatemacro\rowcount{\rowcount-1}

			\foreach \i in {6,...,14,18,19,20,...,\colcount}
			{
				\draw (M-2-\i|-M-2-1.east)--(M-\rowcount-\i|-M-\rowcount-1.east);
			}
			\foreach \i in {2,...,\rowcount}
			{
				\draw (M-\i-1.east)--(M-\i-1-|M-\i-2.west);
				\draw (M-\i-1.east-|M-\i-2.east)--(M-\i-1-|M-\i-3.west);
				\draw (M-\i-1.east-|M-\i-3.east)--(M-\i-1-|M-\i-4.west);
				\draw (M-\i-1.east-|M-\i-4.east)--(M-\i-1-|M-\i-5.west);
				\draw (M-\i-1.east-|M-\i-5.east)--(M-\i-1-|M-\i-15.west);
				\draw (M-\i-1.east-|M-\i-15.east)--(M-\i-1-|M-\i-16.west);
				\draw (M-\i-1.east-|M-\i-16.east)--(M-\i-1-|M-\i-17.west);
				\draw (M-\i-1.east-|M-\i-17.east)--(M-\i-1-|M-\i-\colcount.center);
			}
			\draw (M-2-1.east-|M-2-\colcount.west)-|(M-2-1.south-|M-2-\colcount.south);
			\draw (M-\rowcount-1.east-|M-\rowcount-\colcount.west)-|(M-\rowcount-1.north-|M-\rowcount-\colcount.north);

			\draw ($(M-1-3.north west)+(0.075cm,0.025cm)$)--($(M-1-5.north east)+(-0.075cm,0.025cm)$)			node [midway, above=0.1cm, anchor=base,inner sep=0,outer sep=0] (topo){Actuator topology};
			\draw ($(M-1-6.north west)+(0.075cm,0.025cm)$)--($(M-1-9.north east)+(-0.075cm,0.025cm)$)			node [midway, above=0.1cm, anchor=base,inner sep=0,outer sep=0] (drive){Dr.\&Br.};		
			\draw ($(M-1-10.north west)+(0.075cm,0.025cm)$)--($(M-1-13.north east)+(-0.075cm,0.025cm)$)			node [midway, above=0.1cm, anchor=base,inner sep=0,outer sep=0] (steering){Steer.};	
			\draw ($(M-1-14.north west)+(0.075cm,0.025cm)$)--($(M-1-14.north east)+(-0.075cm,0.025cm)$)			node [midway, above=0.1cm, anchor=base,inner sep=0,outer sep=0] (tire){T.};	
			\draw ($(drive.north west)+(0.075cm,0.025cm)$)--($(M-1-14.east|-drive.north)+(-0.075cm,0.025cm)$)  	node [midway, above=0.1cm, anchor=base,inner sep=0,outer sep=0] {Degradation};		
			\draw ($(M-1-15.north west)+(0.075cm,0.025cm)$)--($(M-1-16.north east)+(-0.075cm,0.025cm)$) 		node [midway, above=0.1cm, anchor=base,inner sep=0,outer sep=0] {Control};
			\draw ($(M-1-18.north west)+(0.075cm,+0.025cm)$)--($(M-1-18.north-|M-1-\colcount.east)+(-0.075cm,0.025cm)$)node [midway, above=0.1cm, anchor=base,inner sep=0,outer sep=0] (reference){Reference};		
			\draw ($(M-1-17.west|-reference.north)+(0.075cm,+0.025cm)$)--($(reference.north east)+(-0.075cm,0.025cm)$)node [midway, above=0.1cm, anchor=base,inner sep=0,outer sep=0] (maneuver){Experiment};		
			\draw [line width=\heavyrulewidth]($(maneuver.north-|M-1-\colcount.east)+(0,0.025cm)$)--($(M-1-1.west|-maneuver.north)+(0,0.025cm)$);
			\draw [line width=\lightrulewidth]($(M-1-1.south west)+(0,-0.025cm)$)--($(M-1-\colcount.east|-M-1-1.south)+(0,-0.025cm)$);
			\draw [line width=\lightrulewidth](M-\rowcount-1.south west)--(M-\rowcount-1.south-|M-1-\colcount.east);
			\input{SteeringDriveBrake2}
			\node at (M-\rowcount-1.south west) [anchor=north west, inner sep = 0cm,outer sep =0](key){\bottomtext} ;
			\draw [line width=\heavyrulewidth](key.south west)--(key.south-|M-1-\colcount.east);
			
			\postProcessEmptyCells
		
		\end{pgfonlayer}
	\end{tikzpicture}

\end{table*}

\begin{table*}
	\centering
	\addtocounter{table}{-1}
	\tablecapcntd%
	\caption{\tablecaption}
	\vspace{-1em}
	\begin{tikzpicture}
		\footnotesize
		
		\pgfdeclarelayer{foreground}
		\pgfdeclarelayer{background}
		\pgfsetlayers{background,foreground}
		
		\begin{pgfonlayer}{foreground}
			\tikzstyle{column 1}=[
				text width = \AuthorWidth,
				anchor=base east,
				align=right
			]	
			\tikzstyle{row 1}=[
				text depth=.5ex
			]	
			\tikzstyle{row 1 column 2}=[
				minimum width = \SourceWidth
			]	
			\tikzstyle{row 1 column 3}=[
				minimum width = \TopologyDriveWidth
			]						
			\tikzstyle{row 1 column 4}=[
				minimum width = \TopologyBrakeWidth
			]						
			\tikzstyle{row 1 column 5}=[
				minimum width = \TopologySteeringWidth
			]
			\tikzstyle{row 1 column 6}=[
				minimum width = \DegradationWidth
			]
			\tikzstyle{row 1 column 7}=[
				minimum width = \DegradationWidth
			]
			\tikzstyle{row 1 column 8}=[
				minimum width = \DegradationWidth
			]
			\tikzstyle{row 1 column 9}=[
				minimum width = \DegradationWidth
			]
			\tikzstyle{row 1 column 10}=[
				minimum width = \DegradationWidth
			]
			\tikzstyle{row 1 column 11}=[
				minimum width = \DegradationWidth
			]
			\tikzstyle{row 1 column 12}=[
				minimum width = \DegradationWidth
			]
			\tikzstyle{row 1 column 13}=[
				minimum width = \DegradationWidth
			]
			\tikzstyle{row 1 column 14}=[
				minimum width = \DegradationWidth
			]
			\tikzstyle{row 1 column 15}=[
				minimum width = \ControlTargetWidth
			]
			\tikzstyle{row 1 column 16}=[
				minimum width = \ControlLawWidth
			]
			\tikzstyle{row 1 column 17}=[
				minimum width = \ExperimentWidth
			]
			\tikzstyle{row 1 column 18}=[
				minimum width = \ManeuverWidth
			]
			\tikzstyle{row 1 column 19}=[
				minimum width = \ManeuverWidth
			]
			\tikzstyle{row 1 column 20}=[
				minimum width = \ManeuverWidth
			]
			\tikzstyle{row 1 column 21}=[
				minimum width = \ManeuverWidth
			]
			\tikzstyle{row 1 column 22}=[
				minimum width = \ManeuverWidth
			]
			\matrix (M) [	
				inner sep=0cm,
				matrix of nodes,
				nodes in empty cells,
				column sep={\columnseparator},
				row sep = 0.01em,
				nodes={	
					inner sep = 0, 
					outer sep = 0,
					anchor=base,
				},
				store number of columns in=\colcount,
				store number of rows in=\rowcount
			]{	
				\input{SteeringDriveBrake3.tab}\\
			};%
		\end{pgfonlayer}
		
		\begin{pgfonlayer}{background}
			\pgfmathtruncatemacro\rowcount{\rowcount-1}

			\foreach \i in {6,...,14,18,19,20,...,\colcount}
			{
				\draw (M-2-\i|-M-2-1.east)--(M-\rowcount-\i|-M-\rowcount-1.east);
			}
			\foreach \i in {2,...,\rowcount}
			{
				\draw (M-\i-1.east)--(M-\i-1-|M-\i-2.west);
				\draw (M-\i-1.east-|M-\i-2.east)--(M-\i-1-|M-\i-3.west);
				\draw (M-\i-1.east-|M-\i-3.east)--(M-\i-1-|M-\i-4.west);
				\draw (M-\i-1.east-|M-\i-4.east)--(M-\i-1-|M-\i-5.west);
				\draw (M-\i-1.east-|M-\i-5.east)--(M-\i-1-|M-\i-15.west);
				\draw (M-\i-1.east-|M-\i-15.east)--(M-\i-1-|M-\i-16.west);
				\draw (M-\i-1.east-|M-\i-16.east)--(M-\i-1-|M-\i-17.west);
				\draw (M-\i-1.east-|M-\i-17.east)--(M-\i-1-|M-\i-\colcount.center);
			}
			\draw (M-2-1.east-|M-2-\colcount.west)-|(M-2-1.south-|M-2-\colcount.south);
			\draw (M-\rowcount-1.east-|M-\rowcount-\colcount.west)-|(M-\rowcount-1.north-|M-\rowcount-\colcount.north);

			\draw ($(M-1-3.north west)+(0.075cm,0.025cm)$)--($(M-1-5.north east)+(-0.075cm,0.025cm)$)			node [midway, above=0.1cm, anchor=base,inner sep=0,outer sep=0] (topo){Actuator topology};
			\draw ($(M-1-6.north west)+(0.075cm,0.025cm)$)--($(M-1-9.north east)+(-0.075cm,0.025cm)$)			node [midway, above=0.1cm, anchor=base,inner sep=0,outer sep=0] (drive){Dr.\&Br.};		
			\draw ($(M-1-10.north west)+(0.075cm,0.025cm)$)--($(M-1-13.north east)+(-0.075cm,0.025cm)$)			node [midway, above=0.1cm, anchor=base,inner sep=0,outer sep=0] (steering){Steer.};	
			\draw ($(M-1-14.north west)+(0.075cm,0.025cm)$)--($(M-1-14.north east)+(-0.075cm,0.025cm)$)			node [midway, above=0.1cm, anchor=base,inner sep=0,outer sep=0] (tire){T.};	
			\draw ($(drive.north west)+(0.075cm,0.025cm)$)--($(M-1-14.east|-drive.north)+(-0.075cm,0.025cm)$)  	node [midway, above=0.1cm, anchor=base,inner sep=0,outer sep=0] {Degradation};		
			\draw ($(M-1-15.north west)+(0.075cm,0.025cm)$)--($(M-1-16.north east)+(-0.075cm,0.025cm)$) 		node [midway, above=0.1cm, anchor=base,inner sep=0,outer sep=0] {Control};
			\draw ($(M-1-18.north west)+(0.075cm,+0.025cm)$)--($(M-1-18.north-|M-1-\colcount.east)+(-0.075cm,0.025cm)$)node [midway, above=0.1cm, anchor=base,inner sep=0,outer sep=0] (reference){Reference};		
			\draw ($(M-1-17.west|-reference.north)+(0.075cm,+0.025cm)$)--($(reference.north east)+(-0.075cm,0.025cm)$)node [midway, above=0.1cm, anchor=base,inner sep=0,outer sep=0] (maneuver){Experiment};		
			\draw [line width=\heavyrulewidth]($(maneuver.north-|M-1-\colcount.east)+(0,0.025cm)$)--($(M-1-1.west|-maneuver.north)+(0,0.025cm)$);
			\draw [line width=\lightrulewidth]($(M-1-1.south west)+(0,-0.025cm)$)--($(M-1-\colcount.east|-M-1-1.south)+(0,-0.025cm)$);
			\draw [line width=\lightrulewidth](M-\rowcount-1.south west)--(M-\rowcount-1.south-|M-1-\colcount.east);
			\input{SteeringDriveBrake3}
			\node at (M-\rowcount-1.south west) [anchor=north west, inner sep = 0cm,outer sep =0](key){\bottomtext} ;
			\draw [line width=\heavyrulewidth](key.south west)--(key.south-|M-1-\colcount.east);
			
			\postProcessEmptyCells
		
		\end{pgfonlayer}
	\end{tikzpicture}

\end{table*}

%% file: SteeringDriveBrake1.tex
\renewcommand{\postProcessEmptyCells}{
\draw (M-2-1.east-|M-2-4.west)--(M-2-1.east-|M-2-4.east);
\draw (M-3-1.east-|M-3-4.west)--(M-3-1.east-|M-3-4.east);
\draw (M-4-1.east-|M-4-4.west)--(M-4-1.east-|M-4-4.east);
\draw (M-5-1.east-|M-5-4.west)--(M-5-1.east-|M-5-4.east);
\draw (M-6-1.east-|M-6-4.west)--(M-6-1.east-|M-6-4.east);
\draw (M-7-1.east-|M-7-3.west)--(M-7-1.east-|M-7-3.east);
\draw (M-7-1.east-|M-7-4.west)--(M-7-1.east-|M-7-4.east);
\draw (M-8-1.east-|M-8-3.west)--(M-8-1.east-|M-8-3.east);
\draw (M-9-1.east-|M-9-4.west)--(M-9-1.east-|M-9-4.east);
\draw (M-10-1.east-|M-10-3.west)--(M-10-1.east-|M-10-3.east);
\draw (M-10-1.east-|M-10-4.west)--(M-10-1.east-|M-10-4.east);
\draw (M-11-1.east-|M-11-4.west)--(M-11-1.east-|M-11-4.east);
\draw (M-12-1.east-|M-12-4.west)--(M-12-1.east-|M-12-4.east);
\draw (M-13-1.east-|M-13-4.west)--(M-13-1.east-|M-13-4.east);
\draw (M-14-1.east-|M-14-4.west)--(M-14-1.east-|M-14-4.east);
\draw (M-15-1.east-|M-15-4.west)--(M-15-1.east-|M-15-4.east);
\draw (M-16-1.east-|M-16-4.west)--(M-16-1.east-|M-16-4.east);
\draw (M-17-1.east-|M-17-4.west)--(M-17-1.east-|M-17-4.east);
\draw (M-18-1.east-|M-18-4.west)--(M-18-1.east-|M-18-4.east);
\draw (M-19-1.east-|M-19-3.west)--(M-19-1.east-|M-19-3.east);
\draw (M-21-1.east-|M-21-3.west)--(M-21-1.east-|M-21-3.east);
\draw (M-22-1.east-|M-22-4.west)--(M-22-1.east-|M-22-4.east);
\draw (M-23-1.east-|M-23-4.west)--(M-23-1.east-|M-23-4.east);
\draw (M-24-1.east-|M-24-4.west)--(M-24-1.east-|M-24-4.east);
\draw (M-25-1.east-|M-25-4.west)--(M-25-1.east-|M-25-4.east);
\draw (M-26-1.east-|M-26-3.west)--(M-26-1.east-|M-26-3.east);
\draw (M-26-1.east-|M-26-4.west)--(M-26-1.east-|M-26-4.east);
\draw (M-27-1.east-|M-27-3.west)--(M-27-1.east-|M-27-3.east);
\draw (M-28-1.east-|M-28-3.west)--(M-28-1.east-|M-28-3.east);
\draw (M-29-1.east-|M-29-4.west)--(M-29-1.east-|M-29-4.east);
\draw (M-30-1.east-|M-30-4.west)--(M-30-1.east-|M-30-4.east);
\draw (M-31-1.east-|M-31-4.west)--(M-31-1.east-|M-31-4.east);
\draw (M-32-1.east-|M-32-4.west)--(M-32-1.east-|M-32-4.east);
\draw (M-33-1.east-|M-33-4.west)--(M-33-1.east-|M-33-4.east);
\draw (M-34-1.east-|M-34-4.west)--(M-34-1.east-|M-34-4.east);
\draw (M-35-1.east-|M-35-4.west)--(M-35-1.east-|M-35-4.east);
\draw (M-36-1.east-|M-36-4.west)--(M-36-1.east-|M-36-4.east);
\draw (M-37-1.east-|M-37-3.west)--(M-37-1.east-|M-37-3.east);
\draw (M-37-1.east-|M-37-5.west)--(M-37-1.east-|M-37-5.east);
\draw (M-38-1.east-|M-38-4.west)--(M-38-1.east-|M-38-4.east);
\draw (M-39-1.east-|M-39-5.west)--(M-39-1.east-|M-39-5.east);
\draw (M-40-1.east-|M-40-4.west)--(M-40-1.east-|M-40-4.east);
\draw (M-41-1.east-|M-41-4.west)--(M-41-1.east-|M-41-4.east);
\draw (M-45-1.east-|M-45-3.west)--(M-45-1.east-|M-45-3.east);
\draw (M-46-1.east-|M-46-3.west)--(M-46-1.east-|M-46-3.east);
\draw (M-48-1.east-|M-48-4.west)--(M-48-1.east-|M-48-4.east);
\draw (M-49-1.east-|M-49-3.west)--(M-49-1.east-|M-49-3.east);
\draw (M-51-1.east-|M-51-3.west)--(M-51-1.east-|M-51-3.east);
\draw (M-52-1.east-|M-52-3.west)--(M-52-1.east-|M-52-3.east);
\draw (M-53-1.east-|M-53-4.west)--(M-53-1.east-|M-53-4.east);
\draw (M-54-1.east-|M-54-3.west)--(M-54-1.east-|M-54-3.east);
\draw (M-55-1.east-|M-55-4.west)--(M-55-1.east-|M-55-4.east);
\draw (M-56-1.east-|M-56-4.west)--(M-56-1.east-|M-56-4.east);
\draw (M-57-1.east-|M-57-4.west)--(M-57-1.east-|M-57-4.east);
\draw (M-58-1.east-|M-58-4.west)--(M-58-1.east-|M-58-4.east);
\draw (M-59-1.east-|M-59-4.west)--(M-59-1.east-|M-59-4.east);
\draw (M-60-1.east-|M-60-4.west)--(M-60-1.east-|M-60-4.east);
\draw (M-61-1.east-|M-61-4.west)--(M-61-1.east-|M-61-4.east);
\draw (M-61-1.east-|M-61-5.west)--(M-61-1.east-|M-61-5.east);
\draw (M-62-1.east-|M-62-4.west)--(M-62-1.east-|M-62-4.east);
\draw (M-63-1.east-|M-63-4.west)--(M-63-1.east-|M-63-4.east);
\draw (M-64-1.east-|M-64-4.west)--(M-64-1.east-|M-64-4.east);
\draw (M-65-1.east-|M-65-3.west)--(M-65-1.east-|M-65-3.east);
}

\renewcommand{\controlLawLegend}{
ACA:~Analytical control allocation; 
D:~Differential; 
DCA:~Direct control allocation; 
FB:~Flatness-based; 
FF:~Feed forward; 
I:~Integral; 
LQR:~Linear–quadratic regulator; 
LT:~Lyapunov theory-based; 
MFA:~Model-free adaptive control; 
MPC:~Model-predictive control; 
OCA:~Optimal control allocation; 
P:~Proportional; 
RB:~Rule-based reconfiguration; 
SF:~State feedback control; 
SM:~Sliding mode.
}

%% file: SteeringDriveBrake2.tex
\renewcommand{\postProcessEmptyCells}{
\draw (M-2-1.east-|M-2-4.west)--(M-2-1.east-|M-2-4.east);
\draw (M-3-1.east-|M-3-4.west)--(M-3-1.east-|M-3-4.east);
\draw (M-5-1.east-|M-5-4.west)--(M-5-1.east-|M-5-4.east);
\draw (M-6-1.east-|M-6-4.west)--(M-6-1.east-|M-6-4.east);
\draw (M-7-1.east-|M-7-4.west)--(M-7-1.east-|M-7-4.east);
\draw (M-8-1.east-|M-8-3.west)--(M-8-1.east-|M-8-3.east);
\draw (M-8-1.east-|M-8-4.west)--(M-8-1.east-|M-8-4.east);
\draw (M-9-1.east-|M-9-4.west)--(M-9-1.east-|M-9-4.east);
\draw (M-10-1.east-|M-10-4.west)--(M-10-1.east-|M-10-4.east);
\draw (M-11-1.east-|M-11-4.west)--(M-11-1.east-|M-11-4.east);
\draw (M-15-1.east-|M-15-4.west)--(M-15-1.east-|M-15-4.east);
\draw (M-16-1.east-|M-16-4.west)--(M-16-1.east-|M-16-4.east);
\draw (M-17-1.east-|M-17-3.west)--(M-17-1.east-|M-17-3.east);
\draw (M-18-1.east-|M-18-4.west)--(M-18-1.east-|M-18-4.east);
\draw (M-19-1.east-|M-19-4.west)--(M-19-1.east-|M-19-4.east);
\draw (M-20-1.east-|M-20-3.west)--(M-20-1.east-|M-20-3.east);
\draw (M-22-1.east-|M-22-4.west)--(M-22-1.east-|M-22-4.east);
\draw (M-23-1.east-|M-23-3.west)--(M-23-1.east-|M-23-3.east);
\draw (M-24-1.east-|M-24-3.west)--(M-24-1.east-|M-24-3.east);
\draw (M-25-1.east-|M-25-4.west)--(M-25-1.east-|M-25-4.east);
\draw (M-26-1.east-|M-26-3.west)--(M-26-1.east-|M-26-3.east);
\draw (M-27-1.east-|M-27-4.west)--(M-27-1.east-|M-27-4.east);
\draw (M-29-1.east-|M-29-4.west)--(M-29-1.east-|M-29-4.east);
\draw (M-30-1.east-|M-30-3.west)--(M-30-1.east-|M-30-3.east);
\draw (M-31-1.east-|M-31-3.west)--(M-31-1.east-|M-31-3.east);
\draw (M-34-1.east-|M-34-3.west)--(M-34-1.east-|M-34-3.east);
\draw (M-34-1.east-|M-34-4.west)--(M-34-1.east-|M-34-4.east);
\draw (M-36-1.east-|M-36-4.west)--(M-36-1.east-|M-36-4.east);
\draw (M-37-1.east-|M-37-4.west)--(M-37-1.east-|M-37-4.east);
\draw (M-38-1.east-|M-38-4.west)--(M-38-1.east-|M-38-4.east);
\draw (M-39-1.east-|M-39-4.west)--(M-39-1.east-|M-39-4.east);
\draw (M-40-1.east-|M-40-4.west)--(M-40-1.east-|M-40-4.east);
\draw (M-41-1.east-|M-41-4.west)--(M-41-1.east-|M-41-4.east);
\draw (M-42-1.east-|M-42-4.west)--(M-42-1.east-|M-42-4.east);
\draw (M-43-1.east-|M-43-4.west)--(M-43-1.east-|M-43-4.east);
\draw (M-44-1.east-|M-44-4.west)--(M-44-1.east-|M-44-4.east);
\draw (M-47-1.east-|M-47-4.west)--(M-47-1.east-|M-47-4.east);
\draw (M-48-1.east-|M-48-4.west)--(M-48-1.east-|M-48-4.east);
\draw (M-49-1.east-|M-49-4.west)--(M-49-1.east-|M-49-4.east);
\draw (M-50-1.east-|M-50-4.west)--(M-50-1.east-|M-50-4.east);
\draw (M-51-1.east-|M-51-4.west)--(M-51-1.east-|M-51-4.east);
\draw (M-52-1.east-|M-52-3.west)--(M-52-1.east-|M-52-3.east);
\draw (M-53-1.east-|M-53-4.west)--(M-53-1.east-|M-53-4.east);
\draw (M-55-1.east-|M-55-4.west)--(M-55-1.east-|M-55-4.east);
\draw (M-56-1.east-|M-56-3.west)--(M-56-1.east-|M-56-3.east);
\draw (M-57-1.east-|M-57-4.west)--(M-57-1.east-|M-57-4.east);
\draw (M-58-1.east-|M-58-4.west)--(M-58-1.east-|M-58-4.east);
\draw (M-59-1.east-|M-59-4.west)--(M-59-1.east-|M-59-4.east);
\draw (M-60-1.east-|M-60-4.west)--(M-60-1.east-|M-60-4.east);
\draw (M-61-1.east-|M-61-4.west)--(M-61-1.east-|M-61-4.east);
\draw (M-62-1.east-|M-62-4.west)--(M-62-1.east-|M-62-4.east);
\draw (M-63-1.east-|M-63-4.west)--(M-63-1.east-|M-63-4.east);
\draw (M-64-1.east-|M-64-4.west)--(M-64-1.east-|M-64-4.east);
}

\renewcommand{\controlLawLegend}{
ACA:~Analytical control allocation; 
D:~Differential; 
DCA:~Direct control allocation; 
FF:~Feed forward; 
I:~Integral; 
LQR:~Linear–quadratic regulator; 
LT:~Lyapunov theory-based; 
MFA:~Model-free adaptive control; 
MPC:~Model-predictive control; 
OCA:~Optimal control allocation; 
P:~Proportional; 
RB:~Rule-based reconfiguration; 
SF:~State feedback control; 
SM:~Sliding mode.
}

%% file: SteeringDriveBrake3.tex
\renewcommand{\postProcessEmptyCells}{
\draw (M-3-1.east-|M-3-4.west)--(M-3-1.east-|M-3-4.east);
\draw (M-4-1.east-|M-4-4.west)--(M-4-1.east-|M-4-4.east);
\draw (M-5-1.east-|M-5-4.west)--(M-5-1.east-|M-5-4.east);
\draw (M-6-1.east-|M-6-4.west)--(M-6-1.east-|M-6-4.east);
\draw (M-7-1.east-|M-7-4.west)--(M-7-1.east-|M-7-4.east);
\draw (M-8-1.east-|M-8-4.west)--(M-8-1.east-|M-8-4.east);
\draw (M-9-1.east-|M-9-3.west)--(M-9-1.east-|M-9-3.east);
\draw (M-10-1.east-|M-10-4.west)--(M-10-1.east-|M-10-4.east);
\draw (M-11-1.east-|M-11-4.west)--(M-11-1.east-|M-11-4.east);
\draw (M-12-1.east-|M-12-4.west)--(M-12-1.east-|M-12-4.east);
\draw (M-13-1.east-|M-13-4.west)--(M-13-1.east-|M-13-4.east);
\draw (M-14-1.east-|M-14-4.west)--(M-14-1.east-|M-14-4.east);
\draw (M-15-1.east-|M-15-4.west)--(M-15-1.east-|M-15-4.east);
\draw (M-16-1.east-|M-16-3.west)--(M-16-1.east-|M-16-3.east);
\draw (M-17-1.east-|M-17-4.west)--(M-17-1.east-|M-17-4.east);
\draw (M-18-1.east-|M-18-4.west)--(M-18-1.east-|M-18-4.east);
\draw (M-19-1.east-|M-19-4.west)--(M-19-1.east-|M-19-4.east);
\draw (M-20-1.east-|M-20-4.west)--(M-20-1.east-|M-20-4.east);
\draw (M-21-1.east-|M-21-4.west)--(M-21-1.east-|M-21-4.east);
\draw (M-22-1.east-|M-22-4.west)--(M-22-1.east-|M-22-4.east);
\draw (M-23-1.east-|M-23-3.west)--(M-23-1.east-|M-23-3.east);
\draw (M-23-1.east-|M-23-8.west)--(M-23-1.east-|M-23-8.east);
\draw (M-23-1.east-|M-23-9.west)--(M-23-1.east-|M-23-9.east);
\draw (M-23-1.east-|M-23-10.west)--(M-23-1.east-|M-23-10.east);
\draw (M-24-1.east-|M-24-4.west)--(M-24-1.east-|M-24-4.east);
\draw (M-24-1.east-|M-24-8.west)--(M-24-1.east-|M-24-8.east);
\draw (M-24-1.east-|M-24-9.west)--(M-24-1.east-|M-24-9.east);
\draw (M-24-1.east-|M-24-10.west)--(M-24-1.east-|M-24-10.east);
\draw (M-25-1.east-|M-25-3.west)--(M-25-1.east-|M-25-3.east);
\draw (M-25-1.east-|M-25-4.west)--(M-25-1.east-|M-25-4.east);
\draw (M-25-1.east-|M-25-8.west)--(M-25-1.east-|M-25-8.east);
\draw (M-25-1.east-|M-25-9.west)--(M-25-1.east-|M-25-9.east);
\draw (M-25-1.east-|M-25-10.west)--(M-25-1.east-|M-25-10.east);
\draw (M-26-1.east-|M-26-3.west)--(M-26-1.east-|M-26-3.east);
\draw (M-26-1.east-|M-26-4.west)--(M-26-1.east-|M-26-4.east);
\draw (M-27-1.east-|M-27-3.west)--(M-27-1.east-|M-27-3.east);
\draw (M-27-1.east-|M-27-8.west)--(M-27-1.east-|M-27-8.east);
\draw (M-27-1.east-|M-27-9.west)--(M-27-1.east-|M-27-9.east);
\draw (M-27-1.east-|M-27-10.west)--(M-27-1.east-|M-27-10.east);
\draw (M-28-1.east-|M-28-3.west)--(M-28-1.east-|M-28-3.east);
\draw (M-28-1.east-|M-28-4.west)--(M-28-1.east-|M-28-4.east);
\draw (M-28-1.east-|M-28-8.west)--(M-28-1.east-|M-28-8.east);
\draw (M-28-1.east-|M-28-9.west)--(M-28-1.east-|M-28-9.east);
\draw (M-28-1.east-|M-28-10.west)--(M-28-1.east-|M-28-10.east);
\draw (M-29-1.east-|M-29-3.west)--(M-29-1.east-|M-29-3.east);
\draw (M-29-1.east-|M-29-4.west)--(M-29-1.east-|M-29-4.east);
\draw (M-29-1.east-|M-29-8.west)--(M-29-1.east-|M-29-8.east);
\draw (M-29-1.east-|M-29-9.west)--(M-29-1.east-|M-29-9.east);
\draw (M-29-1.east-|M-29-10.west)--(M-29-1.east-|M-29-10.east);
\draw (M-30-1.east-|M-30-3.west)--(M-30-1.east-|M-30-3.east);
\draw (M-30-1.east-|M-30-8.west)--(M-30-1.east-|M-30-8.east);
\draw (M-30-1.east-|M-30-9.west)--(M-30-1.east-|M-30-9.east);
\draw (M-30-1.east-|M-30-10.west)--(M-30-1.east-|M-30-10.east);
\draw (M-31-1.east-|M-31-3.west)--(M-31-1.east-|M-31-3.east);
\draw (M-31-1.east-|M-31-4.west)--(M-31-1.east-|M-31-4.east);
\draw (M-31-1.east-|M-31-8.west)--(M-31-1.east-|M-31-8.east);
\draw (M-31-1.east-|M-31-9.west)--(M-31-1.east-|M-31-9.east);
\draw (M-31-1.east-|M-31-10.west)--(M-31-1.east-|M-31-10.east);
\draw (M-32-1.east-|M-32-4.west)--(M-32-1.east-|M-32-4.east);
\draw (M-32-1.east-|M-32-8.west)--(M-32-1.east-|M-32-8.east);
\draw (M-32-1.east-|M-32-9.west)--(M-32-1.east-|M-32-9.east);
\draw (M-32-1.east-|M-32-10.west)--(M-32-1.east-|M-32-10.east);
\draw (M-33-1.east-|M-33-3.west)--(M-33-1.east-|M-33-3.east);
\draw (M-33-1.east-|M-33-8.west)--(M-33-1.east-|M-33-8.east);
\draw (M-33-1.east-|M-33-9.west)--(M-33-1.east-|M-33-9.east);
\draw (M-33-1.east-|M-33-10.west)--(M-33-1.east-|M-33-10.east);
\draw (M-34-1.east-|M-34-3.west)--(M-34-1.east-|M-34-3.east);
\draw (M-34-1.east-|M-34-8.west)--(M-34-1.east-|M-34-8.east);
\draw (M-34-1.east-|M-34-9.west)--(M-34-1.east-|M-34-9.east);
\draw (M-34-1.east-|M-34-10.west)--(M-34-1.east-|M-34-10.east);
\draw (M-35-1.east-|M-35-4.west)--(M-35-1.east-|M-35-4.east);
\draw (M-35-1.east-|M-35-8.west)--(M-35-1.east-|M-35-8.east);
\draw (M-35-1.east-|M-35-9.west)--(M-35-1.east-|M-35-9.east);
\draw (M-35-1.east-|M-35-10.west)--(M-35-1.east-|M-35-10.east);
\draw (M-36-1.east-|M-36-3.west)--(M-36-1.east-|M-36-3.east);
\draw (M-36-1.east-|M-36-4.west)--(M-36-1.east-|M-36-4.east);
\draw (M-36-1.east-|M-36-8.west)--(M-36-1.east-|M-36-8.east);
\draw (M-36-1.east-|M-36-9.west)--(M-36-1.east-|M-36-9.east);
\draw (M-36-1.east-|M-36-10.west)--(M-36-1.east-|M-36-10.east);
\draw (M-37-1.east-|M-37-3.west)--(M-37-1.east-|M-37-3.east);
\draw (M-37-1.east-|M-37-4.west)--(M-37-1.east-|M-37-4.east);
\draw (M-37-1.east-|M-37-8.west)--(M-37-1.east-|M-37-8.east);
\draw (M-37-1.east-|M-37-9.west)--(M-37-1.east-|M-37-9.east);
\draw (M-37-1.east-|M-37-10.west)--(M-37-1.east-|M-37-10.east);
\draw (M-38-1.east-|M-38-3.west)--(M-38-1.east-|M-38-3.east);
\draw (M-38-1.east-|M-38-8.west)--(M-38-1.east-|M-38-8.east);
\draw (M-38-1.east-|M-38-9.west)--(M-38-1.east-|M-38-9.east);
\draw (M-38-1.east-|M-38-10.west)--(M-38-1.east-|M-38-10.east);
\draw (M-39-1.east-|M-39-3.west)--(M-39-1.east-|M-39-3.east);
\draw (M-39-1.east-|M-39-4.west)--(M-39-1.east-|M-39-4.east);
\draw (M-39-1.east-|M-39-8.west)--(M-39-1.east-|M-39-8.east);
\draw (M-39-1.east-|M-39-9.west)--(M-39-1.east-|M-39-9.east);
\draw (M-39-1.east-|M-39-10.west)--(M-39-1.east-|M-39-10.east);
\draw (M-40-1.east-|M-40-3.west)--(M-40-1.east-|M-40-3.east);
\draw (M-40-1.east-|M-40-8.west)--(M-40-1.east-|M-40-8.east);
\draw (M-40-1.east-|M-40-9.west)--(M-40-1.east-|M-40-9.east);
\draw (M-40-1.east-|M-40-10.west)--(M-40-1.east-|M-40-10.east);
\draw (M-41-1.east-|M-41-3.west)--(M-41-1.east-|M-41-3.east);
\draw (M-41-1.east-|M-41-8.west)--(M-41-1.east-|M-41-8.east);
\draw (M-41-1.east-|M-41-9.west)--(M-41-1.east-|M-41-9.east);
\draw (M-41-1.east-|M-41-10.west)--(M-41-1.east-|M-41-10.east);
\draw (M-42-1.east-|M-42-3.west)--(M-42-1.east-|M-42-3.east);
\draw (M-42-1.east-|M-42-8.west)--(M-42-1.east-|M-42-8.east);
\draw (M-42-1.east-|M-42-9.west)--(M-42-1.east-|M-42-9.east);
\draw (M-42-1.east-|M-42-10.west)--(M-42-1.east-|M-42-10.east);
\draw (M-43-1.east-|M-43-3.west)--(M-43-1.east-|M-43-3.east);
\draw (M-43-1.east-|M-43-4.west)--(M-43-1.east-|M-43-4.east);
\draw (M-43-1.east-|M-43-8.west)--(M-43-1.east-|M-43-8.east);
\draw (M-43-1.east-|M-43-9.west)--(M-43-1.east-|M-43-9.east);
\draw (M-43-1.east-|M-43-10.west)--(M-43-1.east-|M-43-10.east);
\draw (M-44-1.east-|M-44-4.west)--(M-44-1.east-|M-44-4.east);
\draw (M-44-1.east-|M-44-8.west)--(M-44-1.east-|M-44-8.east);
\draw (M-44-1.east-|M-44-9.west)--(M-44-1.east-|M-44-9.east);
\draw (M-44-1.east-|M-44-10.west)--(M-44-1.east-|M-44-10.east);
\draw (M-45-1.east-|M-45-3.west)--(M-45-1.east-|M-45-3.east);
\draw (M-45-1.east-|M-45-8.west)--(M-45-1.east-|M-45-8.east);
\draw (M-45-1.east-|M-45-9.west)--(M-45-1.east-|M-45-9.east);
\draw (M-45-1.east-|M-45-10.west)--(M-45-1.east-|M-45-10.east);
}

\renewcommand{\controlLawLegend}{
ACA:~Analytical control allocation; 
D:~Differential; 
DCA:~Direct control allocation; 
FF:~Feed forward; 
I:~Integral; 
LQR:~Linear–quadratic regulator; 
LT:~Lyapunov theory-based; 
MPC:~Model-predictive control; 
OCA:~Optimal control allocation; 
P:~Proportional; 
RB:~Rule-based reconfiguration; 
SF:~State feedback control; 
SM:~Sliding mode.
}

\newcommand{\drawTireSepRule}{22}

%% file: degradations.tex
\label{sec:degradations}

For implementing fault-tolerant control systems, an understanding of how components degrade due to internal faults is key. 
Hence, in this section, we highlight and unify the different ways of how researchers model degradations. %
Based on the unified understanding, we distinguish nine degradation categories, each posing comparable requirements on fault-tolerant vehicle motion control approaches. 

The ways researchers model degradations can be described by means of the general fault-tolerant control problem without feed-through, which also neglects disturbances and uncertainties:
\begin{equation*}
	\vec{\dot{x}}(t)=\vec{f}_\mathrm{f}\left(\vec{x}(t),\vec{u}_\mathrm{f}(t)\right),\quad
	\vec{y}(t)=\vec{g}\left(\vec{x}(t)\right),
\end{equation*}
where $\vec{x}\in \mathbb{R}^n$ denotes the system state vector and $\vec{y}$ the measurement output. 
$\vec{f}_\mathrm{f}$ describes the system dynamic potentially subject to faults. 
Here, the system dynamic~$\vec{f}_\mathrm{f}$ changes in the presence of tire blowouts, which form the first category and are considered in~\numPapersTire\ publications~\citeTirePapers. 

The remaining categories comprise actuator degradations and can be explained by means of the potentially degraded system input $\vec{u}_\mathrm{f}(t)\in \mathbb{R}^m$ with $\vec{u}_\mathrm{f}=(u_{\mathrm{f},1},\ldots u_{\mathrm{f},m})^\top$, where $m$ equals the number of control inputs.
With $\vec{u}_\mathrm{f}(t)$, two ways of describing actuator degradations can be distinguished.
On the one hand, researchers take a set-based perspective, which concentrates on the usable range of $\vec{u}_\mathrm{f}$. 
For the fault-free case, the usable range of control inputs is $\ubar{\vec{u}}\leq\vec{u}_\mathrm{f}\leq\bar{\vec{u}}$, where $\ubar{\vec{u}}$ and $\bar{\vec{u}}$ are the minimum and the maximum value of the control inputs.
In case of faults, the usable range for one or multiple $u_i\in u$, with $i$ indicating the $i$-th actuator, deteriorates such that $u_{\mathrm{f},i}(t)\in[\ubar{u}_{\mathrm{f},i},\bar{u}_{\mathrm{f},i}]\subset[\ubar{u}_{i},\bar{u}_{i}]$, where $\ubar{u}_{i}$ and $\bar{u}_{i}$ denote the physical actuator limitations.
Consequently, actuators are assumed to degrade by providing a reduced range.
For the singleton $\ubar{u}_{\mathrm{f},i}=\bar{u}_{\mathrm{f},i}$ follows $u_{\mathrm{f},i}(t)=\mathrm{const.}$, which is a frequently encountered degradation in literature; in particular $u_{\mathrm{f},i}(t)=0$ is often considered.

On the other hand, several researchers model actuator degradations by defining the system input~$\vec{u}_\mathrm{f}(t)$ as
\begin{equation*}
	\vec{u}_\mathrm{f}(t)=\epsilon(t)\vec{u}(t)+\vec{u}_\mathrm{+}(t),
\end{equation*}
where $\vec{u}(t)$ is the intended input, $\epsilon(t)=\diag\{\epsilon_i\}$ a diagonal matrix with $\epsilon_i\in[0,1]$ denoting a time-varying effectiveness factor, and $\vec{u}_\mathrm{+}(t)$ a time-varying additive fault. 
Common degradations from this perspective are a stuck-at fault ($\epsilon_i(t)=0$, $u_{\mathrm{+},i}(t)=\mathrm{const.}$) and especially a loss-of-effectiveness fault ($\epsilon_i(t)\in(0,1]$, $u_{\mathrm{+},i}(t)=0$).
Other degradations can be summarized as time-varying control inputs 
and can be expressed by means of $u_{\mathrm{+},i}(t)$, too. 
Still, time-varying degradations are rarely encountered in the literature covered in this survey. 
For additive faults $\vec{u}_\mathrm{+}(t)\neq0$, physical actuator limitations are mostly not taken into account explicitly. 	
Limitations are implicitly addressed in the provided experimental examples by choosing fault values that result in an actuator usage that lies in a physically realistic range yielding $u_{\mathrm{f},i}(t)\in[\ubar{u}_{\mathrm{f},i},\bar{u}_{\mathrm{f},i}]\subset[\ubar{u}_{i},\bar{u}_{i}]$. 
\todo{noch mal raussuchen, welche paper zeitlich variierende fehler haben}

It is quite obvious that both ways of describing degradations can be merged to a great extent: 
Both ways can easily model constant control inputs or stuck-at degradations. 
A reduced control input range can be related to the loss-of-effectiveness fault.
The same applies to a constant offset if one explicitly considers physical actuator limitations that bound the actuator usage, which also results in a reduced control input range. 
Only time-varying control inputs are specific to the modeling of faults using loss-of-effectiveness and additive faults. 

Based on these general considerations, we categorize the degradations tolerated by the control algorithms found in the covered literature. 
We summarize degradations of drive and brake actuators as both impact the wheel rotation and create four categories: 
\begin{enumerate}
	\item Zero torque,
	\item Reduced torque range,
	\item Unintended torque, and
	\item Spinning or locking wheel.
\end{enumerate}
In the third category, constant non-zero torques as well as time-varying torques are subsumed that are counteracting the intended vehicle motion.
The fourth category addresses the special yet -- in terms of vehicle dynamics -- interesting case of unintended drive or brake torques that exceed the tire force limits such that the wheel starts spinning or locks when braking.

Also four steering degradation categories can be distinguished in literature:
\begin{enumerate}[resume]
	\item Reduced steering angle range,
	\item Reduced steering rate,
	\item Unintended steering angle, and
	\item Zero steering torque.
\end{enumerate}
The seventh category summarizes unintended steering angles, which can be either constant or time-varying. 
After a degradation according to the last category, the wheel's motion around its $z$-axis is purely determined through the forces at the tire road contact patch, attenuated by friction and other effects in the steering system. 

\begin{figure}
	\centering
	\input{degradationsBarPlot.tex}
	\caption{Number of contributions investigating fault tolerance approaches targeting the nine different degradation categories used in this survey.}
	\label{fig:papersperdegradation}
\end{figure}
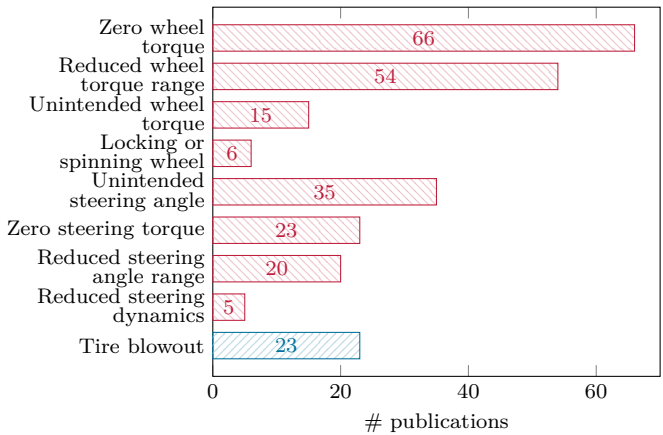

\autoref{fig:papersperdegradation} depicts the number of publications that address each of the nine categories. 
With respect to brake and drive degradations, most publications --~namely \numDegBrakeDriveTorqueZero~-- consider zero torque \citeDegBrakeDriveTorqueZero\ followed by a reduced torque range with \numDegBrakeDriveTorqueReducedRange~publications \citeDegBrakeDriveTorqueReducedRange. 
\numDegBrakeDriveTorqueConst~publications assume a constant torque \citeDegBrakeDriveTorqueConst, whereas only \numberstringnum{\numDegBrakeDriveLockingSpinning} consider locking or spinning wheels~\citeDegBrakeDriveLockingSpinning. 
Last but not least, a sole publication addresses malfunctioning anti-lock and anti-spin control systems~\cite{stolte_2018}, which is not displayed in \autoref{tab:literature}.

Steering degradations are in general less often considered compared to drive and brake degradations.
Fixed steering angles are covered in \numDegSteeringFixed\ publications~\citeDegSteeringFixed. 
\numDegSteeringFree~publications include a free-running steering~\citeDegSteeringFree \todo{hier noch mal gucken, wie realistisch das Lenkverhalten simuliert wird}. 
Fewer publications consider a reduced steering angle range~(\numDegSteeringReducedRange)~\citeDegSteeringReducedRange\ or steering rates~(\numDegSteeringReducedDynamics)~\citeDegSteeringReducedDynamics.

Overall, researchers expect that there are more or less defined ways of how actuators degrade. 
That is, they implicitly assume certain actuator-internal mechanisms that lead to the defined behavior, \eg, a drive motor that is torque-free after a fault.
In contrast to this general black box assumption, a few publications take into account the internal dynamics of an actuator and model the temporal behavior of the actuator during a degradation. 
For instance, \citet{chen_2014a,nguyen_2017} propose approaches to handle drive degradations caused by bearing faults deteriorating the available drive torque range. 
Other examples are \citet{ringdorfer_2013} and \citet{jonasson_2008} who consider the speed-dependent drag torque of synchronous motors caused by an electrical fault of the inverter bridge as a time-varying wheel torque offset in their control schemes.

%% file: degradationsBarPlot.tex
\begin{tikzpicture}
	\footnotesize
	\begin{axis}[
		    xbar stacked,
		    width = .85\linewidth,
		    xlabel={\footnotesize \# publications},
		    symbolic y coords={ 
						    	deg0, %
								deg6, %
		    					deg5, %
		    			    	deg8, %
		    					deg7, %
		    					deg3, %
		    					deg1, %
		    					deg4, %
		    					deg2  %
	    					},
		    ytick={	
					    		deg0, %
						    	deg6, %
						    	deg5, %
						    	deg8, %
						    	deg7, %
						    	deg3, %
						    	deg1, %
						    	deg4, %
						    	deg2  %
		    },
		    yticklabels={	
					    	Tire blowout,
		    				Reduced steering\\dynamics,
		    				Reduced steering\\angle range,
		    				Zero steering torque,
		    				Unintended\\steering angle,
		    				Locking or\\spinning wheel,
		    				Unintended wheel\\torque,
		    				Reduced wheel\\torque range,
		    				Zero wheel\\torque,
	    				},
		    yticklabel style={align=right,font =\footnotesize\linespread{0.8}\selectfont},
		    xticklabel style={font = \footnotesize},
		    ytick style={draw=none},
		    xmin=0,
		    xmax=70,
		    nodes near coords,
	    ]
	    \addplot+[xbar,actuatorbar,label style={font=\footnotesize,fill=, inner sep=0,outer sep=0}, tick label style={font=\footnotesize}] plot coordinates {
	    															(\numDegSteeringFree,deg8) 
							    								   	(\numDegSteeringFixed,deg7)
							    								   	(\numDegSteeringReducedDynamics,deg6)
							    								   	(\numDegSteeringReducedRange,deg5)
							    								   	(\numDegBrakeDriveTorqueReducedRange,deg4)
							    								   	(\numDegBrakeDriveLockingSpinning,deg3)
							    								   	(\numDegBrakeDriveTorqueZero,deg2)
							    								   	(\numDegBrakeDriveTorqueConst,deg1)
							    								   	(0,deg0)
						    								   	};
	   	\addplot+[xbar,blowoutbar,label style={font=\footnotesize,fill=, inner sep=0,outer sep=0}, tick label style={font=\footnotesize}] plot coordinates {
	   																(0,deg8) 
																   	(0,deg7)
																   	(0,deg6)
																   	(0,deg5)
																   	(0,deg4)
																   	(0,deg3)
																   	(0,deg2)
																   	(0,deg1)
																   	(\numPapersTire,deg0)
				    								   			};
	\end{axis}
\end{tikzpicture}

%% file: topologies.tex
\label{sec:topologies}

The effects of degradations as well as the potential to overcome these effects are strongly linked to the available actuator topology.
While compiling this survey it turned out that actuator implementations are not always designated clearly and unambiguously. 
For instance, the term ``front steering'' could refer to either one steering actuator that steers both wheels or to one steering actuator per wheel resulting in wheel-individual steering. 
Therefore, we distinguish between front-axle, rear-axle, and all-wheel topologies where all-wheel refers to separate front and rear-axle actuators.  
It is stated explicitly if wheels of an axle can be controlled individually, otherwise both wheels are affected by the same actuator.
The abbreviation scheme used in \autoref{tab:literature} as well as in \autoref{fig:paperstopologies} is influenced by the work of \citet{jonasson_2010}.

\begin{figure}
	\centering
	\input{topologiesBarPlot.tex}
	\caption{Overview of actuator topologies that are actively involved in the fault-tolerant control schemes  covered in this survey, either as degraded actuator or as actuator to compensate for the degradation
		(Steer.: Steering; 
		\protect\tikz[baseline=-\the\dimexpr\fontdimen22\textfont2\relax ]{\protect\node[anchor=base,actuatorbar] (temp) {};} publications addressing actuator degradations;
		\protect\tikz[baseline=-\the\dimexpr\fontdimen22\textfont2\relax ]{\protect\node[anchor=base,blowoutbar] (temp) {};} publications addressing tire degradations;
		AW:~All-wheel; 
		FA:~Front axle; 
		RA:~Rear axle; 
		-I:~Wheel-individual).}
	\label{fig:paperstopologies}
\end{figure}

Altogether, the available literature presents fault-tolerant control schemes for a wide range of different actuator topologies, which appear with different frequency as can be taken from \autoref{fig:paperstopologies}. 
Concentrating on single actuator types, it can be observed that the most common brake and steering implementations reflect the state of the art in series production vehicles: \numBrakeAWI\ publications involve wheel-individual all-wheel braking and \numSteeringFA\ front-axle steering. 
For steering implementations, all-wheel steering (\numSteeringAW\ publications) and wheel-individual all-wheel steering (\numSteeringAWI) are quite frequently encountered as well.
In contrast, drive implementations found in literature are more projecting into future, for wheel-individual all-wheel drive is by far the most widespread variant (\numDriveAWI\ publications).  
This implementation is frequently motivated by the potential for wheel hub or close-to-wheel motors in electric vehicles as well as the ease of implementation in simulation and control design.

The joint consideration of drive, brake, and steering implementations reveals that the most widely used actuator topology consists of wheel-individual all-wheel drive and front-axle steering~\cite{\citeDriveAWIBrakeUnknownSteeringFA}.
Publications focusing on fault-tolerant control for wheel-individual all-wheel drive only~\cite{\citeDriveAWIBrakeUnknownSteeringUnknown} are considered second most in literature followed by
wheel-individual all-wheel brake together with front-axle steering~\cite{\citeDriveAWIBrakeUnknownSteeringFA}.

In general, \autoref{fig:paperstopologies} reveals that most publications involve topologies that consist of two actuator types, namely steering actuators combined with either drive~%
\cite{
	\citeDriveAWIBrakeUnknownSteeringAWI,%
	\citeDriveAWIBrakeUnknownSteeringAW,%
	\citeDriveAWIBrakeUnknownSteeringFA,%
	\citeDriveAWIBrakeUnknownSteeringFAI,%
	\citeDriveFABrakeUnknownSteeringAW,%
	\citeDriveFABrakeUnknownSteeringFA,%
	\citeDriveFABrakeUnknownSteeringRA,%
	\citeDriveFAIBrakeUnknownSteeringFA,%
	\citeDriveFARAIBrakeUnknownSteeringAW} 
or brake actuators~\cite{
	\citeDriveUnknownBrakeAWISteeringAW,
	\citeDriveUnknownBrakeAWISteeringAWI,
	\citeDriveUnknownBrakeAWISteeringFA,
	\citeDriveUnknownBrakeAWISteeringFAI,
	\citeDriveUnknownBrakeRAISteeringFA
}.

Exceptions from this scheme are a few publications featuring topologies with all three types of actuators or only a single actuator type. 
Topologies featuring three actuator types are the most over-actuated topologies, \ie\ combining wheel-individual all-wheel drive and wheel-individual all-wheel brakes with either wheel-individual all-wheel steering~\cite{\citeDriveAWIBrakeAWISteeringAWI}\ or all-wheel steering~\cite{\citeDriveAWIBrakeAWISteeringAW}.%

Topologies with a sole actuator type are found using drive, brake, or steering in the reviewed literature. 
There are two drive-only topologies, which feature either an all-wheel drive topology~\cite{\citeDriveAWBrakeUnknownSteeringUnknown} or the wheel-individual all-wheel drive topology mentioned above. 
A brake-only topology using wheel-individual brakes at all wheels is encountered as well~\cite{\citeDriveUnknownBrakeAWISteeringUnknown}. 
For these topologies, steering angles act as system parameters or as disturbance if a steering actuators are mentioned but not actively used, \cf\ \autoref{tab:literature}.
Moreover, two different steering-only implementations are found which consist either of all-wheel steering~\cite{\citeDriveUnknownBrakeUnknownSteeringAW}\ or front-axle steering~\cite{\citeDriveUnknownBrakeUnknownSteeringFA}.
Obviously, the approaches using solely front-axle steering are used for compensating effects of tire blowouts.
For the steering-only approaches, vehicle speed serves as system parameter.

Last but not least, we would like to note that some dissertations show integrated vehicle motion control approaches applicable to different actuator topologies, which also investigate the approaches' potential for fault tolerance~\cite{patwardhan_1997,plumlee_2004a,knobel_2009,hoedt_2013}.
For these publications, \autoref{tab:literature} as well as \autoref{fig:paperstopologies} take into account those topologies that are explicitly investigated by the authors with regard to faulty actuators. \todo{hier ncoch mal gucken, ob nicht doch mehrere Varianten aufgenommen werden können}

%% file: topologiesBarPlot.tex
\newlength{\barwidth}
\setlength{\barwidth}{\heightof{\footnotesize 1234567890}}
\addtolength{\barwidth}{0.025cm}
\newlength{\bardistance}
\setlength{\bardistance}{\barwidth}
\addtolength{\bardistance}{0.1cm}

\tikzstyle{labelNodeStyleBrake} = [inner sep=0, outer sep=0, minimum width = \TopologyBrakeWidth, align = center,anchor=base, line width=0]
\tikzstyle{labelNodeStyleDrive} = [labelNodeStyleBrake, minimum width = \TopologyDriveWidth]
\pgfplotsset{topologyaxis/.style={%
		xbar stacked,
		y dir=reverse,
		width = .75\linewidth,
		y=\bardistance,
		bar width=\barwidth,
		xlabel={\footnotesize \# publications},
		yticklabel style={outer sep=0,inner sep=0,align=right,font =\footnotesize\linespread{0.8}\selectfont},
		xticklabel style={font =\footnotesize},
		ytick style={draw=none},
		xmin=0,
		xmax=\topologyXMax,
		nodes near coords,
		clip=false,
		horizontalbar
	}
}

\begin{tikzpicture}
	\footnotesize
	\begin{axis}[topologyaxis,topologyYticks,topologyYtickLabels]
	    \addplot+[xbar,actuatorbar] plot coordinates {\topologyTableData};
	    \addplot+[xbar,blowoutbar] plot coordinates {\topologyTableDataTire};
		\coordinate (topologyHeaderCols) at (axis cs:0,\upmostTopologyTick);
	\end{axis}	
	\node [inner sep=0, outer sep=0, yshift=.75em, anchor=south east, line width=0] (topologyHeaderCols) at(topologyHeaderCols) {%
		\footnotesize\linespread{0.8}\selectfont%
		\tikz[baseline]{\node[labelNodeStyleDrive] at(0,0)(){Drive};}$\vert$%
		\tikz[baseline]{\node[labelNodeStyleBrake] at(0,0)(){Brake};}$\vert$%
		\tikz[baseline]{\node[labelNodeStyleBrake] at(0,0)(){Steer.};}%
	};
	\node [anchor=south] at (topologyHeaderCols.north) (topologyHeader) {Actuator topology};
	\draw [line width=\lightrulewidth]($(topologyHeaderCols.north west)+(0.075cm,.5ex)$) -- ($(topologyHeaderCols.north east)+(-0.075cm,.5ex)$);
\end{tikzpicture}

%% file: approaches.tex
\label{sec:approaches}
In addition to the diversity of actuator topologies and permissible actuator degradations, a plethora of fault-tolerant vehicle motion control approaches exists in the covered literature. 
In order to give an overview, we examine the approaches from three perspectives in the following subsections.  
In \autoref{subsec:controltargets}, as a first perspective, we present the different control targets that researchers aim to achieve with their approaches.
To attain the control targets, researchers use a variety of control techniques, which are summarized as a second perspective in \autoref{subsec:controltechniques}.
The presented control approaches often combine different control techniques.
Therefore, as a third perspective, we give an overview of the resulting control structures in \autoref{subsec:controlstructures}. 
Additionally, \autoref{subsec:steerbydrive} highlights approaches that use brake or drive actuators to steer.

\subsection{Control targets}
\label{subsec:controltargets}

Overall, we distinguish five different control targets. 
An overview of the number of available publications for the different control targets is given in \autoref{fig:papersperctrltarget}.
With \the\numexpr\numCtrlTargetVeD+\numCtrlTargetTireVeD\relax\ publications, the control of vehicle dynamics (comprising longitudinal as well as lateral vehilce motion) is the most common control target in the body of literature considered in this survey.
These are primarily publications addressing actuator degradations~\cite{\citeCtrlTargetVeD}, while only \numberstringnum{\numCtrlTargetTireVeD} publication targets the control of vehicle dynamics in presence of tire blowouts~\cite{\citeCtrlTargetTireVeD}.
These publications are frequently motivated by human-driven cars where reference values for the vehicle dynamics are generated via the driver inputs on the steering wheel as well as the brake and the accelerator pedals. 
\begin{figure}
	\centering
	\input{controlTargetsBarPlot.tex}
	\caption{Number of papers per different control target (%
		\protect\tikz[baseline=-\the\dimexpr\fontdimen22\textfont2\relax ]{\protect\node[anchor=base,actuatorbar] (temp) {};} publications addressing actuator degradations;
		\protect\tikz[baseline=-\the\dimexpr\fontdimen22\textfont2\relax ]{\protect\node[anchor=base,blowoutbar] (temp) {};} publications addressing tire degradations).%
	}
	\label{fig:papersperctrltarget}
\end{figure}

Lateral vehicle dynamics --~\ie, yaw and lateral motion~-- are the second most encountered control target in the surveyed literature.
They are considered in \numCtrlTargetLaD\ publications addressing actuator degradations~\cite{\citeCtrlTargetLaD}\ and \numberstringnum{\numCtrlTargetTireLaD}\ addressing tire blowouts~\cite{\citeCtrlTargetTireLaD}.
In contrast, the reviewed approaches seldom target pure longitudinal dynamics~\cite{\citeCtrlTargetLoD}.

The remaining publications consider the vehicle motion in relation to an external reference.
Among these, path tracking is the biggest group, to which publications considering tire blowouts contribute a significant number with \numCtrlTargetTirePath\ publications~\cite{\citeCtrlTargetTirePath}.
Still, those considering actuator degradations form the majority in this group (\numCtrlTargetPath~publications)~\cite{\citeCtrlTargetPath}.
\todo{LoD bei Pathtracking herausarbeiten?}
Publications that target yaw angle~\cite{\citeCtrlTargetYA,\citeCtrlTargetTireYA} control and pose tracking \cite{\citeCtrlTargetPose,\citeCtrlTargetTirePose} are comparably rare.

\subsection{Control techniques}
\label{subsec:controltechniques}
\todo{wenn Hinf mit SF zusammenkommen macht Hinf.}
\todo{Noch mal alle SF durchschauen}

To reach the control targets, researchers use various control techniques, often in combination.
We categorize the techniques as displayed in \autoref{fig:papersperctrllaw}, which reveals that different techniques are encountered with varying frequency.
Of course, the techniques' implementations differ in the publications, even within one category, which are influenced \ia\ on the actuator topology, tolerated degradations, control target, and control structure. 

\begin{figure}
	\centering
	\input{controlLawsBarPlot.tex}
	\caption{Number of papers per different control techniques (%
		\protect\tikz[baseline=-\the\dimexpr\fontdimen22\textfont2\relax ]{\protect\node[anchor=base,actuatorbar] (temp) {};}~publications addressing actuator degradations;
		\protect\tikz[baseline=-\the\dimexpr\fontdimen22\textfont2\relax ]{\protect\node[anchor=base,blowoutbar] (temp) {};}~publications addressing tire degradations;
		\controlLawOverviewLegend).
	}
	\label{fig:papersperctrllaw}
\end{figure}

\emph{Classic feedback control techniques} are widely used. 
Among these, P, PI, PID, or PD controllers (summarized under PID in \autoref{fig:papersperctrllaw}) form the biggest group~\cite{\citeCtrlLawPID,\citeCtrlLawTirePID}.
To explicitly address actuator limitations, linear quadratic regulators are also frequently used~\cite{\citeCtrlLawLQR,\citeCtrlLawTireLQR}.
In comparison, standard state feedback control is encountered less often~\cite{\citeCtrlLawSF,\citeCtrlLawTireSF}.

Moreover, \emph{advanced feedback control techniques} are employed regularly. 
Here, sliding mode control is most widespread~\cite{\citeCtrlLawSM,\citeCtrlLawTireSM}. 
Further regularly used are Lyapunov theory-based techniques~\cite{\citeCtrlLawHinf,\citeCtrlLawTireHinf}, $\mathcal{H}_\infty$ control~\cite{\citeCtrlLawHinf,\citeCtrlLawTireHinf} as well as model-predictive control~\cite{\citeCtrlLawMPC,\citeCtrlLawTireMPC}. 
$\mathcal{H}_\infty$ control techniques are sometimes enhanced with other norm-optimal techniques, \eg, the $\mathcal{L}_2$-norm is used for disturbance rejection in \cite{\citeCtrlLawLtwo,\citeCtrlLawTireLtwo}.
A few publications also use fuzzy control~\cite{\citeCtrlLawFuzzy,\citeCtrlLawTireFuzzy}, flatness-based~\cite{\citeCtrlLawFB,\citeCtrlLawFB}, as well as model-free adaptive control techniques~\cite{\citeCtrlLawMFA,\citeCtrlLawTireMFA}.

The remaining techniques can be summarized as \emph{non-feedback control techniques}.  
Here, control allocation is the most often used control technique in order to distribute control inputs to the four wheels; sometimes employed as sole control technique, yet in most cases in hierarchical control structures as control layer underlying a feedback control technique.
\the\numexpr\numCtrlLawOCA+\numCtrlLawTireOCA\relax~publications solve the optimization problem underlying the control allocation online~\cite{\citeCtrlLawOCA,\citeCtrlLawTireOCA} and \the\numexpr\numCtrlLawACA+\numCtrlLawTireACA\relax~analytically~\cite{\citeCtrlLawACA,\citeCtrlLawTireACA}.
Among the latter, solutions based on pseudo-inverses are frequently encountered~\cite{\citeCtrlLawPCA,\citeCtrlLawTirePCA}.
Additionally, researchers use simpler control allocation techniques, which are subsumed as direct control allocation in \autoref{tab:literature} and \autoref{fig:papersperctrllaw}.
These techniques allocate forces either evenly among the wheels~\cite{\citeCtrlLawAvCA,\citeCtrlLawTireAvCA} or proportionally to the tire normal force distribution~\cite{\citeCtrlLawFzCA,\citeCtrlLawTireFzCA}.
In case of tire blowouts, also uneven distributions are presented in order to account for the specific effects~\cite{\citeCtrlLawTireTBCA}. 
Commonly, all direct control allocation techniques cannot directly take into account physical tire limits. 
Thus, sometimes rule-based redistribution is proposed if maximum tire forces are exceeded. 

Other non-feedback control techniques are feed forward control and rule-based reconfiguration.
Feed forward control comprise techniques that reduce control efforts of a parallel feedback controller in closed loop applications or control the desired vehicle motion in open-loop manner~\cite{\citeCtrlLawFF,\citeCtrlLawTireFF}.
In a broader sense, many of the control approaches can be described as rule-based since controller reconfiguration necessary for fault tolerance normally follows a certain set of rules.
However, we summarize those approaches as rule-based where discrete and simple reconfiguration takes place~\cite{\citeCtrlLawRB,\citeCtrlLawTireRB}, \eg, in a four-wheel individual drive application, disabling the non-faulty drive of an axle when the axle's other drive motor stops providing torque to a wheel.

\todo{Game theory based approaches: \citet{sakthivel_2018,sakthivel_2020,zhang_2020a,zhang_2020b}}
\subsection{Control structures}
\label{subsec:controlstructures}

As stated, different control techniques are often combined to reach the control targets. 
The resulting control structures are either parallel or hierarchical, sometimes a combination of both. %
Hierarchical control structures with usually two control layers are found in \the\numexpr\numCtrlStructureHierarchical+\numCtrlStructureTireHierarchical\relax\ publications~\cite{\citeCtrlStructureHierarchical,\citeCtrlStructureHierarchicalTire}.
\the\numexpr\numCtrlStructureParallel+\numCtrlStructureTireParallel\relax\ publications feature  parallel control structures~\cite{\citeCtrlStructureParallel,\citeCtrlStructureParallelTire}.
Here, either multiple controlled variables are addressed by distinct control laws or a single controlled variable is addressed by a combination of different control techniques.%
\todo{Noch mal die parallelen Strukturen angucken}
The remaining portion of the examined publications uses a sole control technique~\cite{\citeCtrlStructureSingle,\citeCtrlStructureSingleTire}.

It is worth mentioning that some publications also present the control techniques used at the actuator level. 
As we focus purely on vehicle motion control (\cf\ \autoref{sec:focus}), these are neglected in the statements of this subsection and in the overviews of \autoref{tab:literature}.
The same applies to fault estimation techniques, which are included in some publications. 
Last but not least, we would like to highlight that a few publications feature alternative control techniques on selected control layers~\cite{\citeCtrlAlternativeApproaches,\citeCtrlAlternativeApproachesTire}.

\subsection{Steering redundancy through drive or brake actuation at front axle} %
\label{subsec:steerbydrive}

In presence of degradations, the majority of control approaches controls the vehicle motion directly by means of the remaining healthy actuators. 
However, indirect approaches can be found too, primarily regarding degradations of front-axle steering whose actuator has stopped providing steering torque.
Then, these approaches control the steering angle by either wheel-individual front-axle drives~\citeSteerByDrive\ or brakes~\citeSteerByBrake\ in order to gain the desired vehicle motion.
The drive or brake actuation causes a moment around the tires' vertical axis, which is used to induce the desired steering motion --~given that certain prerequisites with regard to suspension kinematics such as a positive mechanical trail are fulfilled.
These approaches are referred to as drive or brake assisted steering as well as steer-by-drive or steer-by-brake. 

Of course, using wheel torques as backup steering mechanism influences the overall vehicle dynamics as these are strongly coupled with the dynamics of the steering system. 
Consequently, \citet{zhang_2017} investigate possible consequences of their steer-by-brake approach on longitudinal dynamics while \citet{kirli_2017} focus on demonstrating limitations for a steer-by-drive approach.

\todo{\subsubsection{Brake-by-Steering aka Snow plough brake}
	\citet{feng_2013,reinold_2010}, ggf. \citet{jansen_2010} erwähnen, da dort das Potenzial gezeigt wird
}

%% file: controlTargetsBarPlot.tex
\setlength{\barwidth}{\heightof{\footnotesize 1234567890}}
\addtolength{\barwidth}{0.025cm}
\setlength{\bardistance}{\barwidth}
\addtolength{\bardistance}{0.35cm}

\begin{tikzpicture}
	\footnotesize
	\begin{axis}[
	    xbar stacked,
	    width = .95\linewidth,
	    y=\bardistance,
	    bar width=\barwidth,
	    xlabel={\footnotesize \# publications},
	    symbolic y coords={	
	    	deg1, %
	    	deg6, %
	    	deg5, %
	    	deg2, %
	    	deg4, %
	    	deg3  %
    	},
	    ytick={	
	    	deg1, %
	    	deg6, %
	    	deg5, %
	    	deg2, %
	    	deg4, %
	    	deg3  %
    	},
	    yticklabels={	
	    				Pose tracking,
	    				Yaw angle,
	    				Longitudinal\\dynamics,
	    				Path tracking,
	    				Lateral\\dynamics,
	    				Vehicle\\dynamics
    				},
	    yticklabel style={align=right,font =\footnotesize\linespread{0.8}\selectfont},
	    xticklabel style={font = \footnotesize},
	    ytick style={draw=none},
	    xmin=0,
	    xmax=70,
	    horizontalbar
	    ]
	    \addplot+[xbar,actuatorbar] plot coordinates {
							    								   (\numCtrlTargetYA,deg6)
							    								   (\numCtrlTargetLoD,deg5)
							    								   (\numCtrlTargetLaD,deg4)
							    								   (\numCtrlTargetVeD,deg3)
							    								   (\numCtrlTargetPath,deg2)
							    								   (\numCtrlTargetPose,deg1)
						    								   	};
	   \addplot+[xbar,blowoutbar] plot coordinates {
							    								   	(\numCtrlTargetTireYA,deg6)
							    								   	(\numCtrlTargetTireLoD,deg5)
							    								   	(\numCtrlTargetTireLaD,deg4)
							    								   	(\numCtrlTargetTireVeD,deg3)
							    								   	(\numCtrlTargetTirePath,deg2)
							    								   	(\numCtrlTargetTirePose,deg1)
						    								   	};
	\end{axis}
\end{tikzpicture}

%% file: controlLawsBarPlot.tex
\setlength{\barwidth}{\heightof{\footnotesize 1234567890}}
\addtolength{\barwidth}{0.025cm}
\setlength{\bardistance}{\barwidth}
\addtolength{\bardistance}{0.05cm}

\pgfplotsset{controlLawYAxis/.style={%
		xbar stacked,
		y dir=reverse,
		width = \linewidth,
		y=\bardistance,
		bar width=\barwidth,
		xlabel={\footnotesize \# publications},
		yticklabel style={align=right,font =\footnotesize\linespread{0.8}\selectfont},
		xticklabel style={font = \footnotesize},
		ytick style={draw=none},
		xmin=0,
		xmax=45,
		horizontalbar
	}
}

\begin{tikzpicture}
	\footnotesize
	\begin{axis}[controlLawYAxis,controlLawYtickLabels]
		\addplot+[xbar,actuatorbar, every node near coord/.append style= {font=\footnotesize,fill=,inner sep=0,outer sep=0}] plot coordinates {\controlLawTable};
	   	\addplot+[xbar,blowoutbar, every node near coord/.append style= {font=\footnotesize,fill=,inner sep=0,outer sep=0}] plot coordinates {\controlLawTableTire};
	\end{axis}
\end{tikzpicture}

%% file: experiments.tex
\label{sec:experiments}

For demonstrating the suitability of fault-tolerant control approaches, experimental validation is key. 
Just as with the perspectives presented in the previous sections, the experimental setups used for validation differ considerably in the covered literature and make comparisons between approaches difficult, even when similar topologies and degradations are used. 

\subsection{Type of experiments}
A first distinguishing point is the type of experiments. 
Researchers use model-in-the-loop experiments  in the vast majority of publications~\cite{\citeExpMiL,\citeExpMiLTire}. 
Rare exceptions are publications showing hardware-in-the-loop~\cite{\citeExpHiL,\citeExpHiLTire} or even vehicle-in-the-loop experiments~\cite{\citeExpViL,\citeExpViLTire}. 

For model-in-the-loop experiments, commercial simulation environments are quite frequently employed. 
Among these, CarSim is most often used~\cite{\citeSimToolCarSim,\citeSimToolCarSimTire}, followed by 
veDYNA~\cite{\citeSimToolVeDYNA,\citeSimToolVeDYNATire}, 
CarMaker~\cite{\citeSimToolCarMaker,\citeSimToolCarMakerTire}, 
ADAMS~\cite{\citeSimToolAdams,\citeSimToolAdamsTire},
Cruise~\cite{\citeSimToolAVL,\citeSimToolAVLTire},
Dyna4~\cite{\citeSimToolDyna,\citeSimToolDynaTire}, and
AMESim~\cite{\citeSimToolAMESim,\citeSimToolAMESimTire}.
Altogether, the vehicle models' parametrizations are selected more or less individually in each publication and also depend strongly on the actuator topology as well as on the considered degradations.

\subsection{Reference trajectories}
In particular, the experiments differ in terms of the chosen reference trajectories, which are the second distinguishing point.
Although often oriented towards standard vehicle testing maneuvers, the references are specific for each publication and vary widely in speed, yaw rate, lateral, as well as longitudinal acceleration. 
As a result, they pose different challenges for vehicle motion control, especially in the presence of degradations.

Based on the standard vehicle testing maneuvers, we divide the reference trajectories into five categories to give an overview.  
Reference trajectories oriented to lane change maneuvers are often used, either in terms of single lane changes~\cite{\citeTrajSLC,\citeTrajSLCTire} or double lane changes~\cite{\citeTrajDLC,\citeTrajDLCTire}. 
Moreover, reference trajectories that are curved in only one direction are found regularly~\cite{\citeTrajCurved,\citeTrajCurvedTire}, for instance J-turns or circular drives, which cannot be sharply distinguished in the literature. 
Usually simpler to handle from a vehicle dynamics point of view, trajectories going straight are employed in~\cite{\citeTrajStraight,\citeTrajStraightTire}.
Other trajectories are of arbitrary shape~\cite{\citeTrajOther,\citeTrajOtherTire}.
Please note that the maneuvers overall are not sharply distinguishable, \eg, we categorize curved reference trajectories with large curvatures as straight since they evoke negligible lateral acceleration and yaw motion.

\todo{speed profiles extrahieren!? ggf. Path vs geschwindigkeitsprofil als Plot}

%% file: conclusion.tex
\label{sec:conclusion}

Motivated by the need for a high degree of fault tolerance in SAE level~4+ automated vehicles, we survey fault-tolerant vehicle motion control approaches targeting drive, brake, steering, and tire degradations in this paper. 
Altogether, we are surprised by the plethora of available publications, which comes along with a huge variety of different control approaches.%

Basically, the literature reveals that different actuator degradations can be handled in vehicles with varying actuator topologies even in dynamic driving situations.
This implies a high potential for exploiting functional redundancies as part of safety concepts for future series production vehicles. 
However, further research is required to support the mandatory thorough safety argumentation.
First, advanced real world testing would increase the validity when arguing real-time capability and suitability since the majority of publications uses simulation experiments.

Furthermore, explicitly highlighting the limits of an approach would contribute to the state of the art.
While good case experiments basically show that an approach is working, investigating the limits would allow a statement under which conditions a fault-tolerant control approach fails.
This would enable a comparison of the functional range that an approach can ensure with the functional range required by a specific application.
For example, we introduce a first concept to investigate systematically whether fault-tolerant vehicle motion control is suitable under normal driving conditions for a given automated vehicle application in~\cite{stolte_2019}. \todo{\citet{patil_2020}}

Investigating the functional limits could be one approach that would allow for a comparison of different fault-tolerant control approaches. 
However, a comparison based on the current state of the art, in general, is hardly possible due to the manifold of varying conditions used in the reviewed literature. 
Hence, a general framework to test different approaches under otherwise identical conditions (\ie, standardized vehicles as well as reference trajectories) would be of high value for researchers and practitioners. 
In this context, an approach for comparing trajectory tracking controllers for automated vehicles is presented by \citet{calzolari_2017}, though not considering actuator faults.
The authors also introduce a scoring mechanism, which is based on the deviation from a reference trajectory and allows an easy comparison of different control approaches. 
Such an approach could be extended for comparing different fault-tolerant vehicle motion controllers.

%% file: stolte.tex
studied Automation Technologies at Universität Lüneburg (Diplom (FH) 2008) and Electrical Engineering at Technische Universität Braunschweig (M.\,Sc. 2011). 
Since 2011 he is a research assistant at the Institute of Control Engineering of Technische Universität Braunschweig. 
Parallely, he worked in a collaboration with Porsche Engineering as functional safety engineer from 2011 to 2014. 
His research interest is safety of automated vehicles. He investigates the potential of fault-tolerant vehicle motion control towards safety of steering, brake, and drive actuators. 